\documentclass[twocolumn]{aa}  

\usepackage{graphicx}
\usepackage{mathrsfs}
\usepackage{comment}
\usepackage{array}
\usepackage{xcolor}
\newcolumntype{L}[1]{>{\raggedright\let\newline\\\arraybackslash\hspace{0pt}}m{#1}}
\newcolumntype{C}[1]{>{\centering\let\newline\\\arraybackslash\hspace{0pt}}m{#1}}
\newcolumntype{R}[1]{>{\raggedleft\let\newline\\\arraybackslash\hspace{0pt}}m{#1}}
\usepackage{txfonts}
\usepackage{amstext}

\begin{document} 

    \titlerunning{The GALAH survey: Multiple stars and our Galaxy. I}
   \title{The GALAH survey: Multiple stars and our Galaxy. I. A comprehensive method for deriving properties of FGK binary stars.}

   \subtitle{}

  \author{G. Traven \inst{1} \and 
S. Feltzing \inst{1} \and
T. Merle \inst{2} \and
M. Van der Swaelmen \inst{3} \and
K. Čotar \inst{4} \and
R. Church \inst{1} \and
T. Zwitter \inst{4} \and
Y.--S. Ting \inst{5,6,7} \and
C. Sahlholdt \inst{1} \and
M. Asplund \inst{8,9} \and
J. Bland-Hawthorn \inst{8,10} \and
G. De Silva \inst{11} \and
K. Freeman \inst{9} \and
S. Martell \inst{8,12} \and
S. Sharma \inst{8,10} \and
D. Zucker \inst{11,13} \and
S. Buder \inst{8,9,14} \and
A. Casey \inst{15,16} \and
V. D'Orazi \inst{17} \and
J. Kos \inst{4} \and
G. Lewis \inst{10} \and
J. Lin \inst{9} \and
K. Lind  \inst{14,18} \and
J. Simpson \inst{12} \and
D. Stello \inst{12} \and
U. Munari \inst{19} \and
R.~A. Wittenmyer \inst{20}
}

\institute{Lund Observatory, Department of Astronomy and Theoretical Physics, Box 43, SE-221 00 Lund, Sweden \and 
Institut d’Astronomie et d’Astrophysique, Université Libre de Bruxelles, CP. 226, Boulevard du Triomphe, 1050 Brussels, Belgium \and
INAF - Osservatorio Astrofisico di Arcetri, Largo E. Fermi, 5, I50125 Firenze, Italy \and
Faculty of Mathematics and Physics, University of Ljubljana, Jadranska 19, 1000 Ljubljana, Slovenia \and
Institute for Advanced Study, Princeton, NJ 08540, USA \and 
Department of Astrophysical Sciences, Princeton University, Princeton, NJ 08544, USA \and
Observatories of the Carnegie Institution of Washington, 813 Santa Barbara Street, Pasadena, CA 91101, USA \and
ARC Centre of Excellence for All Sky Astrophysics in Three Dimensions (ASTRO-3D), Canberra, ACT 2611, Australia \and
Research School of Astronomy \& Astrophysics, Australian National University, Canberra, ACT 2611, Australia \and
Sydney Institute for Astronomy, School of Physics, The University of Sydney, Sydney, NSW 2006, Australia \and 
Department of Physics \& Astronomy, Macquarie University, Sydney, NSW 2109, Australia \and
School of Physics, University of New South Wales, Sydney, NSW 2052, Australia \and
Research Centre in Astronomy, Astrophysics \& Astrophotonics, Macquarie University, Sydney, NSW 2109, Australia \and
Max Planck Institute for Astronomy (MPIA), Koenigstuhl 17, 69117 Heidelberg, Germany\and
Monash Centre for Astrophysics (MoCA) and School of Physics and Astronomy, Monash University, Clayton Vic 3800, Australia \and  
Faculty of Information Technology, Monash University, Clayton 3800, Victoria, Australia \and
INAF, Vicolo dell’Osservatorio, 5, 35122, Padova, PD, Italy \and
Department of Physics and Astronomy, Uppsala University, Box 516, SE-751 20 Uppsala, Sweden \and
INAF Astronomical Observatory of Padova, I-36012 Asiago (VI), Italy \and
University of Southern Queensland, Centre for Astrophysics, West Street, Toowoomba, QLD 4350 Australia
}

  \abstract
    % context heading (optional)
  % {} leave it empty if necessary  
   {Binary stellar systems form a large fraction of the Galaxy's stars. They are useful as laboratories for studying the physical processes taking place within stars, and must be correctly taken into account when observations of stars are used to study the structure and evolution of the Galaxy. The advent of large-scale spectroscopic and photometric surveys allows us to obtain large samples of binaries that permit characterising their populations.
 }
  % aims heading (mandatory)
{
        We aim to obtain a large sample of double-lined spectroscopic binaries (SB2s) by analysis of spectra from the GALAH survey in combination with photometric and astrometric data. A combined analysis will provide stellar parameters of thousands of binary stars that can be combined to form statistical observables of a given population. We aim to produce a catalogue of well-characterised systems, which can in turn be compared to models of populations of binary stars, or to follow-up individual systems of interest.
}
  % methods heading (mandatory)
   {
We obtained a list of candidate SB2 systems from a t-distributed stochastic neighbour embedding (t-SNE) classification and a cross-correlation analysis of GALAH spectra. To compute parameters of the primary and secondary star, we used a Bayesian approach that includes a parallax prior from \textit{Gaia} DR2, spectra from GALAH, and apparent magnitudes from APASS, \textit{Gaia}, 2MASS, and WISE. We used a Markov chain Monte Carlo approach to sample the posterior distributions of the following model parameters for the two stars: $T_{\rm eff[1,2]}$, $\log g_{[1,2]}$, [Fe/H], $V_{r[1,2]}$, $v_{\rm mic[1,2]}$, $v_{\rm broad[1,2]}$, $R_{[1,2]}$, and $E(B-V)$.
}
  % results heading (mandatory)
   {
We present results for 12\,760 binary stars detected as SB2s. We construct the statistical observables $T_1/T_2$, $\Delta V_r$, and $R_1/R_2$, which demonstrate that our sample mostly consists of dwarfs, with a significant fraction of evolved stars and several dozen members of the giant branch. The majority of these binary stars is concentrated at the lower boundary of the $\Delta V_r$ distribution, and the $R_1/R_2$ ratio is mostly close to unity. The derived metallicity of our binary stars is statistically lower than that of single dwarf stars from the same magnitude-limited sample. 
}
  % conclusions heading (optional), leave it empty if necessary 
   {
Our sample of binary stars represents a large population of well-characterised double-lined spectroscopic binaries that are appropriate for statistical studies of the binary populations. The derived stellar properties and their distributions show trends that are expected for a population of close binary stars (a $<$ 10 AU) detected through double lines in their spectra. Our detection technique allows us to probe binary systems with mass ratios $0.5 \leq q \leq 1$.
}


\maketitle

\section{Introduction}

Stars form in smaller or larger clusters. Some of them remain today as open clusters or loosely bound stellar associations, but the vast majority of the star-forming regions have dissolved and spread their stars into the Milky Way field. One major aim of the GALactic Archaeology with HERMES survey \citep[GALAH; ][]{2015MNRAS.449.2604D} is to quantify this process and to trace individual stars back to the regions where they originally formed \citep{2002ARA&A..40..487F,2010ApJ...713..166B}. A main tool for this is the so-called chemical tagging, where elemental abundances are used to trace stars with a common origin \citep[examples where chemical tagging has been applied, developed, and tested include][]{2007AJ....133..694D,2014MNRAS.438.2753M,2015A&A...577A..47B,2015MNRAS.450.2354Q,2015A&A...580A.111L,2016ApJ...816...10T,2018MNRAS.473.4612K,2019ApJ...871...42A,2019MNRAS.482.5302S}. To this end, efficient and robust pipelines for the analysis of large numbers of single stars have been developed \citep{2015ApJ...808...16N,2016AJ....151..144G,2018MNRAS.478.4513B,2019ApJ...879...69T}.

However, we know from  studies of the stellar populations near to the Sun that a large fraction of stars is in binaries. To understand star formation, we need to know the properties of the binary population. Until now, the binary population in the field (i.e. not in stellar clusters, where photometric displacement in the cluster Hertzsprung-Russel, HR, diagram points to binarity) has been very difficult to study, and we are uncertain about the percentage of stars that are in binary systems. The fraction could be as low as 20\% or as high as 80\%, and might also vary with stellar type \citep[see the review by][]{2013ARA&A..51..269D}. The prime targets of the large spectroscopic surveys, FGK stars, can have a binary fraction as high as 50\% \citep{2010ApJS..190....1R}. Additionally, binary population statistics are important for surveys targeting exoplanets, and they in turn contribute to understanding of binarity through serendipitous detections \citep[e.g. ][]{2016A&A...593A.133B,2020MNRAS.491.5248W}.

Input catalogues for spectroscopic surveys are built from all-sky photometric surveys \citep[see e.g.][]{2017AJ....154...94M,2016MNRAS.460.1131S,2019Msngr.175...30C,2019Msngr.175...35B}, and they therefore include many as yet undiscovered spectroscopic non-eclipsing binary systems. This opens new unbiased possibilities for studying binary stars and their population statistics across the whole Milky Way and not in the solar neighbourhood alone. Because the \textit{Gaia}-ESO Survey \citep{2012Msngr.147...25G} reaches  faint magnitudes (down to $V\sim 19$), it provides a good example of how we can extend the knowledge about the binary population far beyond the solar neighbourhood. \citet{2017A&A...608A..95M} found 185 confirmed double-lined binaries and 5 triple systems in the field populations and across the \textit{Gaia}-ESO magnitude range (amounting to $\sim$ 0.7~\% of the analysed internal data-release 4 sample). 

All of this means that to fully understand the stellar populations in the Milky Way, we also need to be able to analyse the spectra from binary systems that are observed with these large surveys. Most analyses of spectra belonging to binary systems have been made in dedicated studies with multi-epoch spectroscopy \citep[see e.g. the collection of spectroscopic binary orbits by][]{2004A&A...424..727P}. Large spectroscopic surveys re-observe only a small and random fraction of their targets, primarily for internal consistency checks, and they offer limited opportunities for multi-epoch radial velocity studies \citep[see e.g.][]{2019AJ....158..155B}. Apache Point Observatory Galactic Evolution Experiment \citep[APOGEE; ][]{2017AJ....154...94M}, and to some extent, the RAdial Velocity Experiment \citep[RAVE; ][]{2006AJ....132.1645S} and \textit{Gaia}-ESO Survey, are the only surveys that have multi-epoch spectroscopy that is rich enough for a time-series analysis \citep{2018MNRAS.476..528E,2018AJ....156...45S,2011AJ....141..200M,2017A&A...608A..95M}.
Because many surveys do not have multi-epoch spectroscopy, it becomes important to develop robust tools for analysing large sets of double-lined stellar spectra. In this study we develop such a method. Our approach is to first detect double-lined spectra and then use all available information for the associated stellar systems, such as spectra, photometry, and parallaxes, in a combined analysis. We develop a pipeline based on Bayesian inference that explicitly takes uncertainties in the observed data into account. The framework allows for the derivation of a large number of stellar properties. In this work we focus on deriving the stellar parameters, extinction, and overall metallicities. The inclusion of the parallax enables the derivation of radii of both stars in the binary. We apply our general framework to the GALAH survey.

The GALAH survey is an on-going spectroscopic survey, observing a million stars in the magnitude range 12 -- 14 in $V$ across the Southern sky. Data products from the analysis of the single-star spectra include elemental abundances for as many as 32 different elements \citep{2018MNRAS.478.4513B}. The selection of targets is simple: a magnitude cut is used, and the Galactic plane is avoided. The survey thus provides excellent opportunities for the observation of binary stars. Because the stars are relatively bright and the dwarfs therefore are nearby, they have excellent \textit{Gaia} parallaxes \citep{2018A&A...616A...1G}, which we exploit in the analysis. 

Another non-trivial task is identifying the potential binary systems in a large survey data-set before their subsequent analysis \citep[see e.g.][]{2017A&A...608A..95M}. Several approaches are possible, and we show that a combination of methods yields a larger catch of binaries. We start with using the t-distributed stochastic neighbour embedding (t-SNE) classification to detect binaries \citep[][]{2017ApJS..228...24T}, but also employ a method that uses cross-correlation with a spectral template \citep[e.g.][]{2017A&A...608A..95M}. The two approaches are highly complementary. 

This paper is organised as follows: Sect.\,\ref{sec:data} introduces the data, Sect.\,\ref{sec:method} describes the comprehensive framework that we develop, Sect.\,\ref{sec:validation} describes our validation of the method; Sect.\,\ref{sec:detection} describes and discusses our detection of binary stars in the GALAH data-set, Sect.\,\ref{sec:selection} describes the cuts that we applied to obtain our final sample, Sect.\,\ref{sec:results} describes our results in some detail, Sect.\,\ref{sec:discussion} provides the discussion, and Sect.\,\ref{sec:conclusions} contains our conclusions. A catalogue of the binaries and their derived stellar parameters is provided in Appendix\,\ref{app:cat} (the full catalogue is made publicly available on-line), and additional ancillary figures are shown in Appendix\,\ref{app:unc}.

\section{Data} \label{sec:data}

We here analyse a specific data-set: the extended GALAH data-set. This consists of stellar spectra from the GALAH survey \citep[reduced as explained in ][]{2017MNRAS.464.1259K}, apparent magnitudes from a variety of photometric catalogues (AAVSO Photometric All Sky Survey - APASS; \citealt{2016yCat.2336....0H}, \textit{Gaia} DR2; \citealt{2018A&A...616A...1G}, Two Micron All Sky Survey - 2MASS; \citealt{2006AJ....131.1163S}, Wide-field Infrared Survey Explorer - WISE; \citealt{2010AJ....140.1868W}), and the parallax measurements from \textit{Gaia} DR2. We applied a zero-point correction to \textit{Gaia} DR2 parallaxes of +0.029 mas. This correction is based on the estimate that the parallaxes are on the whole too small, following from the \textit{Gaia} investigation of quasars and validation solutions \citep[see ][]{2018A&A...616A...2L}.

Our analysis was performed on a per-spectrum basis (some targets have spectra from different epochs), therefore we chose the identifier {\em sobject\_id} (see Table~5 in \citealt{2018MNRAS.478.4513B}) as a unique reference to each observed spectrum in the GALAH survey, accompanied by corresponding information that is included in the extended GALAH data-set. Hereafter, we refer to all investigated entities identified by {\em sobject\_id} as GALAH objects, which means that repeated observations of the same target are treated independently and are compared only at the level of final products.  

\begin{table} 
\caption{GALAH spectral bands.}
\label{tab:galbands}
\centering
\begin{tabular}{l c l}
\hline\hline
Band & Wavelength region [\AA] & Dominating feature\\
\hline
1 Blue & 4718--4903 & H$\beta$\\
2 Green & 5649--5873 & \\
3 Red & 6481--6739 & H$\alpha$\\
4 IR & 7590--7890 & O~I triplet\\
\hline
\end{tabular}
\end{table}

GALAH spectra are recorded in three visual bands and in one infra-red band (see Table~\ref{tab:galbands}), with a resolution of R $\approx 28\,000$ and a typical signal-to-noise ratio (S/N) per resolution element $\sim$ 100 \citep{2018MNRAS.478.4513B}. The stellar spectra used in this study come from the GALAH internal data release 2 (iDR2). We only included spectra that successfully passed the reduction pipeline up to the point where the normalised spectrum, shifted to the rest frame, is produced (for more details, see \citealt{2017MNRAS.464.1259K} and \citealt{2018MNRAS.478.4513B}). GALAH iDR2 also includes sources targeted by the GALAH pilot \citep{2018MNRAS.476.5216D}, K2-HERMES \citep{2018AJ....155...84W}, and TESS-HERMES \citep{2018MNRAS.473.2004S} surveys, which are reduced using the same reduction pipeline as the main GALAH survey.

\section{Method} \label{sec:method}

In order to exploit all available information about the binary sources, we have designed our binary pipeline to incorporate constraints from spectroscopic, photometric, and astrometric measurements. The binary pipeline presented here is specific to the data-set we used, but it generalises to include all types of observational data and prior information. All observational data are fit simultaneously, which allows us to avoid converging to local minima. To provide a probability distribution function (PDF) for all parameters of interest, we employ Bayesian inference in tandem with a Monte Carlo Markov chain (MCMC) sampler \citep{gregory_2005,2010arXiv1008.4686H,2013PASP..125..306F,2017ARA&A..55..213S}. This approach allows us to a) define and incorporate a generative model of our data, b) apply prior information about the model parameters, c) explore a relatively large parameter space, and d) estimate the model parameters and their statistical uncertainties.

\subsection{Bayesian scheme}

In order to estimate the model parameters and their uncertainties within the binary pipeline, we used Bayes's theorem \citep[see e.g.][]{2010arXiv1008.4686H,2017ARA&A..55..213S}. We write the probability distribution of the model parameters $\theta$ (see Table~\ref{modpar}) given the data $X$ and their uncertainties $\sigma_X$, $p(\theta | X, \sigma_X)$, as
\begin{equation} \label{PDF}
p(\theta | X, \sigma_X) = \frac{p(X|\theta, \sigma_X) \cdot p(\theta)}{p(X)},
\end{equation}
where $\mathscr{L} \equiv p(X|\theta, \sigma_X)$ is the {\it \textup{likelihood}} of the parameters or the probability of the observed data given the model, $\mathscr{P} \equiv p(\theta)$ is the {\it \textup{prior}} knowledge about the model parameters, and $p(X)$ is a constant  (known as the {\it \textup{evidence}}) that ensures that the posterior probability distribution $p(\theta | X, \sigma_X)$ is normalised to unity.

\begin{table}
\caption{Model parameters $\theta$ used in the spectroscopic and photometric model of the binary pipeline (see Sects. \ref{galphot} and \ref{galspec}). Indices 1 and 2 denote the primary and secondary star in the binary system. The line broadening parameter incorporates the effects of stellar rotation and macro-turbulence (see \citealt{2018MNRAS.478.4513B}). Columns Sp. and Ph. indicate whether a parameter is included in the spectroscopic or photometric model, respectively.}
\label{modpar}
\centering
\begin{tabular}{l l L{3.4cm} c c} 
\hline\hline
Parameter & Unit & Description & Sp. & Ph.\\
\hline
$T_{\rm eff[1,2]}$ & K & effective temperature & \checkmark & \checkmark\\
$\log g_{[1,2]}$ & dex & surface gravity & \checkmark & \checkmark\\
$\mathrm{[Fe/H]}$ & dex & iron abundance & \checkmark & \checkmark\\
$V_{r[1,2]}$ & km\,s$^{-1}$ & radial velocity & \checkmark & \\
$v_{\rm mic[1,2]}$ & km\,s$^{-1}$ & micro-turbulence & \checkmark & \\
$v_{\rm broad[1,2]}$ & km\,s$^{-1}$ & line broadening & \checkmark & \\
$R_{[1,2]}$ & $R_{\odot}$ & stellar radius & & \checkmark\\
$E(B-V)$ &  & interstellar reddening & & \checkmark\\
$\varpi$ & mas & parallax & & \checkmark\\
\hline
$\eta_n$ &  & luminosity ratio (see Eq.\,\eqref{eqlum}) & & \checkmark\\
\hline
\end{tabular}
\end{table}

We assume Gaussian uncertainties for all measured quantities, and express the likelihood as the product of conditional probabilities for each data point in the following form:
\begin{equation} \label{likelihood}
\mathscr{L} = \prod_{i=1}^N p(X_i|\theta, \sigma_{X,i}) = \prod_{i=1}^N \frac{1}{\sqrt{2 \pi \sigma_{X,i}^2}} \exp \left(- \frac{\left[ X_i - \mathcal{M}(\theta)_i  \right]^2}{2 \sigma_{X,i}^2} \right),
\end{equation}
where $N$ is the number of data points and $\mathcal{M}(\theta)_i$ represents the model value for a corresponding data point $i$ given some independent variable (e.g. the wavelength in the case of spectroscopic data). 

Similarly, when we have prior informative knowledge about the model parameters $\theta$ in the form of their value and standard deviation, $\sigma$, the prior can be written as 
\begin{equation} \label{priors}
\mathscr{P} = \prod_{j=1}^M p(\theta_j) = \prod_{j=1}^M \frac{1}{\sqrt{2 \pi \sigma_{\theta,j}^2}} \exp \left(-\frac{\left[ \theta_j - \theta_{j,0} \right]^2}{2 \sigma_{\theta,j}^2} \right),
\end{equation}
where $M$ is the number of model parameters, while $\theta_{j,0}$ and $\sigma_{\theta,j}$ are the informative value and standard deviation for a given parameter $\theta_j$.

In this scenario, we seek to maximise $\mathscr{L} \cdot \mathscr{P}$, which is for computational convenience commonly implemented as maximising the natural logarithm of the same expression, yielding
\begin{equation} \label{loglikelihood}
\ln \left( \mathscr{L} \cdot \mathscr{P} \right) = K - \sum_{i=1}^N \frac{[X_i - \mathcal{M}(\theta)_i]^2}{2 \sigma_{X,i}^2} - \sum_{j=1}^M \frac{[\theta_j - \theta_{j,0}]^2}{2 \sigma_{\theta,j}^2},
\end{equation}
where $K$ contains all the terms that remain constant in the process of inference or the exploration of the parameter ($\theta$) space (the evidence term is independent of $\theta$). Eq.\,\eqref{loglikelihood} shows that maximising $\ln \left( \mathscr{L} \cdot \mathscr{P} \right)$ is in our case identical to minimising a sum of different $\chi^2$ terms.

\subsection{Binary star model and prior information} \label{sec:model}

In accordance with our philosophy of simultaneously applying information from the spectroscopic, photometric, and astrometric domain to characterise binary stars, our model of the data and prior information represent three distinct entities: 1) a model for the spectrum of the binary system, 2) a model for apparent magnitudes of the binary system, and 3) a prior on the parallax of the binary system. When we take this into account and remove the constant terms from Eq.\,\eqref{loglikelihood}, the objective function that we aim to minimise in our Bayesian inference becomes

\begin{align} 
\chi^2_{\rm obj} &= \chi^2_{\rm spec} + \chi^2_{\rm phot} + \chi^2_{\rm astro} \label{lnobj} \\ 
\chi^2_{\rm spec} &= \frac{1}{2} \sum_{i=1}^{N_{\rm spec}} \frac{[X_{{\rm spec},i} - \mathcal{M}_{{\rm spec},i}(\theta)]^2}{\sigma_{X_{{\rm spec},i}}^2} \label{lnspec} \\ 
\chi^2_{\rm phot} &= \frac{1}{2} \sum_{i=1}^{N_{\rm phot}} \frac{[X_{{\rm phot},i} - \mathcal{M}_{{\rm phot},i}(\theta)]^2}{\sigma_{X_{{\rm phot},i}}^2} \label{lnphot} \\ 
\chi^2_{\rm astro} &= \frac{1}{2} \frac{[\varpi - \varpi_0]^2}{\sigma_{\varpi}^2}. \label{allbayes2}
\end{align}

Subscripts $_{\rm spec}$, $_{\rm phot}$, and $_{\rm astro}$ refer to the spectroscopic, photometric, and astrometric parts of the objective function and also indicate the source and type of data (for details, see Sects. \ref{galphot} and \ref{galspec}). $N_{\rm spec}$ and $N_{\rm phot}$ represent the number of data points, $X_{{\rm spec},i}$ and $X_{{\rm phot},i}$ are individual values of data points, $\mathcal{M}_{{\rm spec},i}(\theta)$ and $\mathcal{M}_{{\rm phot},i}(\theta)$ are the corresponding values from the spectroscopic and photometric model, and $\sigma_{X_{{\rm spec},i}}$ and $\sigma_{X_{{\rm phot},i}}$ represent all pertaining uncertainties. The parallax $\varpi$ is one of the model parameters $\theta$ , and $\varpi_0$ and $\sigma_{\varpi}$ represent the prior information on the parallax and its standard deviation obtained from \textit{Gaia} DR2.
In an ideal situation, our models would be accurate representations of the physical systems we study, and all uncertainties pertaining to observed data would be properly accounted for. This is almost never the case, therefore the uncertainties $\sigma_{X_{\rm spec}}$, $\sigma_{X_{\rm phot}}$, and $\sigma_{\varpi}$ must be scaled appropriately if they are under- or overestimated. The under- or overestimation of uncertainties in observational data, along with any systematic biases, can be estimated in the Bayesian scheme with additional $\theta$ parameters. However, doing so simultaneously for different data-sets while also including uncertainties of the spectroscopic and photometric model can be challenging. We plan to address this problem in a future study. For now, $\sigma_{X_{\rm spec}}$, $\sigma_{X_{\rm phot}}$, and $\sigma_{\varpi}$ are included in the binary pipeline as reported in the respective sources of observational data.

\subsubsection{Priors} \label{sec:priors}

All $\theta$ parameters except for the parallax are assigned uniform or flat priors, that is, we only fixed their allowed range. The values that we chose are given in Table~\ref{galahpar}. In the absence of strong prior information for these parameters, the flat priors used here do not hinder our inference because of the strongly identifiable likelihood. The prior on parallax is Gaussian, with the mean and standard deviation given by the \textit{Gaia} DR2 parallax value and its uncertainty. 

We set a generous prior for the absolute radius of a star because this quantity directly depends on the assumed distance to the binary system given by the inverse of the sampled parallax. Because matching the synthetic and observed apparent magnitudes effectively determines the angular size of the star ($2 \cdot R/d$) through luminosity and distance consideration, a potentially erroneous value of the parallax might produce unrealistically small or large radii, even though the other parameters are well determined. Therefore, we allow for such (rare) exceptions with a generous prior on the absolute radius and advise the user to confirm the quality estimates of the parallax before using the values of absolute radii (see e.g. the RUWE parameter in Table~\ref{tab:cat}).

\begin{table} 
\caption{Prior distributions for $\theta$ parameters listed in Table \ref{modpar} when the binary pipeline is applied to GALAH objects.}
\label{galahpar}
\centering
\begin{tabular}{l c c}
\hline\hline
Parameter & Prior & Unit\\
\hline
$T_{\rm eff[1,2]}$ & 3800--7300 & K\\
$\log g_{[1,2]}$ & 0.5--5.0 & dex\\
$\mathrm{[Fe/H]}$ & -2.5--0.5 & dex\\
$V_{r[1,2]}$ & -500--500 & km\,s$^{-1}$\\
$v_{\rm mic[1,2]}$ & 0.85--2.25 & km\,s$^{-1}$\\
$v_{\rm broad[1,2]}$\tablefootmark{a} & 4--50 & km\,s$^{-1}$\\
$R_{[1,2]}$ & 0.05--100 & $R_{\odot}$\\
$E(B-V)$ & 0.0--2.0 &\\
$\varpi$ & $\mathcal{N}$($\varpi$,\,$\sigma_{\varpi}^2$) & \\
\hline
$\eta_n$ & unconstrained &  \\
\hline
\end{tabular}
\tablefoot{
\tablefoottext{a}{See end of Sect.\,\ref{galspec} for a modification of this prior when binary spectra are fitted.}
}

\end{table}

\subsubsection{Photometric model} \label{galphot}

Synthetic apparent magnitudes are obtained by convolving a model of the spectral energy distribution (SED) of the stars in the binary system with the desired passband transmission functions, taking the radius of each star and the parallax of the system into account. The SED of the binary is constructed by summing the stellar SEDs, after which reddening is applied. These computed apparent magnitudes form $\mathcal{M}_{{\rm phot},i}(\theta)$ in Eq.~\eqref{lnphot}. In addition, the SEDs are used to compute $\eta_n$ as the luminosity ratio of the two stars (contribution of light from the secondary compared to the primary star) in a binary system for each of the four GALAH bands,
\begin{equation} \label{eqlum}
\eta_n = 10^{0.4[M_{p,n} - M_{s,n}]}, \qquad n \in (1,2,3,4)
,\end{equation}
where $M_{p,n}$ and $M_{s,n}$ are absolute magnitudes of the primary and the secondary star in a given GALAH band (see Table~\ref{tab:galbands}).

\begin{figure}[!htp]
   \centering
   \includegraphics[width=\linewidth]{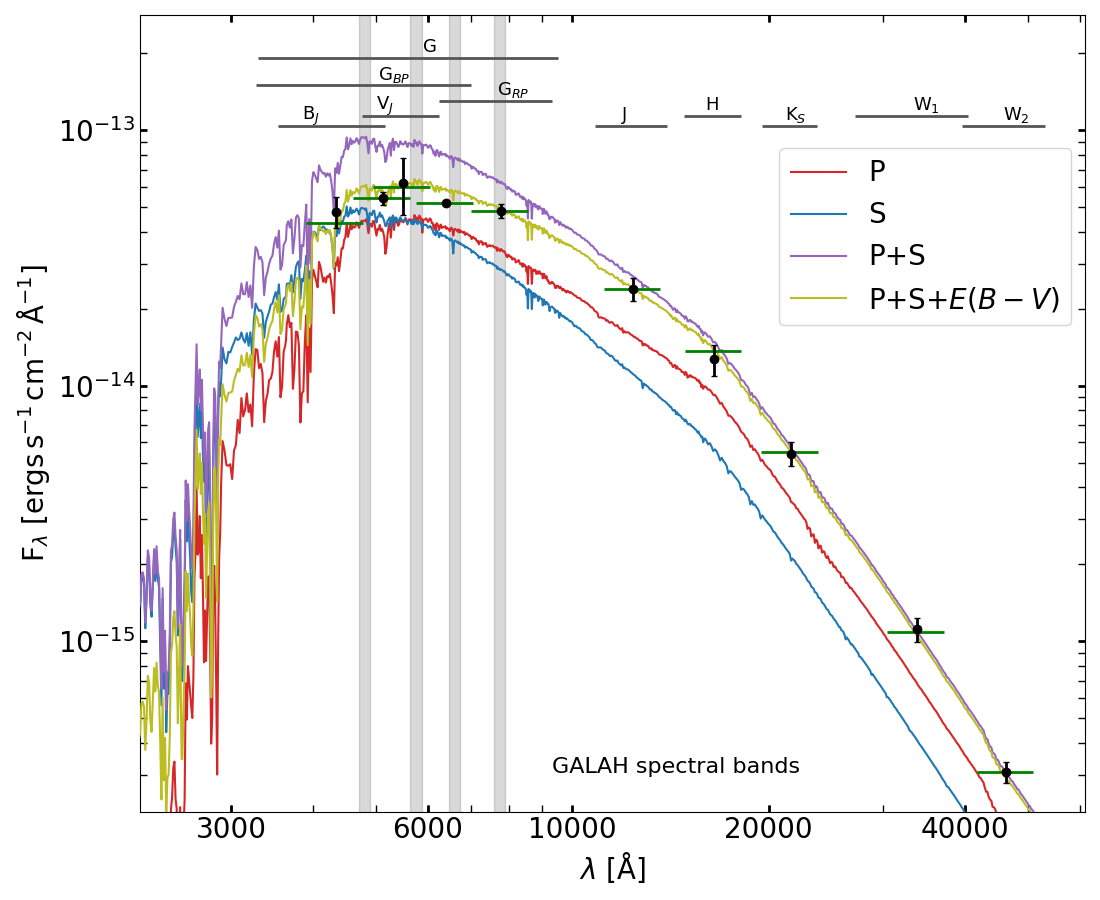}
   \caption{Spectral energy distributions for an example of a binary system ($T_{\rm eff,1} = 5018$ K, $T_{\rm eff,2} = 5715$ K, $R_1 = 1.9 R_{\odot}$, $R_2 = 1.3 R_{\odot}$, $E(B-V) = 0.13$). The legend explains which SEDs are plotted (P is the primary star, and S is the secondary star). The four GALAH spectral bands are indicated by vertical shaded regions, while the width of the transmission functions of photometric bands is indicated on the top. The black points are observed integrated fluxes in the following order of photometric bands: B$_J$, G$_{BP}$, V$_J$, G, G$_{RP}$, J, H, K$_S$, W$_1$, and W$_2$ (see Sect.~\ref{galphot}), and the short green lines indicate the synthetic integrated fluxes at the same position. The error bars of observed fluxes are exaggerated by a factor of 10 for clarity. Despite this, the uncertainty for the \textit{Gaia} G value is smaller than the point size.}
   \label{evolved}
\end{figure}

We impose a common [Fe/H] for both stars in a binary (see the discussion in Sect.~\ref{sec:diffusion}). Because both stars are on the main-sequence (MS), turn-off, or sub-giant branch, the more massive star always has the larger radius. Therefore we define the primary star as the one with the larger radius. We prefer this definition to the traditional one that is based on the luminosity in a given passband because when a star evolves away from the MS, its temperature drops and its radius increases, which may lead to a configuration in which one of the stars contributes more light at the red end of the spectrum, while the other star dominates in the blue. Figure\,\eqref{evolved} depicts such a binary system, where the secondary star contributes more light in the GALAH blue band, and less in the others. In the rare case of detecting binary systems harbouring two giants, the primary star by our definition might no longer be the more massive one because of the complex evolution on the giant branch. Results for such systems should therefore be handled with care.

The extended GALAH data-set includes observed apparent magnitudes from \textit{Gaia} DR2 and three other photometric surveys (APASS, 2MASS, and WISE) that span the wavelength range from the visible to the infrared regime. These magnitudes and their uncertainties are converted into fluxes for all computations in the binary pipeline ($X_{{\rm phot},i}, \sigma_{X_{{\rm phot},i}}$ in Eq.\,\eqref{lnphot}). The synthetic fluxes ($\mathcal{M}_{{\rm phot},i}(\theta)$ in Eq.\,\eqref{lnphot}) are based on the SEDs generated from the ATLAS9 stellar atmosphere models \citep{2004astro.ph..5087C}, using the Python packages PyPhot \citep{morgan} and PySynPhot \citep{2013ascl.soft03023S} for their construction and convolution with given passband transmission functions. 

The ATLAS9 stellar atmosphere models were chosen because they are incorporated in PySynPhot and because they agree excellently with the MARCS models for FGK stars, as discussed in \citet{2008A&A...486..951G}. The latter is especially important because our aim is that the underlying physical models that we used both in the SME spectroscopic fit of the training set (see {\it The Cannon} model below) and in the photometric model described here are the same or very similar because some of the parameters (e.g. the effective temperature) are constrained by the spectroscopic and the photometric fit. Different extinction curves are available for PySynPhot, and we adopted {\sl mwavg}, which is modelled for the Milky Way diffuse medium \citep[standard R(V) = 3.1 extinction law by][]{1989ApJ...345..245C}. 

The modelled SEDs (with $T_{\rm eff}, \log g,$ and [Fe/H] as input) were convolved with the following ten photometric passbands\footnote{Passband transmission curves in PyPhot are mostly identical to Synphot from STScI, while others come from the Spanish Virtual Observatory (SVO) and directly from the observational facilities (M. Fouesneau, private communication). \textit{Gaia} DR2 passbands are from \citet{2018A&A...616A...4E}.}: B$_J$, V$_J$, G$_{BP}$, G, G$_{RP}$, J, H, K$_S$, W$_1$, and W$_2$ from the APASS, \textit{Gaia} DR2, 2MASS, and WISE photometric catalogues, respectively. These passbands were chosen because corresponding observational data for GALAH stars is available, whose intrinsic quality is also discussed in \citet{2019arXiv190404841C}, and because their transmission curves are readily available in the PyPhot package.

\subsubsection{Spectroscopic model} \label{galspec}

In the spectroscopic model (Eq.\,\eqref{lnspec}) we used spectral templates derived from observed spectra. This brings a number of advantages compared to using synthetic spectra: the majority of spectral lines are accounted for, the effects of the instrument are embedded automatically, the resolution is identical, and the spectral parameters directly relate to the main survey results. The downside is that some of the properties are less well determined, such as the exact continuum placement, but the continuum normalisation is effectively the same for the observed and modelled spectra (templates). Additionally, some parameters may be under-sampled, for instance, the higher rotational velocities, which are common to stars in close binary systems but not to their single counterparts. 

The main GALAH pipeline incorporates {\it The Cannon} \citep{2015ApJ...808...16N}, a data-driven interpolation method for propagating astrophysical information within the same observed data-set of stellar spectra. For consistency with the main pipeline, we also adopt {\it The Cannon} as a generative model for the GALAH spectra. {\it The Cannon} model is based on the assumption that for a given spectrum $n$, we can approximate its flux value at some wavelength ($f_{n,\lambda}$) by a linear combination of the label terms pertaining to that spectrum, $l_n$, the corresponding coefficients for those terms, ($\Theta_{\lambda}$), and the noise at that wavelength, summarised as
\begin{align}
    f_{n,\lambda} &= \Theta_{\lambda}^T \cdot l_n + {\rm noise}_{\lambda} \label{cannon}\\ 
    l_n &= (T_{{\rm eff},n}, \log g_n, \mathrm{[Fe/H]}_n, T_{{\rm eff},n}^2, ...),
\end{align}
where the label terms can be any parameters and their combinations to an arbitrary order, and the noise term accounts for the observational noise ($\sigma_{n,\lambda}$) as well as for the inability of the model to reproduce the data ($s_{\lambda}$). The $\Theta_{\lambda}$ and $s_{\lambda}$ finally represent {\it The Cannon} model, and they are obtained by fitting Eq.~\eqref{cannon} to a training set \citep[more details are provided in ][]{2015ApJ...808...16N}.

For consistency with the main GALAH pipeline, we employed the same training set as was used to train {\it The Cannon} model there (see Sect.~3.2 in \citealt{2018MNRAS.478.4513B}). It consists of 10\,605 spectra of dwarf and giant stars, but here we excluded 150 spectra that were recognised as binary star candidates in Sect.\,\ref{sec:detection}, and are thus left with 10\,455 spectra, to which we refer hereafter as the {\it \textup{training set}}. The parameters for the \textup{{\it \textup{training set}}} were estimated using the detailed spectrum synthesis code Spectroscopy Made Easy \citep[SME;][]{2017A&A...597A..16P}, and the spectra were selected to be of the highest quality and to sufficiently cover the Hertzsprung-Russell diagram \citep[for details, see ][]{2018MNRAS.478.4513B}. 

We produce spectral templates by feeding {\it The Cannon} model with a desired label vector (a set of parameters as described in Table \ref{modpar}). These templates thus become the model part of Eq.\,\eqref{lnspec}---$\mathcal{M}_{{\rm spec},i}(\theta)$. The scatter of {\it The Cannon} model ($s_{\lambda}$) is added in quadrature to the uncertainty of observed flux values of GALAH spectra $\sigma_{X_{{\rm spec},i}}$, and $X_{{\rm spec},i}$ contains flux values from all four GALAH spectral bands (see Table~\ref{tab:galbands}), which contribute approximately equally to the final number of spectroscopic data points $N_{\rm spec}$ of 12\,230. 

In the construction of spectral templates, we set the $A_{Ks}$ parameter of {\it The Cannon} model to 0, unlike in the main GALAH pipeline, where it is fitted. The $A_{Ks}$ parameter encapsulates the spectral signature of diffuse interstellar bands \citep[see Sect.~3.3.1 in][]{2018MNRAS.478.4513B}, and if not nullified, would produce duplicated interstellar absorption lines in the modelled spectrum of a binary star because we combine two templates at different RV shifts. This parameter is therefore not listed in Table \ref{modpar}, and although it is related to interstellar absorption, it carries different information than $E(B-V)$, which describes the influence of extinction on the overall SED.

It is difficult to determine the S/N of templates constructed by {\it The Cannon} model, but we estimate that it is much higher than the median S/N of the {\it \textup{training set}} spectra ($\sim 200$ per resolution element in the GALAH spectral band 2) because {\it The Cannon} effectively averages over them. The S/N of templates is also parameter dependent because it will change with the number of {\it \textup{training set}} spectra in a given bin of the parameter space. The quality of the fit will depend on the S/N of templates, but also on the ability of {\it The Cannon} to accurately reproduce the {\it \textup{training set}} spectra, which in turn also depends on the complexity of {\it The Cannon} model. Most of the literature employs {\it The Cannon} model with labels in the second order, arguing that the level of complexity thus obtained is sufficient. We follow this here, largely because we wish to be consistent with the main GALAH pipeline, but we nevertheless performed a brief investigation of proceeding to the third and fourth order in labels that can be found in Appendix\,\ref{app:cannon}. 

The final synthetic spectrum of the stars in a binary system is constructed by summing together the model spectra for the primary and secondary star, using the luminosity ratio $\eta_n$. Before summation, both the primary and secondary spectra are shifted according to their radial velocities (RVs) given by $\theta$. 

We identified a specific problem when {\it The Cannon} model was used to produce both spectral components of the synthetic binary template. In the regime of higher $v_{\rm broad}$ (e.g. >$\sim$ 20 km\,s$^{-1}$) and with a certain combination of other parameters, all within the prior ranges given in Table \ref{galahpar}, the model can produce double lines in a single-component spectrum. This leads to a rejection of solutions for genuine double-lined spectra at the low $\Delta V_r$ end of the distribution of binary systems, with the primary component mimicking double lines with small velocity separations and the secondary component ending up aligned with the primary and thus not satisfying the minimum $\Delta V_r$ for a reliable solution. 

We have checked the relatively small number of spectra of the {\it \textup{training set}} with high values of $v_{\rm broad}$ and cannot safely reject any of them as binaries, which might resolve this problem. We therefore imposed a more strict prior in the analysis of binary spectra by changing the upper limit of the flat prior on $v_{\rm broad}$ to 15 km\,s$^{-1}$ (which still includes the majority of the {\it \textup{training set}} $v_{\rm broad}$ values). This limit therefore represents the highest $v_{\rm broad}$ of the modelled spectrum, and can limit the accuracy of the spectroscopic fit in the hotter regime of the HR diagram or in the case of close binaries with synchronous rotation (that are normally spun up), but it does not significantly influence the majority of our results. Because our solution is not ideal, further attention to this problem is required in future work, while in the current catalogue of results, values of $v_{\rm broad}$ close to 15 km\,s$^{-1}$ should be considered with care.

\subsubsection{Posterior probability distributions of model parameters} \label{sec:sample}

The model parameters $\theta$ and their uncertainties $\sigma_{\theta}$ are obtained by sampling their posterior probability distributions based on Eq.\,\eqref{lnobj} and using the {\sl emcee} code, an implementation of the affine-invariant MCMC ensemble sampler \citep[][version 2.2.1]{2013PASP..125..306F}. This scheme offers a way to explore a given parameter space and very elegantly integrates (marginalises) posterior probability distributions. 

We report the results for the model parameters $\theta$ by the statistics of their posterior probability distributions, each of which contains 25600 samples (64 walkers times last 400 steps of the MCMC chain). We calculated the mean value and extracted the 16th$^{\rm }$, 50th$^{\rm }$ (the median), and 84th$^{\rm }$ percentile for each posterior distribution. Additionally, we provide the mode as the centre of the largest bin by binning the posterior distribution in 30 equal bins.

Because the MCMC scheme automatically provides a sampling of posterior probability distributions, which can be carried forward to subsequent inferences (useful for propagating uncertainty), or standing in as a prior, we used this feature in subsequent steps of our method, as discussed in Sect.~\ref{walk}.
We list the model parameters in Table \ref{modpar}. The additional hidden parameter $\eta_n$ is listed separately at the bottom of Table \ref{modpar} because it is not directly sampled from the posterior probability distribution, but it is determined by the photometric model and used by the spectroscopic model.

\subsubsection{MCMC sampling process} \label{walk}

By trial and error, we developed an MCMC procedure for posterior sampling that produces converged solutions close to global optima while minimising computation time. The MCMC process for each system was initialised with 1024 walkers in order to cover the relatively high dimensional parameter space. The initial parameter values for walkers were assigned randomly from the prior distributions (see Table~\ref{galahpar}). First of all, 250 steps were used to obtain priors on RVs of both binary components, using only the spectroscopic fit. Hereafter, the full model was used, starting with 100 steps, after which the upper 10\,\% of best solutions were kept, and from then on, the MCMC walk continued using 64 walkers. In this second part, 750 steps were made, after which the upper 25\,\% of best solutions were kept, followed by 500 more steps. At this point, the convergence of solutions was checked using a custom convergence criterion, and if convergence was not achieved, the MCMC walk continued in batches, each of which contained 500 steps and a convergence check. Taking only the upper best solutions of subsequent batches potentially excludes multimodal solutions, but close to convergence, the results are not highly multimodal. The process stopped either with achieved convergence or a maximum of ten repeated batches.

The convergence criterion consists of two custom rules for comparison of the scatter and slope of the MCMC chain in subsequent batches, which proved to be efficient and faster than standard convergence diagnostics such as auto-correlation. The first rule checks that the median values of $\chi^2_{\rm obj}$ at the start and end of the batch do not differ by more than one standard deviation of $\chi^2_{\rm obj}$ at the end. The second rule effectively checks that there are no multiple chains by comparing 10\% of the best and 10\% of the worst $\chi^2_{\rm obj}$ at the end of the batch, imposing that their average standard deviation multiplied by 20 is always larger than the difference of their medians.

The MCMC setup used in this work takes on average 8 CPU hours to arrive at a converged solution for each observed spectrum. Evaluating $\chi^2_{\rm phot}$ is approximately ten times slower than evaluating $\chi^2_{\rm spec}$; the computation of SEDs using PySynPhot is by far the most expensive calculation.

\section{Validation of the method} \label{sec:validation}

We first validated our method for deriving the properties of stars in a binary system by applying it to single stars. This makes the process of validation transparent and comprehensible because the single-star model is less complex, and also lets us benchmark selected derived values against external sources. The photometric and spectroscopic models change such that the synthetic apparent magnitudes are derived using only one SED, and the synthetic spectra are composed of only one component. 
We used different sets of GALAH objects for two different validation tests: {\it The Cannon} validation sample (Sect.~\ref{sec:spec_valid}), and the benchmark validation sample (Sects. \ref{photonly} and \ref{fuse}).

\subsection{Spectroscopic validation} \label{sec:spec_valid}

We verified that the binary pipeline, when using only the spectroscopic fit, produced similar results to the main GALAH pipeline. We expect this to be the case because a) the generative model for stellar spectra is produced by the same technique ({\it The Cannon}), b) the training set for {\it The Cannon} is the same as was used in the main GALAH pipeline, and c) the objective function for comparing the model to data (see Eq.\,\eqref{lnspec}) is essentially the same as in the main GALAH pipeline. The only significant difference is the process of minimisation of the objective function. The test was made on {\it The Cannon} validation sample, which is a subset of $\sim$ 2000 random stars from the training set\textup{}. This sample covers the HR diagram in the same way as the {\it \textup{training set}}, but more sparsely. It contains dwarf as well as giant stars, and their positions on the sky are indicated in Fig.\,\ref{fig:foot}.

\begin{figure}[!htp]
   \centering
   \includegraphics[width=\linewidth]{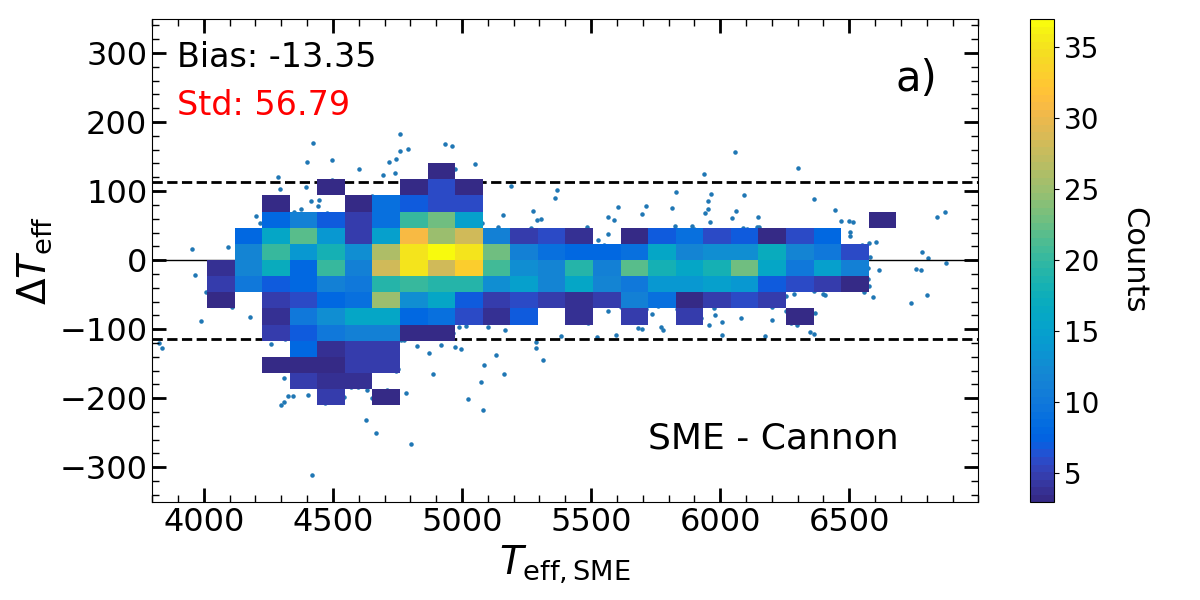}
   \includegraphics[width=\linewidth]{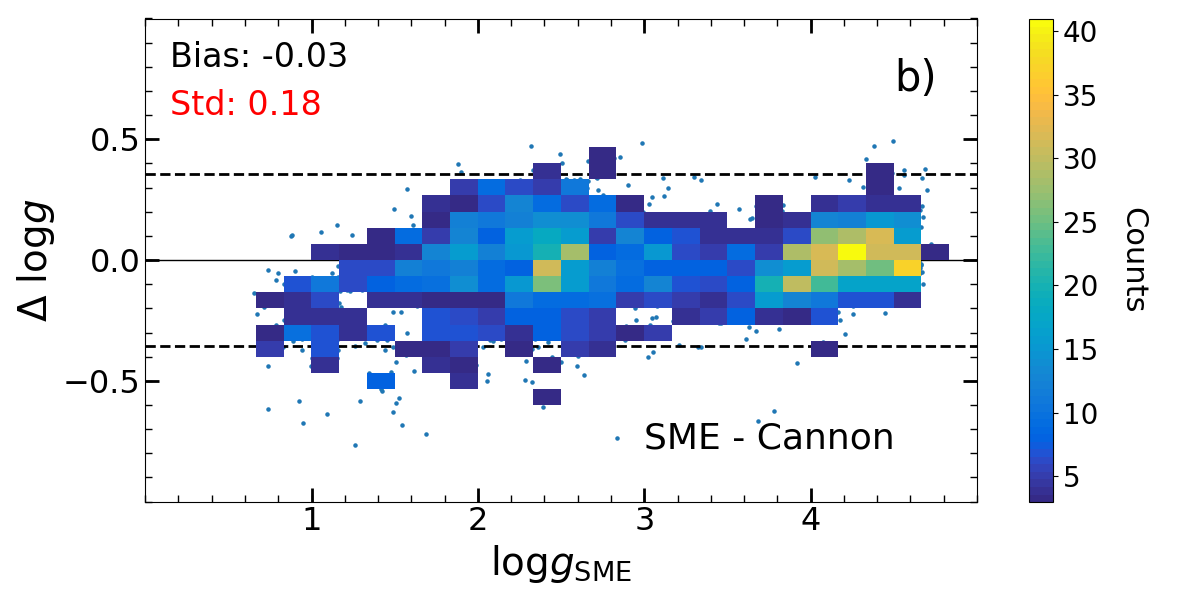}
   \includegraphics[width=\linewidth]{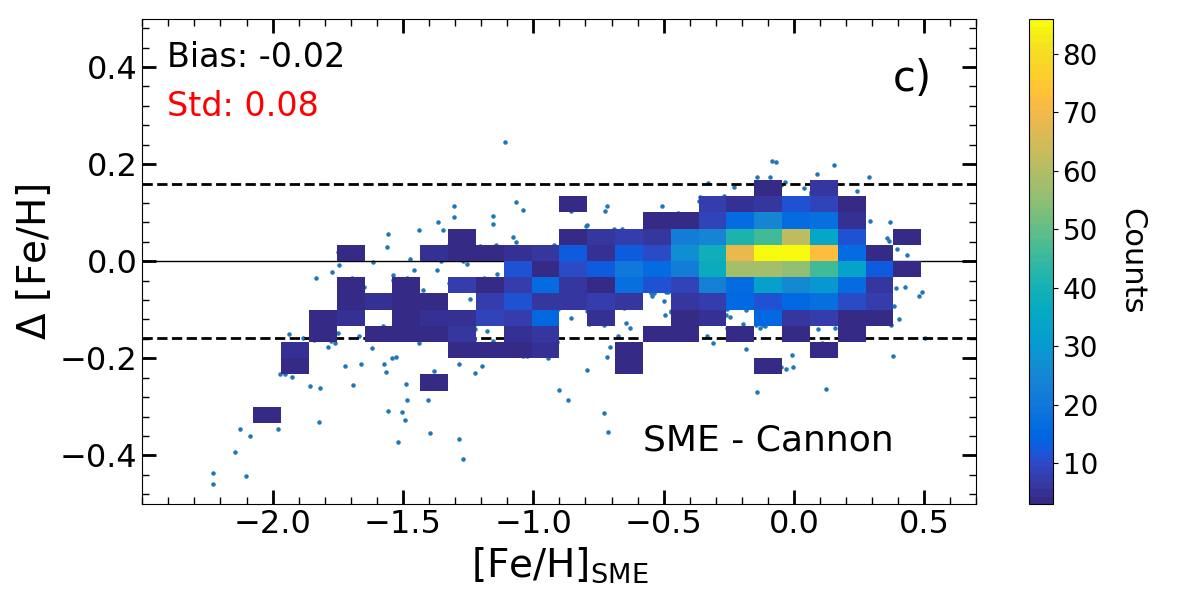}\\
   \caption{Comparison of SME and {\it The Cannon} labels for {\it The Cannon} validation sample. {\it The Cannon} values produced by the binary pipeline are subtracted from the SME values.}
   \label{BPsGTS}
\end{figure}  

Figures \ref{BPsGTS} and \ref{cannons} show the comparison of SME values (see \citealt{2018MNRAS.478.4513B}) to the results of both the binary pipeline (Fig.\,\ref{BPsGTS}) and the main GALAH pipeline (Fig.\,\ref{cannons}), confirming their close agreement. We observe some trends and biases in these figures, such as the inconsistency for the lowest metallicities in the bottom panel of Fig.\,\ref{BPsGTS}, which are a consequence of both the flexibility of the {\it The Cannon} model and the adequacy of the {\it \textup{training set}}, where it is difficult to ensure equal representation of spectra in all bins of the modelled parameter space \citep[see Sect.~3.3.1 in ][]{2018MNRAS.478.4513B}.

\subsection{Photometric validation} \label{photonly}

In this validation check and in Sect.~\ref{fuse}, we used the benchmark validation sample, which is a subset of GALAH objects with external information about the temperature from the infrared flux method (IRFM; \citealt{2019MNRAS.482.2770C}) and radii (asteroseismology; \citealt{2017ApJ...835...83S,2019arXiv190412444S}). This sample of 620 giant stars is therefore suitable for benchmark testing, and is dominated by stars from K2 campaigns C1 and C6 \citep[][see also Fig.\,\ref{fig:foot}]{2019arXiv190412444S}. Compared to the previous spectroscopic validation, this sample has almost no overlap with the {\it \textup{training set}} and hence no SME parameters to compare to.

\begin{figure}[!htp]
   \centering
   \includegraphics[width=\linewidth]{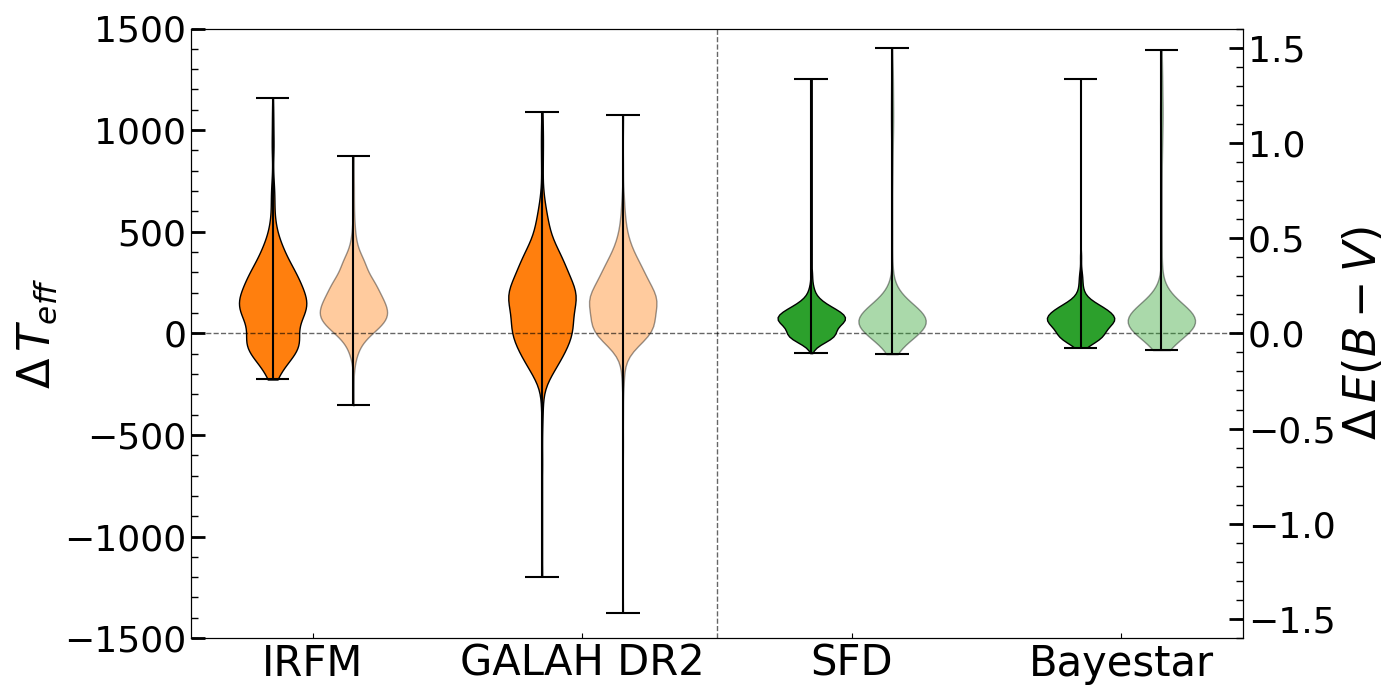}
   \caption{Comparison of $T_{\rm eff}$ (orange) and $E(B-V)$ (green) parameters as derived by the binary pipeline (using only the photometric fit) to the literature values (indicated on the horizontal axis). The literature values are subtracted from those given by the binary pipeline. The lighter shaded distributions correspond to results from the binary pipeline when the \textit{Gaia} magnitudes are excluded from the fit.}
   \label{degent}
\end{figure}

We took the results from the binary pipeline, using only the photometric fit (Eq.\,\eqref{lnphot}), and compared them to the IRFM temperature values and reddening values from SFD (Schlegel, Finkbeiner, David) reddening maps \citep{1998ApJ...500..525S} and 3D reddening maps based on Pan-STARRS~1 photometry (\citealt{2018MNRAS.478..651G}; hereafter the Bayestar maps). This comparison is shown in Fig.\,\ref{degent}, where we display results with and without the \textit{Gaia} magnitudes in the photometric fit. We distinguish the \textit{Gaia} magnitudes from other photometric passbands used in this work because their uncertainties are approximately an order of magnitude lower. They therefore affect the fit more strongly.

Fig.\,\ref{degent} shows some important problems. The most striking is a bias towards higher temperatures as compared to both the GALAH DR2 values and those from the IRFM. An additional test that we performed using a larger validation sample (without the need for external information about the radii) also shows strong grid effects, which diminish considerably, however, when we exclude \textit{Gaia} magnitudes. The grid effect has a spacing of about 250 K, and it might originate from the SED model in combination with small uncertainties of \textit{Gaia} data. Excluding \textit{Gaia} magnitudes from the fit also slightly reduces the bias towards higher temperature and reddening values. In any case, this effect is caused by the degeneracy between the two parameters, where an increase in the modelled temperature is compensated for by the increased extinction (and therefore reddening). More details of this comparison, which is summarised in Fig.\,\ref{degent}, are shown in Fig.\,\ref{fig:degent_ext}, where the x-axis of the panels spans the literature values to which we compare our results. The grid effect shown in Fig.\,\ref{fig:degent_ext} is reflected in the bimodality of distributions in Fig.\,\ref{degent}. 

The photometric validation test implies that our photometric data alone cannot constrain the shape of the SED to an extent where we could produce reliable estimates of the interstellar reddening and temperature of the stars. The photometric fit therefore has to be complemented by another constraint on the temperature (or reddening), which in this work is provided by the spectroscopic fit. 

\subsection{Overall validation for single stars} \label{fuse}

We now investigate the results by fusing the spectroscopic and photometric part of the binary pipeline and the prior on parallax. By including the photometric fit in the binary pipeline, we can determine the radii and reddening for our target stars. The results for stellar radii do not depend significantly on the degeneracy between the temperature and extinction. However, this degeneracy is detrimental to determining $E(B-V)$, as illustrated in Fig.\,\ref{degent}. It arises from the interplay between the effects of temperature and extinction on the modelled SED. By increasing the temperature, the blackbody SED experiences a relatively larger increase of flux in the blue part of the spectrum. This can be compensated for by an increase in reddening (extinction), which decreases the flux more strongly at shorter wavelengths. Thus the approximate shape of the SED and the apparent synthetic magnitudes are conserved, while the modelled radius is largely unaffected.  

\begin{figure}[!htp]
   \centering
   \includegraphics[width=\linewidth]{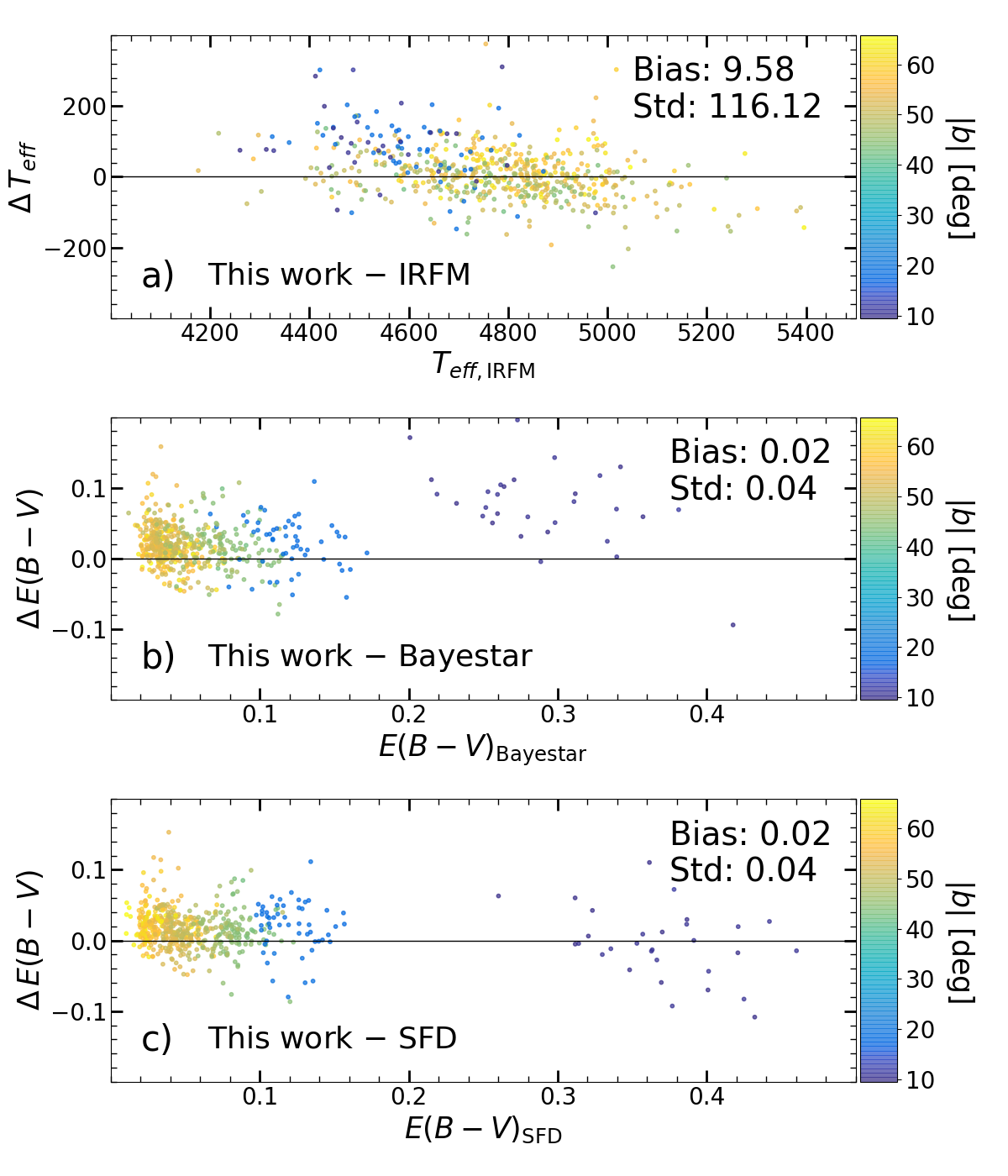}
   \caption{Temperature and reddening for the benchmark validation sample derived by the binary pipeline compared to the literature values from the IRFM method and Bayestar and SFD reddening maps (see also Fig.\,\ref{degent}). The colour-coding is by absolute Galactic latitude.}
   \label{fig:teff_ebv}
\end{figure}

Figure\,\ref{fig:teff_ebv} demonstrates that the bias towards higher temperatures and reddening, which was observed in Fig.\,\ref{degent}, is now largely overcome with the aid of the spectroscopic part of the fit in the binary pipeline. Figure\,\ref{fig:teff_ebv}a, which compares the temperature values, is consistent with Fig.\,16 from \citet{2018MNRAS.478.4513B}, indicating a good agreement for the temperature parameter between the binary pipeline and the main GALAH pipeline. The reddening parameter shown in Figs. \ref{fig:teff_ebv}b and \ref{fig:teff_ebv}c is still somewhat overestimated, however, and in the comparison with SFD reddening maps, there seems to be a better agreement at higher extinction regions than in the case of Bayestar maps. Nevertheless, we might expect the Bayestar maps to be more realistic because they are in 3D, and thus it would be plausible that the binary pipeline overestimates reddening in low and high extinction regions, as is indicated in Fig.\,\ref{fig:teff_ebv}b. 

\begin{figure}[!htp]
   \centering
   \includegraphics[width=\linewidth]{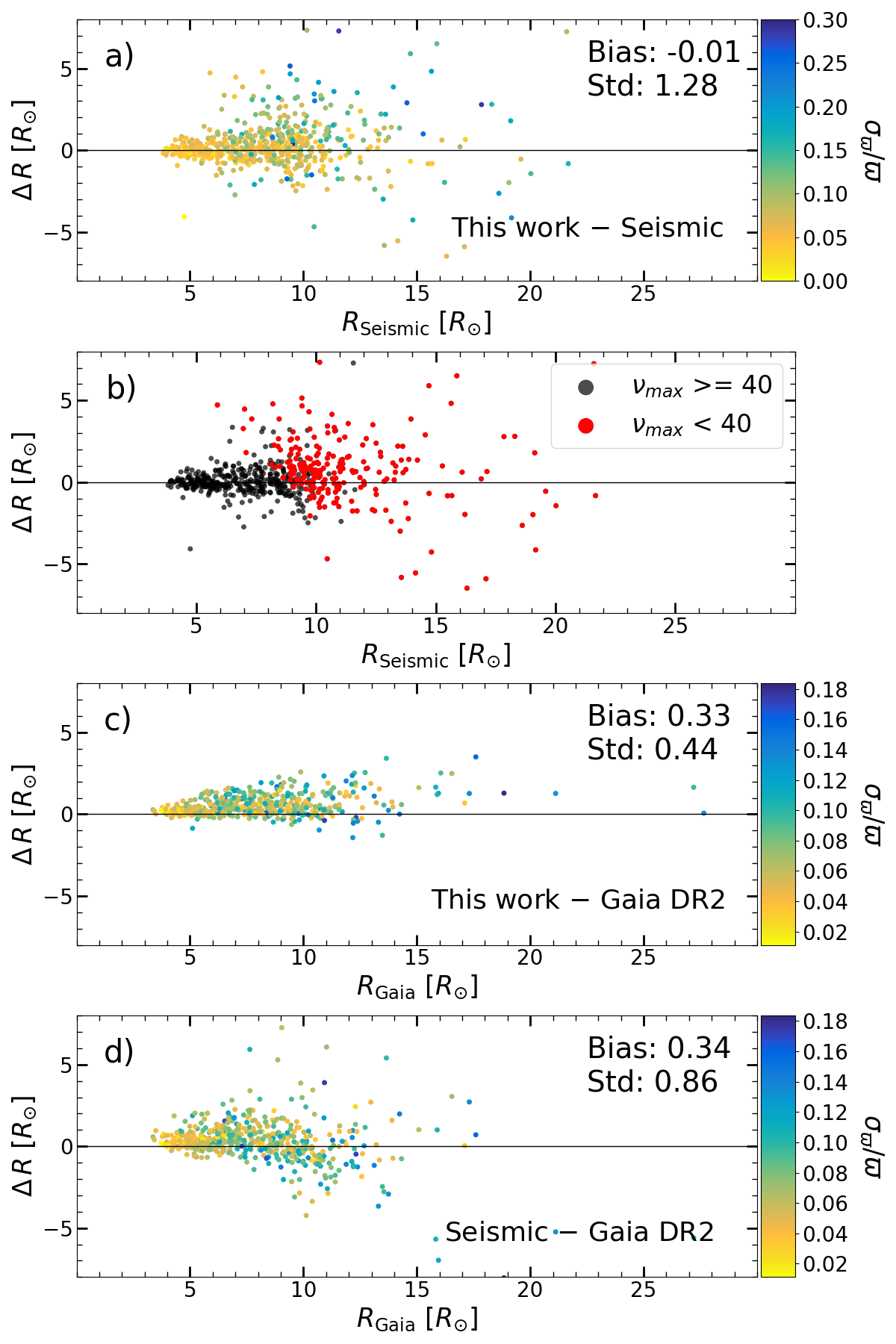}
   \caption{Comparison of radii for the benchmark validation sample derived by the binary pipeline, \textit{Gaia} DR2, and computed from asteroseismic parameters (more details in Sect.~\ref{fuse}). The colour-coding is by relative parallax uncertainty except in panel b), which is a copy of panel a), but colour-coded to indicate stars with less reliable seismic information ($\nu_{max}$ < 40; S. Sharma, private communication). The more reliable part of the sample is used to calculate the bias and scatter as reported in panels a), c), and d). The colour scale in panel a) is clipped for clarity, while panels b) and c) have a smaller range of $\sigma_{\varpi}/\varpi$ because some \textit{Gaia} DR2 values for radii were lacking.}
   \label{radii}
\end{figure}

Figure \ref{radii} shows a comparison of radii derived by the binary pipeline, \textit{Gaia} DR2, and those computed from asteroseismic parameters $\nu_{max}$ and $\Delta \nu$ using scaling relations adopted from \citet[][see their Eq. 3]{2017ApJ...844..102H}. We applied a $\Delta \nu$ correction following \citet{2016ApJ...822...15S}. The constants $\nu_{max \odot}$ and $\Delta \nu_{\odot}$ are taken from \citet{2017ApJ...844..102H}, whereas T$_{\rm eff,\odot}$ is set to 5772 K \citep{2015arXiv151007674M} and the T$_{\rm eff}$ is the one computed by the binary pipeline. The \textit{Gaia} DR2 radii are taken as is, computed based on luminosity (obtained using temperature-only dependent bolometric correction) and effective temperature as determined by \textit{Gaia} \citep[for more details, see ][]{2018A&A...616A...8A}. We did not correct \textit{Gaia} radii for extinction, the parallax zero-point, or the difference in temperature scales employed by \textit{Gaia} and this work. The comparison shown in panels c--d in Fig.~\ref{radii} should therefore be taken as approximate because thorough corrections have to be applied for a more faithful benchmarking against \textit{Gaia} radii, such as the one performed by \citet{2019ApJ...885..166Z} for seismic radii of \textit{Kepler} field stars. 

Because we did not correct for \textit{Gaia} DR2 radii, both the binary pipeline and seismic values show an offset towards larger radii in our crude comparison. In the case of seismic--\textit{Gaia} comparison, the discrepancy is thus larger than the reported agreement for \textit{Kepler} field stars \citep{2019ApJ...885..166Z}. However, the agreement between the binary pipeline and seismic values is encouraging, showing a negligible bias and a median relative difference in radii of 5.7\,\%. The stars with the largest absolute differences ($\Delta R$) in all panels of Fig.\,\ref{radii} also have larger relative parallax uncertainties ($\sigma_{\varpi}/\varpi$; up to 40\,\%).

In summary, the validation of the binary pipeline on single stars indicates that we can achieve results that are consistent with selected literature sources to $\sim 45$ K in effective temperature, $\sim 6\,\%$ in stellar radii, and $\sim 40\,\%$ in reddening (measured in magnitudes), keeping in mind that this holds true for a subset of giant stars that constitute the benchmark validation sample. We do not speculate on how this comparison translates into accuracy because different methods can be more reliable for different sub-populations of investigated objects. However, in the case of binary stars, where two components constrain the overall shape of the SED or the combined spectrum, we expect less degeneracy between model parameters and thus lower uncertainties on temperatures, radii, or reddening, for instance.

\subsection{Validation of the binary pipeline on artificial binary stars} \label{sec:valbin}

In order to estimate the performance of the binary pipeline in producing reliable parameters for SB2 systems, we constructed an artificial sample of 2000 binary systems based on single stars from the {\it \textup{training set}}. This sample was produced in a simplified approach such that the properties of both stars roughly represented realistic combinations of stars in binary systems. A more thorough consideration of stellar properties based on the mass, mass ratio, and evolutionary stage of the binary system, for example, is beyond scope of this validation, and we are additionally limited by the choice of stars in the training set\textup{}. 

The simplified approach of combining the primary and secondary star differs in the way we treat dwarf and giant systems, and we only produced dwarf-dwarf and giant-giant combinations (we followed the dwarf-giant distinction given by \citealt{2018MNRAS.481..645Z}). For every star in the {\it \textup{training set}} (the primary), we imposed three conditions for possible secondary candidates: 1) a matching metallicity ($|\Delta \mathrm{[Fe/H]}| < 0.02$), 2) the surface gravity ($\log g$) of the secondary must be higher than or equal to that of the primary, and 3) $T_{\rm eff2} \leq T_{\rm eff1}$ for dwarf-dwarf pairs and $T_{\rm eff2} \geq T_{\rm eff1}$ for giant-giant pairs. We then randomly selected five candidates for the secondary among the training set{\it } stars that satisfied the three conditions. 

To obtain the necessary properties for combining primary and secondary spectra, we assigned a mass ratio to each pair. For giant-giant pairs we assumed a mass ratio of one because both stars need to have approximately equal mass in order to simultaneously evolve to the giant stage, whereas for dwarfs, we adopted an empirical relation between mass and effective temperature ($m_2/m_1 = [T_{\rm eff2}/T_{\rm eff1}]^{1/0.7}$) based on the study by \cite{Eker_2015}. The primary was then assigned a radius of 1 $R_{\odot}$, while the radius of the secondary was computed from the previously determined mass ratio and the measured surface gravity of both stars. We set a parallax of 1 mas for all systems, after which artificial observed magnitudes, together with luminosity ratios for combining spectra of both components, were computed using the same tools as presented in Sect.~\ref{sec:method}. When the luminosity ratio in any GALAH spectral band was below 0.1, the system was rejected. The $E(B-V)$ value for each system was then drawn randomly from an exponential distribution (scale = 0.15) that closely reflected the overall distribution of reddening for GALAH targets.

The primary star in each pair was kept at rest frame, whereas we shifted the secondary by randomly sampling its RV from an exponential distribution (scale = 50 km\,s$^{-1}$) with a random sign. Then we combined the spectra of the two components. Finally, we randomly chose 1500 out of $\sim 140k$ produced pairs with 15\% giant-giant and the remaining dwarf-dwarf systems. 

The results of the comparison between the input (SME parameters, additional assigned properties) and the parameters recovered by the binary pipeline are characterised in Table~\ref{tab:bin_val_tab}. The overall distributions largely resemble the uncertainty of the spectroscopic model, as illustrated in Figs. \ref{BPsGTS} and \ref{cannons}, with slightly amplified biases that indicate the degeneracy between the primary and secondary star. The scatter of [Fe/H] in this validation is smaller than that in Fig.~\ref{BPsGTS} and might result from the constraint on this parameter by both components. 

\begin{table} 
\caption{Comparison between the input parameters and those derived by the binary pipeline for the sample of 2000 binary stars that were artificially constructed by combining single stars from the {\it \textup{training set}}. The binary pipeline parameters are subtracted from the input ones and the quantiles describe the resulting distributions.}
\label{tab:bin_val_tab}
\centering
\begin{tabular}{l c c c c}
\hline\hline
Parameter & $\eta.16$ & $\eta.50$ & $\eta.84$ & Unit\\
\hline
$T_{\rm eff,1}$ & -59 & -14 & 24 & K\\ \
  $T_{\rm eff,2}$ & -52 & 3 & 72 & K\\ \
  $\log g_{1}$ & -0.13 & -0.02 & 0.10 & dex\\ \
  $\log g_{2}$ & -0.11 & 0.02 & 0.19 & dex\\ \
  $\mathrm{[Fe/H]}$ & -0.04 & 0.01 & 0.05 & dex\\ \
  $V_{r,1}$ & -0.3 & 0.1 & 0.5 & km\,s$^{-1}$\\ \
  $V_{r,2}$ & -0.4 & 0.0 & 0.4 & km\,s$^{-1}$\\ \
  $v_{\rm mic,1}$ & -0.04 & -0.00 & 0.03 & km\,s$^{-1}$\\ \
  $v_{\rm mic,2}$ & -0.04 & 0.00 & 0.05 & km\,s$^{-1}$\\ \
  $v_{\rm broad,1}$ & -0.7 & 0.2 & 1.1 & km\,s$^{-1}$\\ \
  $v_{\rm broad,2}$ & -0.5 & 0.8 & 2.1 & km\,s$^{-1}$\\ \
  $R_{1}$ & -0.01 & -0.00 & 0.01 & $R_{\odot}$\\ \
  $R_{2}$ & -0.01 & 0.00 & 0.02 & $R_{\odot}$\\ \
  $E(B-V)$ & -0.01 & -0.00 & 0.00 & \\
\hline
\end{tabular}
\end{table}

\section{Identification of multiple star candidates} \label{sec:detection}

A number of techniques to identify stellar binaries from their integrated spectra have been developed, mostly based on the analysis of the cross-correlation function (CCF). We complemented this technique with a sample of binaries detected by applying a dimensionality reduction algorithm to classify the spectrum.  We identified binaries using the procedure described by \citet{2017ApJS..228...24T} and updated in \citet{2018MNRAS.478.4513B}, and confirmed and extended the sample with a CCF analysis following \citet{2017A&A...608A..95M}. 

\subsection{Detection through a dimensionality reduction} \label{sec:detect}

The detection and selection of multiple star candidates was first performed with the help of a dimensionality reduction technique, the t-SNE, which is very efficient at grouping similar objects in a visually manageable 2D map (Fig.~\ref{sol}). The selection process includes a clustering algorithm, DBSCAN \citep{citeulike:3509601}, that alleviates the detection of boundaries for different morphological groups of spectra in the t-SNE map, but it can also be complemented by manual selection of smaller regions or individual spectra. We refer to \citet{2018MNRAS.478.4513B} for more details on the iterative semi-automatic classification procedure. 

We selected the spectra that were flagged as binary stars in \citet{2018MNRAS.478.4513B}, but we also added a few hundred spectra that were classified as binaries in a previous study \citep{2017ApJS..228...24T}, which are not flagged as such in the new classification. This already implies that our classification procedure is not perfect (see discussion in \citealt{2017ApJS..228...24T}), but the majority of the double-lined binary stars are detected. A group of binary stars that is generally harder to isolate in the map are the most massive and therefore hotter stars, where the spectral features caused by the high temperature (strong, broad lines) dominate the binary signature (double lines). The hotter binary stars are better accounted for in the second detection procedure discussed in the following section. 

\subsection{Detection through the cross-correlation function} \label{sec:ccf}

The second selection of multiple star candidates was performed with the CCF technique and automated identification of multiple RV components developed by \citet{2017A&A...608A..95M}. Because the observed stars are mainly of FGK spectral type (and most of the GALAH binaries are nearby dwarf systems), we used a single spectral template for the CCF computation following \citet{2010AJ....140..184M}. We used the MARCS model atmosphere \citep{2008A&A...486..951G} of the Sun ($T_\mathrm{eff}=5777$~K, $\log{g}=4.44$ and [Fe/H$]=0$) produced by the 1D local thermal equilibirum (LTE) radiative transfer code Turbospectrum \citep{2012ascl.soft05004P}. 

\begin{figure}
\includegraphics[width=\linewidth]{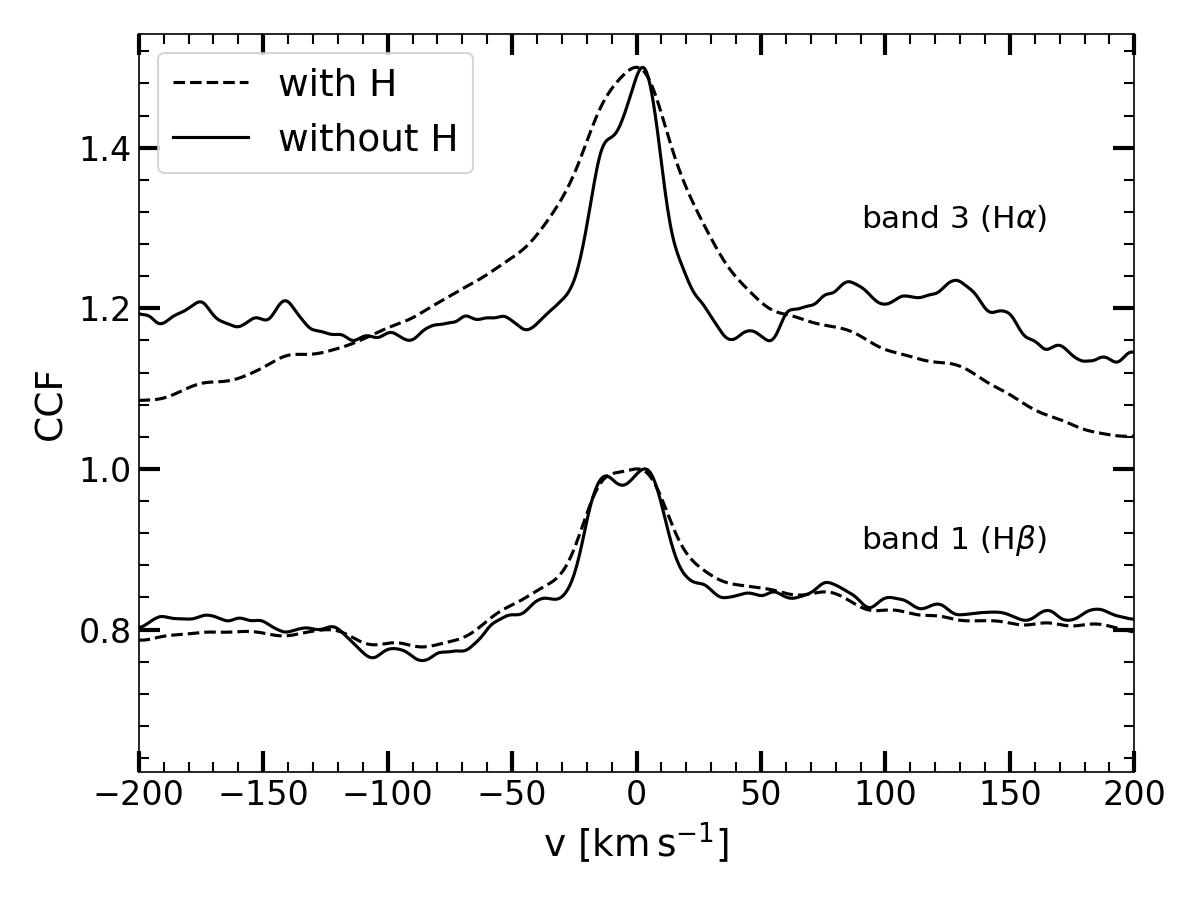}
\caption{Effect of excluding H$\alpha$ (in band 3) and H$\beta$ (in band 1) in the solar template on the width of the CCF. The decrease in CCF peak width reveals a second close component that is difficult to detect when H$\alpha$ and H$\beta$ are included in the template.}\label{fig:ccf_comp}
\end{figure}

The most prominent absorption lines in GALAH spectra are the hydrogen lines H$\alpha$ and H$\beta$, which are usually either saturated or very strong in late-type stars. They produce a very broad peak in the CCF that can hide close secondary components. We show an example in Fig.~\ref{fig:ccf_comp}, where we plot CCFs that are computed with (dashed lines) and without (plain lines) H$\alpha$ (band~3) and H$\beta$ (band~1). By excluding  H$\alpha$ and H$\beta$ from the template, the typical width of a single RV component CCF is similar in all four bands of the GALAH spectra because bands~2 and 4 do not include saturated or strong absorption lines. Hence we chose to exclude the hydrogen lines when we computed the CCF. 

\begin{figure}[t]
    \centering
    \includegraphics[width=\linewidth]{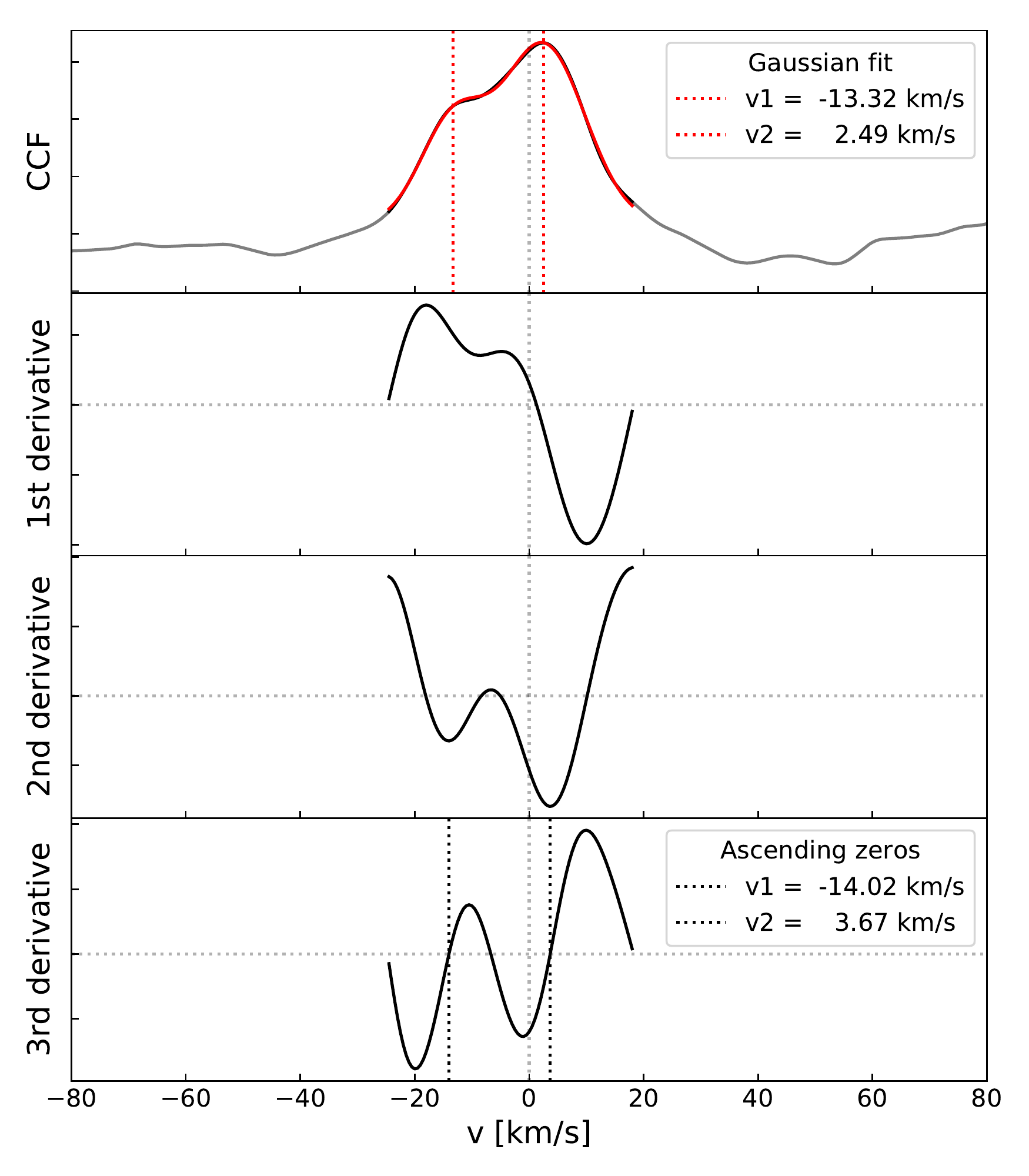}
    \caption{Identification of the RV components in a CCF of a close SB2 candidate. The top panel shows the CCF. The three panels below correspond to the three successive derivatives. The vertical black lines show the RV components detected by taking the ascending zeros of the third derivative (bottom panel). The vertical red lines give the positions of RV components obtained by fitting the multi-peaked CCF.}
    \label{fig:doe_481}
\end{figure}

The solar templates (one for each band) used in the computation of the CCFs were calculated by Turbospectrum with a wavelength step equal to half of the original sampling of GALAH spectra (0.020, 0.025, 0.030, and 0.035~\AA), corresponding to resolving powers larger than 220\,000. The templates were not convolved to the resolution of the instrument. Each CCF was computed over a RV range of 700~km\,s$^{-1}$ centred on the zero velocity with a velocity step of 0.5~km\,s$^{-1}$. The number and positions of RV components were computed using the code detection of extrema, DOE\footnote{The DOE parameters used for the GALAH spectra are THRES0=0.5, THRES2=0.1, and SIGMA=3 km\,s$^{-1}$ (see \citealt{2017A&A...608A..95M} for an explanation of these parameters).} \citep[ see][]{2017A&A...608A..95M}. For each CCF, the number of components were obtained using the first three derivatives of the CCF. The use of the first derivative does not allow us to automatically identify two close components when a local minimum between the two components is missing. However, the minima of the second derivative provide the positions of the CCF peaks. The ascending zeros of the third derivative were used to identify these positions. When the number of components were identified, we performed a Gaussian fit of the core of each component. \citet{2017A&A...608A..95M} provide more details on the process. An example of an SB2 candidate for which the first derivative of the CCF is not sufficient to detect the second component is given in Fig.~\ref{fig:doe_481}. The RVs obtained by the Gaussian fits are slightly different than the zeros of the third derivative because the successive derivatives are smoothed (by a Gaussian kernel with $\sigma = 8$~km/s) to decrease the numerical noise.

\subsection{SB2 candidate sample from combined detections}

\begin{figure}[!htp]
   \centering
   \includegraphics[width=\linewidth]{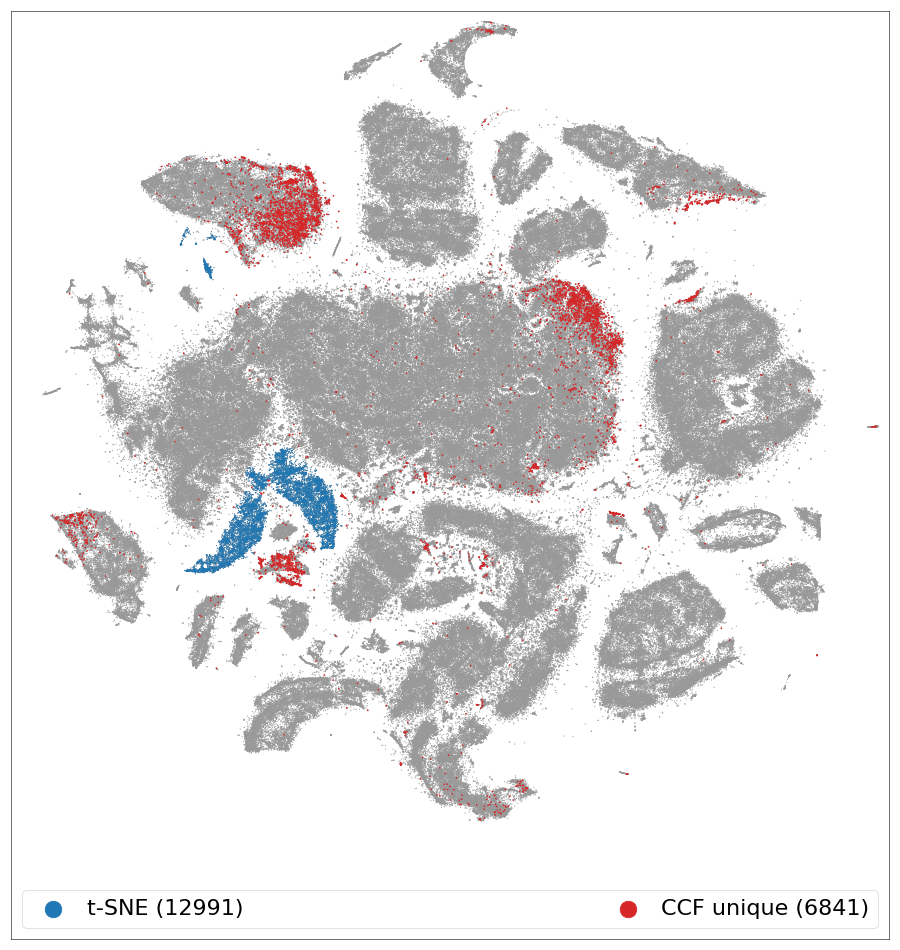}                        
   \caption{t-SNE projection map of 587\,153 spectra as presented in \citet{2018MNRAS.478.4513B}. The points (spectra) are colour-coded by all t-SNE detected SB2s (blue) and the unique additional detections by CCF (red). The count of spectra for each group is given in the legend.}
   \label{comp_tsne_ccf}
\end{figure}

We combined the SB2 spectra detected with t-SNE and CCF and obtained a final list of 19\,773 unique candidates, where the intersection between t-SNE and CCF detections amounts to 7556 objects. The CCF approach reproduces the majority of t-SNE identifications in the region of isolated groups of binary stars (blue points in Fig.\,\ref{comp_tsne_ccf}), but some candidates well inside these groups and most of those in the transition regions are missed. On the other hand, the CCF approach additionally contributes 6841 candidates from different parts of the parameter space (red points in Fig.\,\ref{comp_tsne_ccf}). These identifications with CCF are, as expected and clearly shown in Fig.\,\ref{comp_tsne_ccf}, mostly outliers or edge members in groups of spectra from distinct regions of the parameter space. 

\begin{figure*}[!htp]
   \centering
   \includegraphics[width=0.49\linewidth]{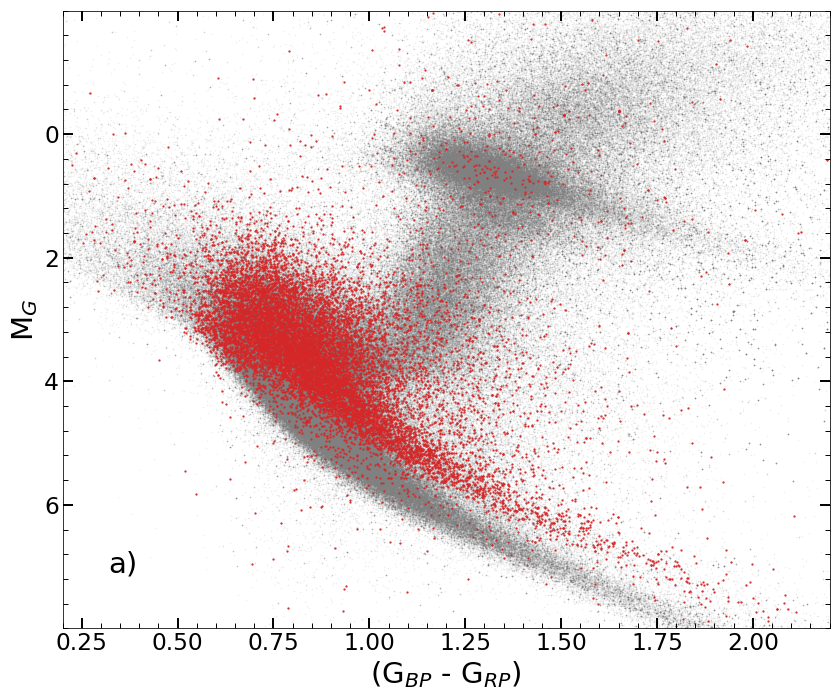}    
   \includegraphics[width=0.49\linewidth]{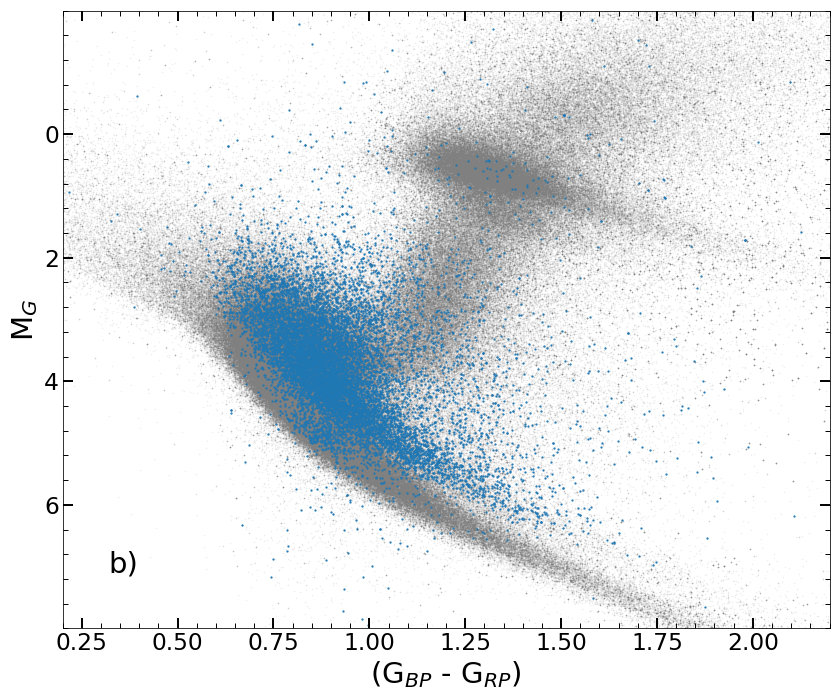}    
   \caption{Colour-magnitude diagram of SB2 identifications. a) All identifications with CCF, b) all identifications with t-SNE. The grey points are all GALAH objects considered in this work. The absolute magnitudes and colours are taken from the \textit{Gaia} DR2 catalogue, and no reddening corrections have been applied.}
   \label{cand_HR}
\end{figure*}

The colour-magnitude diagram in Fig.\,\ref{cand_HR} displays all t-SNE and CCF detected SB2 candidates against the background of all GALAH spectra considered here. We can distinguish the binary main-sequence, which is offset to brighter absolute magnitudes. The over-density of detections just below this binary main-sequence shown in Fig.\,\ref{cand_HR}b points to false positives due to a smooth transition from well-distinguished double lines to a mixture of unresolved binary and single-star spectra. Figure \ref{cand_HR}a also shows that CCF extends the detection towards cooler and hotter temperatures than t-SNE.  

\section{Selecting the final sample of SB2s} \label{sec:selection}

\begin{figure*}[!htp]
   \centering
   \includegraphics[width=\linewidth]{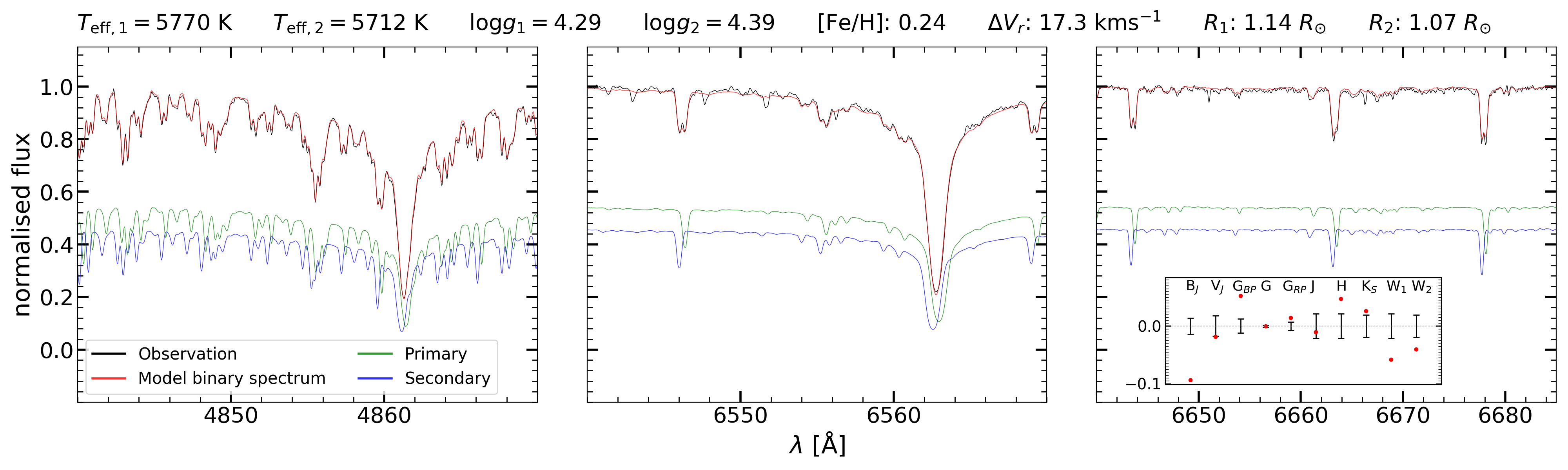} \\   
   \includegraphics[width=\linewidth]{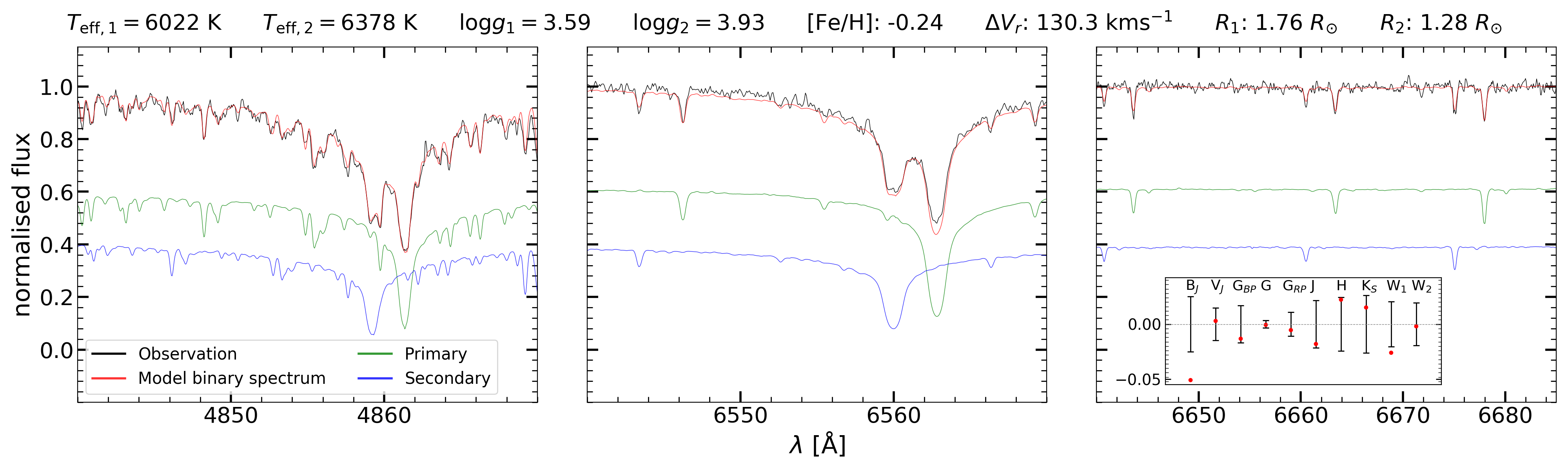}    
   \caption{Example fits of binary stars from the final sample. The top and bottom panels show a binary system at low- and high-velocity separation of double lines, respectively. The spectral region around H$\beta$ and H$\alpha$ is shown in the left and middle, respectively, while a region of a few well-isolated lines is shown on the right. Some of the derived $\theta$ parameters are indicated for each binary system, and the bottom example shows that the primary star has already evolved from the main-sequence. The primary and secondary component of the modelled binary spectrum are shown separately at a continuum level in accordance with their contribution. The black error bars in the inset plot of the photometric fit represent the relative uncertainty of the observed fluxes in the indicated photometric passbands, while the red points represent the relative offset of synthetic fluxes. The error bars of \textit{Gaia} photometry are multiplied by a factor of 10 for clarity.}
   \label{fig:exs}
\end{figure*}

In this section we define the final sample of binary stars for our investigation. We consider different aspects of the results produced by the binary pipeline, and summarise the four criteria that we used to define the final sample (Table \ref{masks}). It is important to recognise that these selection criteria by themselves produce a selection effect, adding to others, for example, from the identification process described above. After applying the selection criteria to the sample of 19\,773 analysed objects, we are left with 12\,760. From this final sample, we show two examples of the fit performed by the binary pipeline in Fig.\,\ref{fig:exs}.

\subsection{Selection criteria}

The main aim of the selection presented here is to reject physically or technically erroneous solutions. We note that the cuts based on these selection criteria significantly reduce the number of binary stars in our sample. This is mostly due to false identifications and the conservative nature of our selection criteria.
Below we discuss the four criteria listed in Table \ref{masks} in more detail.

\begin{table}
\caption{Selection criteria used to remove the results of the binary pipeline that are identified as being of poor quality or unphysical. The number of rejected objects is evaluated on a sample that was previously cleaned by all criteria listed prior to the one in question.}
\label{masks}
\centering
\begin{tabular}{c L{5.5cm} C{1.5cm}}
\hline\hline
Crit. & Short description & Number of rejected objects\\
\hline
{\bf 1} & binary pipeline solution does not converge &  {\bf 1425} \\
\hline
{\bf 2} & $\Delta V_r$ < 15 km\,s$^{-1}$ &  {\bf 3465} \\
\hline
{\bf 3} & luminosity ratio (secondary/primary) in any band < 0.1 &  {\bf 1857} \\
\hline
{\bf 4} & mismatch of $\Delta V_r$ between CCF and binary pipeline, > 20 km\,s$^{-1}$ &  {\bf 266} \\
\hline
\end{tabular}
\end{table}

{\bf Criterion 1:} The convergence of the solution to the global minimum in the binary pipeline, or at least to a minimum as defined by our custom convergence criterion (Sect.~\ref{walk}), can fail for several reasons. By letting the MCMC sampling process run much longer, more objects would arrive at convergence, but it is not guaranteed that these would really be good solutions because convergence issues are often connected to data of poor quality, multiple minima, and thus degenerate solutions, and other issues that might make a lengthy convergence doubtful in any case. When we apply each criterion from Table \ref{masks} on the full sample instead of in sequence, we find that the overlap of objects rejected by criterion 1 and the others is large, indicating that criterion 1 is connected to specific issues that the rest of them address.

{\bf Criterion 2:} The restriction on the difference of RVs of the two stars, or the velocity separation of double lines, is based on an estimate of the limit to which GALAH spectra can be used to resolve almost blended double lines. It is difficult to quantify the smallest reliable velocity difference where we can be certain that double lines are indeed present, and it depends on the flux ratio, stellar type, stellar rotation, varying resolution across the spectral range, and other effects. By visual inspection of the $\chi^2_{\rm spec}$ fits and the degeneracies of stellar parameters in the case of severely blended lines, we set a conservative limiting value of 15 km\,s$^{-1}$. An example of partially blended lines at this velocity separation is shown in Fig.\,\ref{fig:exs}. Nevertheless, some line profiles that seem double-lined but heavily blended can originate from spotted surfaces or pulsation (either radial or non-radial), which is present in stars from many places of the HR diagram, causing diverse line asymmetries. 

{\bf Criterion 3:} In the case of spectra where the lines of the secondary star are very weak compared to those of the primary star, the spectroscopic fit of the secondary component becomes unreliable. For a main-sequence binary with the Sun as the primary, $\eta \sim 0.1$ translates into $q \sim 0.5$. It is also possible that our method fits noisy features of observed spectra with the model of the secondary star when the actual lines of the secondary star are comparable to the noise level or when the spectrum actually belongs to a single star. We avoided such unreliable solutions by the criterion on the luminosity ratio ($\eta_n$ > 0.1), which has to be met in all of the four GALAH spectroscopic bands. In principle, the condition on the most extreme luminosity ratio is S/N dependent, meaning that in spectra with higher S/N, we can fit the lines of the secondary star more reliably, but we defer these considerations to future studies.

{\bf Criterion 4:} We expect an approximate agreement between the velocity separation of double lines ($\Delta V_r$) as determined by the binary pipeline and the one produced by the CCF detection method (see Sect.\,\ref{sec:ccf}). Upon inspecting the difference between the two values, we decided to use the large discrepancy between the two (> 20 km\,s$^{-1}$) as a criterion for rejection of poor solutions, which are mostly due to binary pipeline fit of the secondary component at arbitrarily large velocity separations in the case of noisy or otherwise peculiar spectra, where the binary signature is less evident.

\subsection{Ensemble properties} \label{sec:cor}

\begin{figure*}[!htp]
   \centering
   \includegraphics[width=\linewidth]{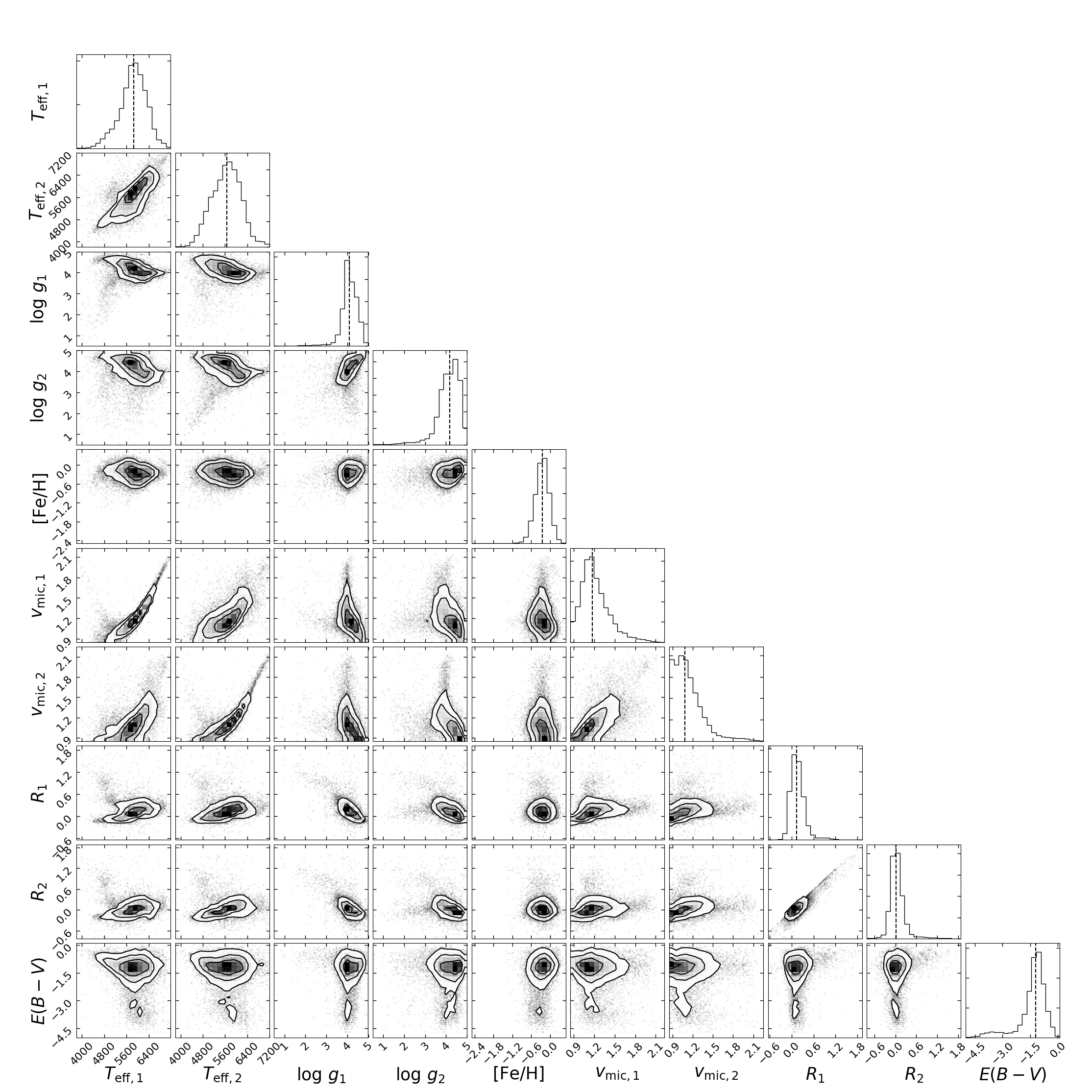}
   \caption{Ensemble of derived parameters for the final sample of analysed binary systems. Distributions of $\theta$ parameter values for 12\,760 analysed objects are shown with diagonal panels, whereas the other panels indicate correlations between these model parameters. The associated uncertainty distributions are displayed in Fig.\,\ref{mcmceres}. For clarity, we show the logarithm (base 10) of values for $R_1$, $R_2$, and $E(B-V)$. The parameter units are given in Table~\ref{modpar}.}
   \label{mcmcres}
\end{figure*}

As an additional confirmation of our results and to verify the adequacy of the selection criteria, we investigated ensemble distributions of model parameters $\theta$ derived by the binary pipeline and their interdependence for our final sample. First of all, Fig.\,\ref{mcmcres} shows some expected correlations such as the connection between the temperature and surface gravity for individual stars (illustrating the main-sequence) and the strong connection between temperature and microturbulence (see Eqs. (1) and (2) in \citealt{2018MNRAS.478.4513B}). The strong positive correlation between the temperatures of the primary and the secondary star stems from the constraint on the luminosity ratio, meaning that we cannot observe a hot primary in combination with a very cool secondary if we are to detect double lines in the spectrum. For the same reason, the radii of both stars in a binary system are also positively correlated.  

We also note the negative correlation between the temperature of the two stars and [Fe/H]. Concerning this anti-correlation, we checked the GALAH DR2 sample of reliable results for single dwarf stars (following the dwarf-giant distinction given by \citealt{2018MNRAS.481..645Z}), and we found a similar trend, with a measured Pearson coefficient of $-0.33$. This anti-correlation might be explained by the shortcomings of 1D hydrostatic models of stellar structure, which can overestimate the temperature of stellar atmospheres, especially in the lower metallicity regime (\citealt{2012MNRAS.427...27B} and references therein). Alternatively, this might be a feature of the underlying stellar population because when we consider equal-mass stars of different metallicities on the main-sequence, the more metal-poor stars have a smaller radius and higher surface temperature than their metal-rich counterparts.

\section{Results} \label{sec:results}

In this section we present the results of our analysis of binary stars. The catalogue contains the final sample as defined in Sect.~\ref{sec:selection}, and its contents are described in Appendix \ref{app:cat} (Table \ref{tab:cat}).

\subsection{Kiel diagram of binary stars}

\begin{figure*}[!htp]
   \centering
   \includegraphics[width=\linewidth]{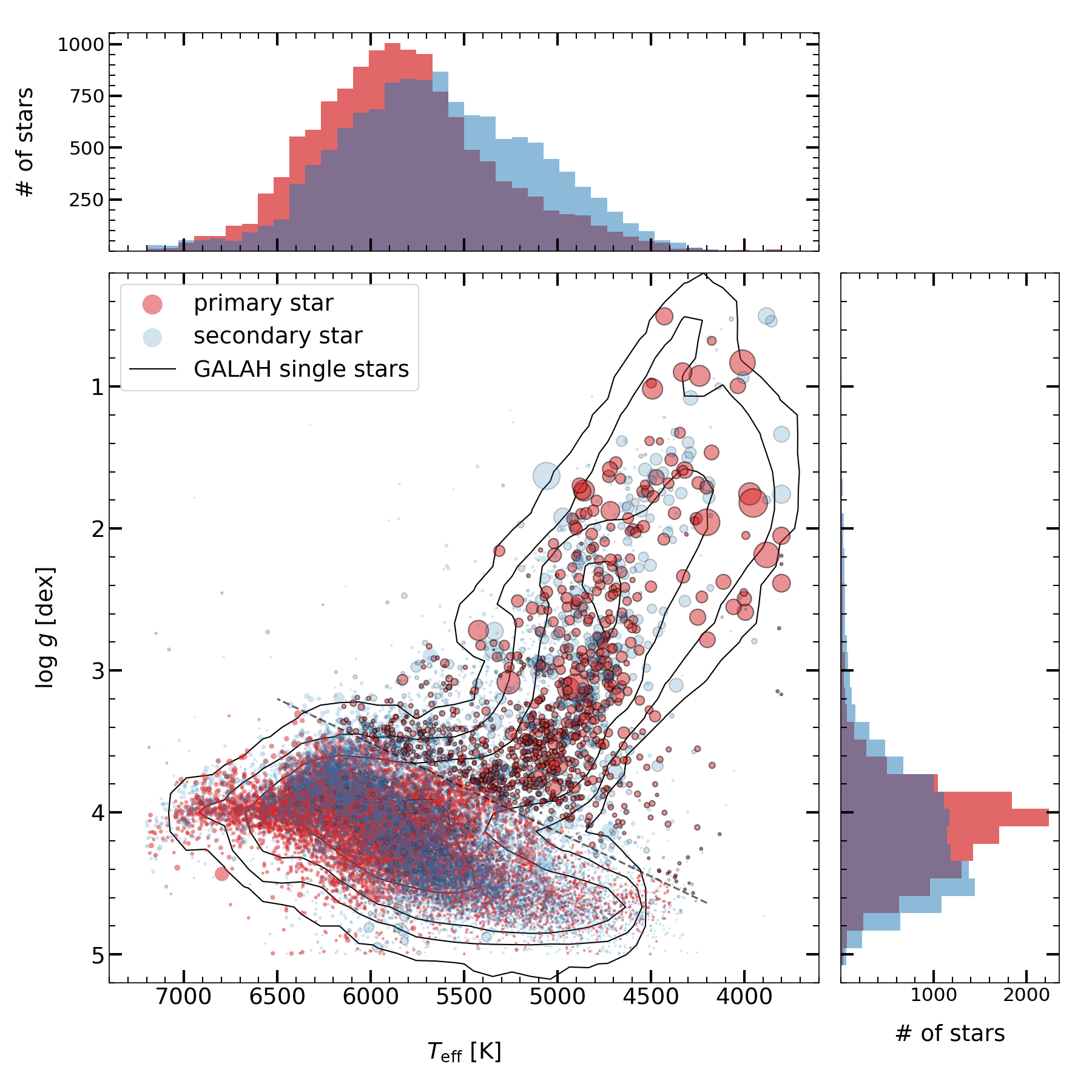}                        
   \caption{Kiel diagram of $T_{\rm eff}$ and $\log g$ parameters as derived by the binary pipeline for the final sample of binary stars. The dashed line denotes a dwarf-giant distinction given by \cite{2018MNRAS.481..645Z}, and the black edge of circles indicates systems where the primary lies above the line. The size of the circles relates to the radius of each star as derived by the binary pipeline, and the contours correspond to the density of the GALAH DR2 results for single stars (see \citealt{2018MNRAS.478.4513B}; we only used their results with {\sl flag\_cannon} = 0).}
   \label{kiel1}
\end{figure*}

Figure \ref{kiel1} shows the Kiel diagram for our final sample of binary stars. The positions of the primary stars agree well with the background distribution of the parameter values for single stars from the GALAH DR2 published results. As expected, the secondary stars are overall cooler and of higher surface gravity than the primary stars. However, the measured $T_{\rm eff}$ and $\log g$ values of secondary stars have larger uncertainties than those of primary stars (compare Fig.\,\ref{mcmceres}), which causes the secondaries to be scattered more around their true position in the different parts of the Kiel diagram. 

The poorly determined $\log g$ values of many secondary stars, whose radii indicate a main-sequence membership but which are instead positioned on the giant branch in Fig.\,\ref{kiel1}, cannot be simply filtered out by applying cuts on the RV separation of double lines or the luminosity ratio. This issue might be alleviated if the spectra used here included more diagnostic lines for $\log g$ determination such as Mg~b triplet or the Ca~I 6162 {\AA} lines \citep[][and references therein]{2016MNRAS.461.2174R}. It may also be that this is a chance alignment of a foreground dwarf (primary) and a background giant (secondary), in which case the binary pipeline would declare a small radius for the giant.

In general, we do not expect to find many double-lined binaries among giants for several reasons, for example, the constraint on similar luminosity and high enough orbital velocity. However, based on investigating all systems with a primary star situated above the dashed line in Fig.\,\ref{kiel1}, which marks the distinction between dwarfs and giants as defined in \citealt{2018MNRAS.481..645Z}, we confirm their true binary nature in approximately half of the cases, whereas for others, the double lines in spectra are either not conclusive or the fit is clearly problematic. No general rule for distinguishing between the true giant binary systems and anomalous ones in this part of the Kiel diagram exists, therefore we advise caution in using such solutions.

\subsection{Spatial distribution of binary stars}

\begin{figure*}[!htp]
   \centering
   \includegraphics[width=\linewidth]{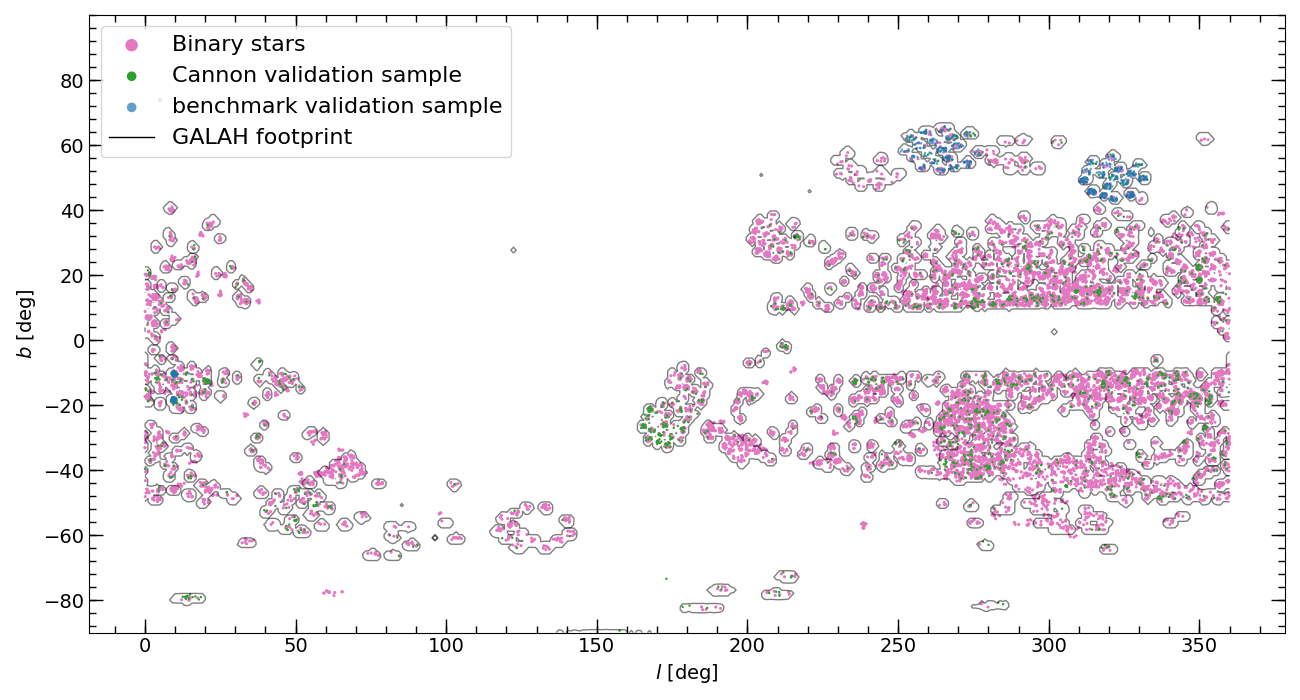}       
   \caption{Footprint of GALAH DR2 stars (black contours; single dwarf stars according to \citealt{2018MNRAS.481..645Z}), the final sample of binary stars (see Sect.~\ref{sec:selection}), {\it The Cannon} validation sample, and the benchmark validation sample (see Sect.~\ref{sec:validation}). GALAH stars include sources targeted by the GALAH pilot \citep{2018MNRAS.476.5216D}, K2-HERMES \citep{2018AJ....155...84W}, and TESS-HERMES \citep{2018MNRAS.473.2004S} surveys, which undergo the same reduction pipeline as the main GALAH survey.}
   \label{fig:foot}
\end{figure*}

\begin{figure}[!htp]
   \centering
   \includegraphics[width=\linewidth]{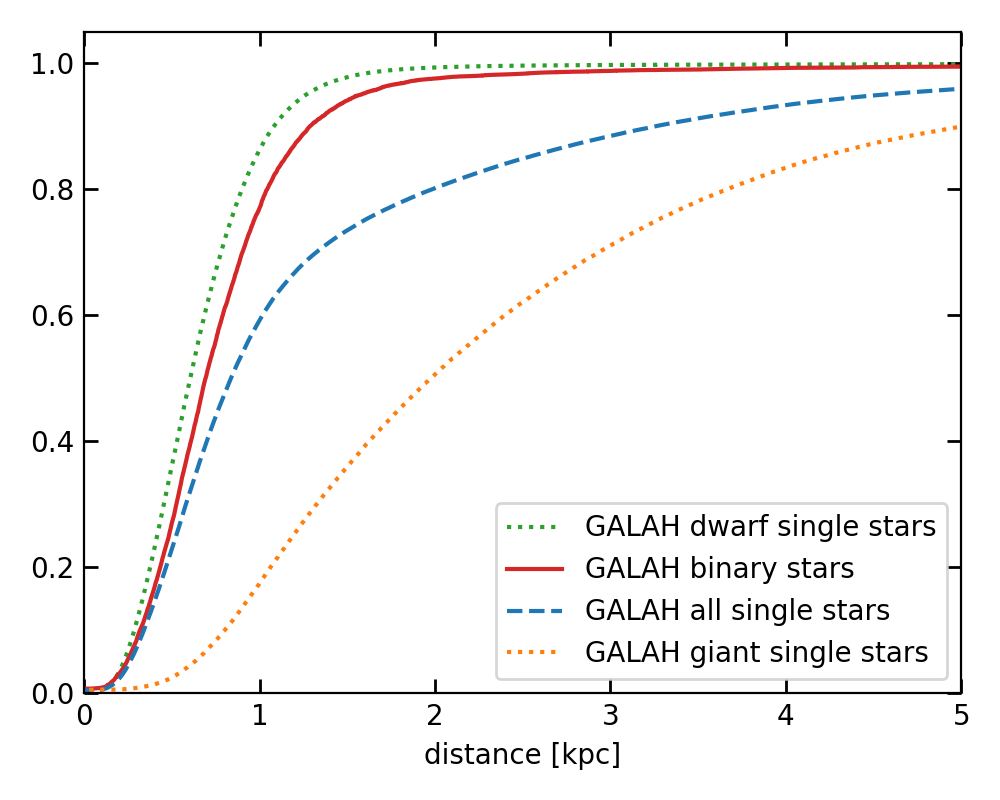}                        
   \caption{Cumulative distribution of the distances towards sources based on \textit{Gaia} DR2 parallaxes. The binary star sample is the sample defined in Sect.~\ref{sec:selection}, while the GALAH DR2 single-star samples are those with {\sl flag\_cannon} = 0 \citep{2018MNRAS.478.4513B} and the dwarf-giant distinction following \cite{2018MNRAS.481..645Z}.}   
   \label{fig:dist}
\end{figure}

The volume of the Galaxy probed by the GALAH survey is described in \cite{2018MNRAS.478.4513B}, while Figs. \ref{fig:foot} and \ref{fig:dist} place our sample of binary stars in the perspective of the survey footprint. The sky map shows a fairly homogeneous location of binary stars compared to all GALAH DR2 stars. The cumulative distributions of the distances in Fig.\,\ref{fig:dist} demonstrate that the binary stars are observed to larger distances than their single-dwarf counterparts because of the magnitude limit of the survey and because binary stars are intrinsically brighter.

\subsection{Extinction}

\begin{figure}[!htp]
   \centering
   \includegraphics[width=\linewidth]{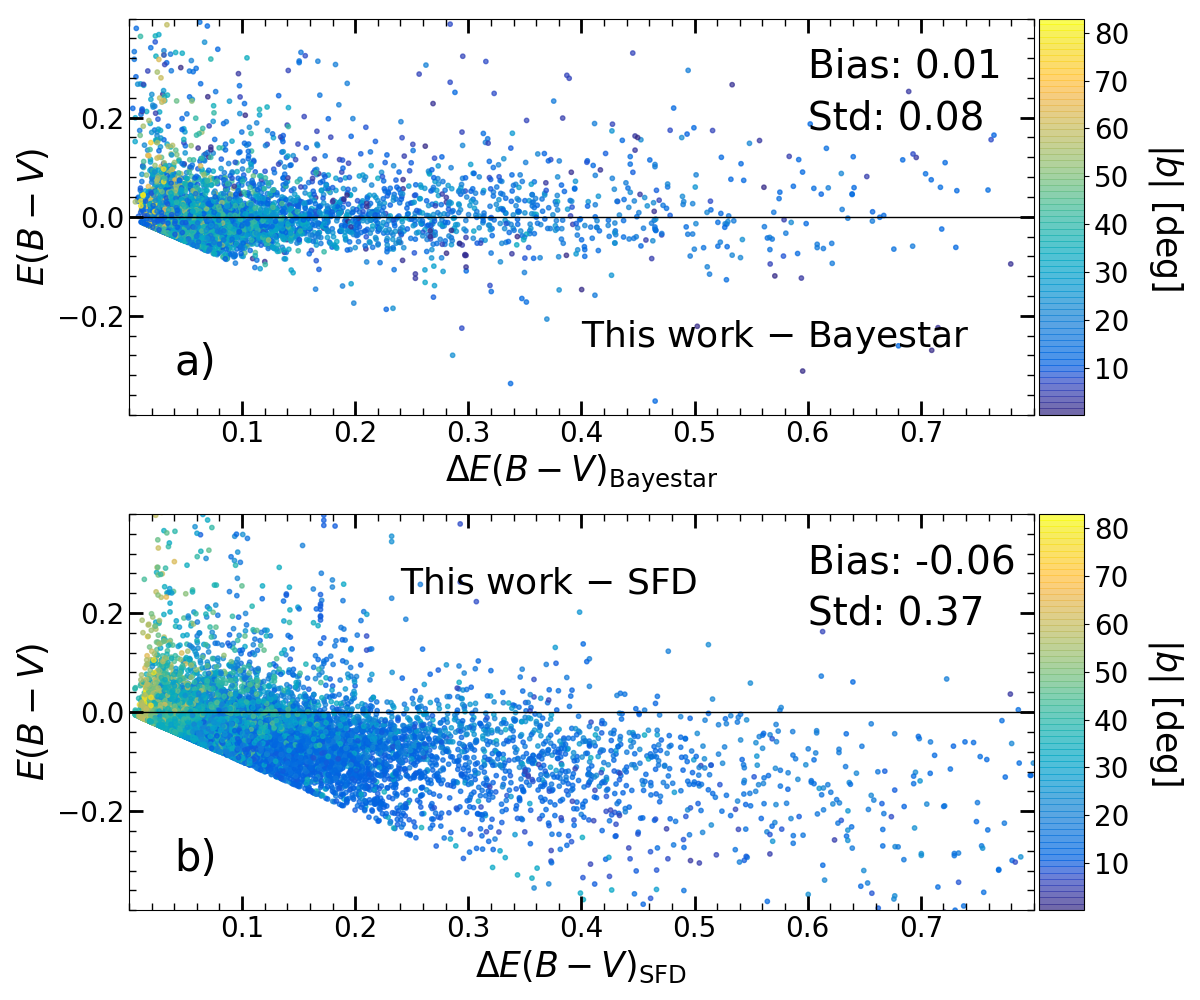}                        
   \caption{Comparison of $E(B-V)$ for the final sample of binary stars derived by the binary pipeline to a) Bayestar and b) SFD reddening maps. The colour-coding is by absolute Galactic latitude.}
   \label{extres}
\end{figure}

Following the same procedure as in Sect.~\ref{sec:validation}, we compared $E(B-V)$ values of the final sample of binary stars with the SFD and Bayestar reddening maps. Figure \ref{extres} shows a very good agreement with the Bayestar values for our sample, and the bias for the SFD values is systematic, which is expected for mostly nearby binary systems because SFD  integrates extinction along the line of sight. We emphasise that in comparison to the results of the binary pipeline for single stars shown in Fig.\,\ref{fig:teff_ebv}, the reddening parameter $E(B-V)$ appears to be better constrained in the case of binary stars because the photometric fit has less freedom for compensating the change in temperature by a corresponding change in $E(B-V)$. 

\subsection{Population statistics} \label{sec:popstat}

\begin{figure}[!htp]
   \centering
   \includegraphics[width=\linewidth]{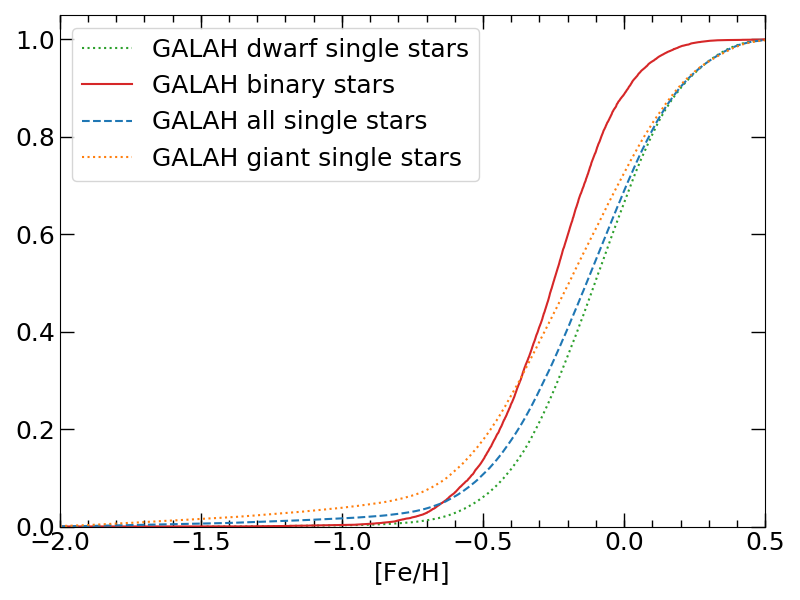}             
   \caption{Cumulative distribution function of ([Fe/H]) for different samples of stars as indicated in the legend. The binary star sample is the one defined in Sect.~\ref{sec:selection}, while the GALAH DR2 single-star samples are those with {\sl flag\_cannon} = 0 \citep{2018MNRAS.478.4513B} and the dwarf-giant distinction following \cite{2018MNRAS.481..645Z}.}
   \label{fig:mdf}
\end{figure}

We first examine the metallicity distribution function (MDF) of our final sample of binary stars. Fig.\,\ref{fig:mdf} shows the difference between the GALAH dwarf, giant, and binary stars. As expected, the giant stars have overall lower metallicities with a larger spread because they are located in a larger volume that reaches farther away from the Galactic plane than dwarf stars. The binary stars exhibit a narrow MDF reminiscent of the MDF for dwarf stars, but with lower metallicity values similar to those of giants. Several aspects need to be considered in trying to explain this, and we discuss them in Sect.~\ref{sec:discussion}.

\begin{figure*}[!htp]
   \centering
   \includegraphics[width=0.245\linewidth]{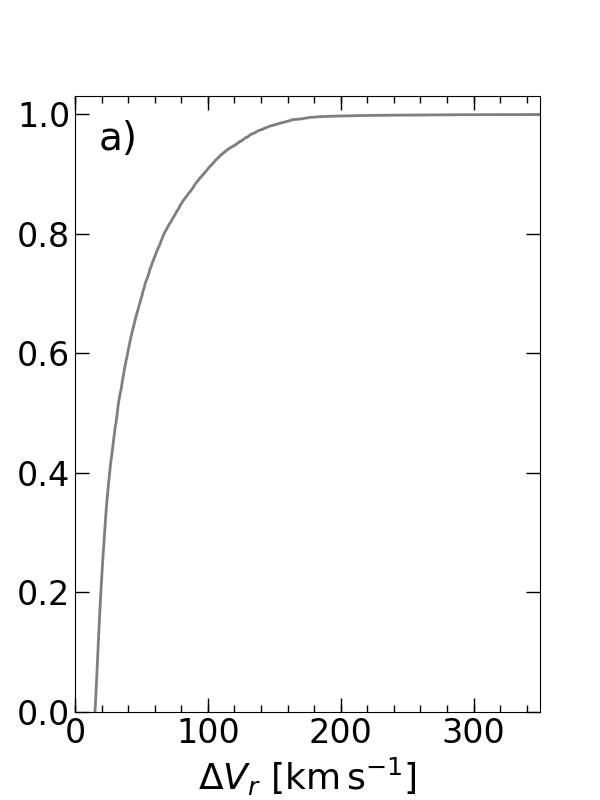}   
   \includegraphics[width=0.245\linewidth]{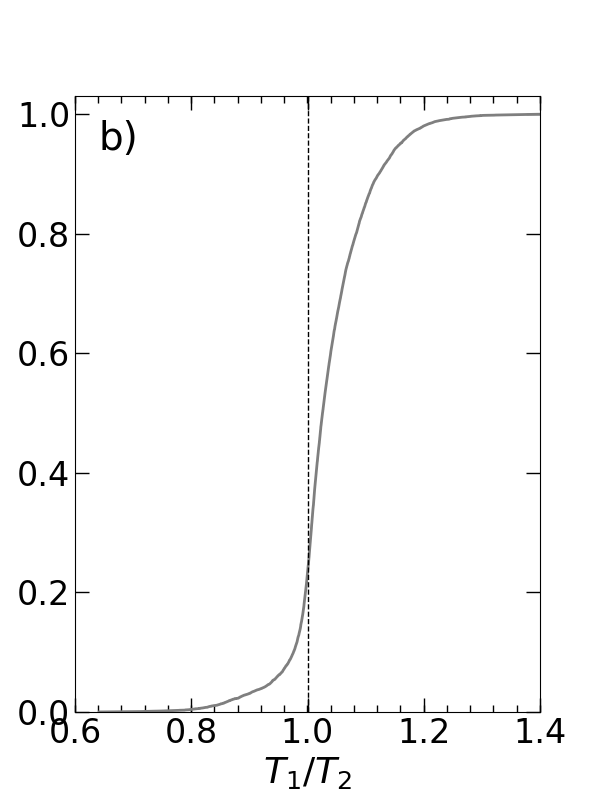}   
   \includegraphics[width=0.245\linewidth]{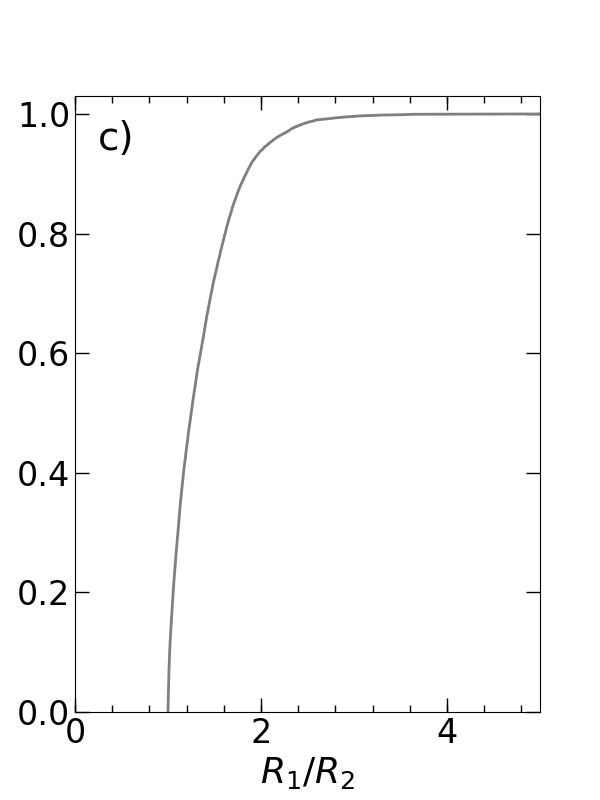}   
   \includegraphics[width=0.245\linewidth]{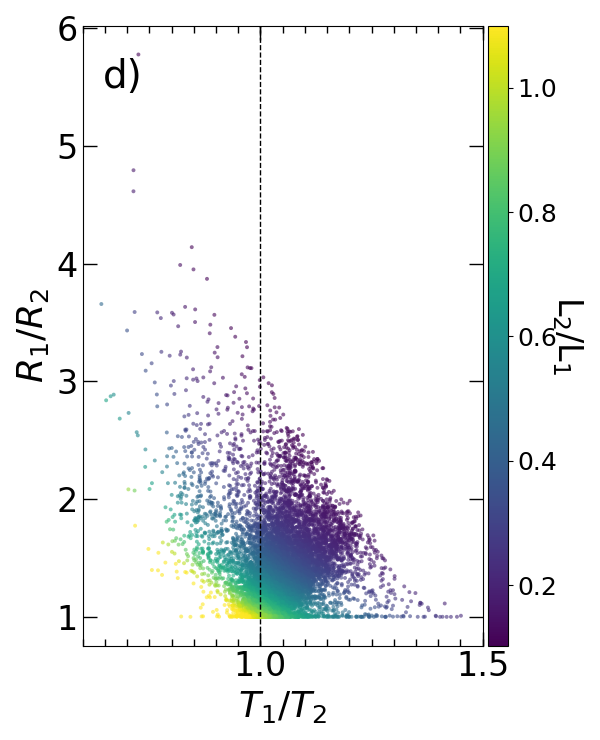}   
   \caption{Cumulative distribution functions of the RV separation ($\Delta V_r$) between the two stars in a binary system, ratio of temperatures ($T_1/T_2$), ratio of radii ($R_1/R_2$), and the relation between $T_1/T_2$ and $R_1/R_2$ in the fourth panel. The distribution of $\Delta V_r$ is cut at the lower level of 15 km\,s$^{-1}$ as explained in Sect.~\ref{sec:selection}, while $R_1/R_2$ starts at unity by construction because we define the primary as the star with a larger radius.}
   \label{fig:cumu}
\end{figure*}

The $\Delta V_r$ distribution in Fig.\,\ref{fig:cumu}a shows a strong decline of binary stars with increasing RV separation. This is expected in general because for larger $\Delta V_r$, the orbital period of the binary system must be short and the inclination high (edge-on view), and we have to catch the system at just the right time (close to quarter phase). Otherwise, these factors will greatly reduce the line-of-sight projected velocities of the two stars in the binary system and we will not see it.

The ratio of temperatures can indicate the evolutionary stage of stars in a binary system. Following our definition of the primary and secondary star, which states that the primary is the larger (Fig.~\ref{fig:cumu}c) and hence more massive star, and assuming that both are on the main-sequence, the condition $T_1 \geq T_2$ is true. It can also be true for systems in which one or both stars are already evolved, but $T_1 < T_2$ indicates that at least the  primary star has already evolved, and therefore the $T_1 < T_2$ condition sets the lower limit of binary stars with at least one component already evolved from the main-sequence. Fig.\,\ref{fig:cumu}b shows that there are $\sim$25\,\% of such binary stars in our final sample. When we assume that the majority of the remaining $\sim$75\,\% of systems (and both stars in them) lie on the main-sequence, and using standard relations (i.e. luminosity $\propto R^2 T^4 \propto M^{3.8}$), we estimate that we  probe systems of binary stars with a mass ratio down to 0.5. 

The relatively large fraction ($\sim$25\,\%) of binary stars with at least one evolved component can be partly attributed to the Malmquist bias because evolved stars are intrinsically brighter than their counterparts on the main-sequence and are therefore over-represented in a magnitude-limited sample of stars. On the other hand, it is also possible that the temperature estimate for the primary and secondary star has a different bias or systematics, which might result in a shift of the division that is indicated by Fig.\,\ref{fig:cumu}b.

Combining the ratio of temperatures and radii in Fig.\,\ref{fig:cumu}d, we note the high density of points that represents the main-sequence, while the largest relative difference in radii is observed for systems with already evolved stars. There is a clear indication that a diagonal line connects the highest ratio of radii and the highest ratio of temperatures, which suggests a forbidden region in the upper right part of the plot. This is due to the luminosity ratio restriction that we imposed in Sect.~\ref{sec:selection} (criterion 4).

\subsection{Note on the accuracy of the results}

All model parameters $\theta$ reported in the catalogue are accompanied by the corresponding uncertainties as given by our Bayesian scheme (see Sect.~\ref{sec:sample}). The distributions of these formal uncertainties are displayed in Fig.\,\ref{mcmceres}. However, as discussed in Sect.~\ref{sec:model}, our method ensures that these uncertainties are realistic only when all observational and model uncertainties are properly accounted for. 

While it would be possible to account for the reliability or uncertainty of the spectroscopic model given its internal precision (see Figs. \ref{BPsGTS}, \ref{cannons}, and \ref{mcmceres}) and the uncertainty of the parameters of the {\it \textup{training set}} spectra, the evaluation of the reliability of the photometric model represents a greater challenge. One reason for this is the lack of flux-calibrated observed spectra that would span the FGK parameter space considered in this work and could be compared to model SEDs. Another obstacle is the assessment of validity of passband transmission curves because they are defined for different observing locations, different reference systems, with varying availability and quality of information in the literature \citep[for more information on this, see][]{2000A&AS..147..361M,2003A&A...401..781F}. Additionally, the photometric observations, which cover a wide spectral range from visible to near-IR, suffer more from a complex interplay of diverse extinction effects than the spectroscopic data in the form of normalised spectra.

The method we presented here incorporates only observational uncertainties as reported in the source catalogues, but it is general enough to allow inclusion of model uncertainties or systematic biases in observational data. We provide columns of additional information in our catalogue that indicate the performance of our method and quality of the results, and the user is encouraged to take these into account (see also Sect.~\ref{sec:priors}). They are explained in Appendix~\ref{app:cat}. Additionally, some indication of the performance of our method is given in the following sections. 

\subsubsection{Internal consistency check with repeat observations} \label{sec:repeats}

In order to better evaluate the internal consistency of our method, we compared the derived binary parameters for systems that have repeat observations in GALAH. Distinct GALAH objects (as defined in Sect.~\ref{sec:data}) can represent repeat observations of the same star (or binary system), of which we have 341 binary systems with two repeats, followed by 36, 3, 1, and 2 systems with three, four, six, and seven repeats, respectively.

Although there are physical reasons for a possible change of parameters between repeat observations due to a different phase of the binary system (e.g. eclipses or reflection effects), and different results can be obtained due to a different quality of repeat observations (spectra), we expect that for the majority of cases, these differences are small.

We compute the maximum difference between repeat observations for each $\theta$ parameter and list the quantiles of the corresponding distributions in Table \ref{tab:diffs}. This shows, for example, that for the majority of binary systems with repeat observations, the agreement of parameter values that are constrained by the spectroscopic fit is comparable to the indicated uncertainty of the spectroscopic model (see Fig.~\ref{BPsGTS}). 

It is expected that the RV of the secondary component will change between repeats as a result of a change in binary phase, whereas the RV of the primary is expected to remain constant because spectra are predominantly shifted to the rest frame of the strongest spectral component in the main GALAH pipeline. Both these assumptions are confirmed by inspecting Table \ref{tab:diffs}. However, exceptions are possible in the case of equally massive components.

\begin{table} 
\caption{Quantiles of the distribution of absolute differences between $\theta$ parameter values for repeat observations of the same binary systems.}
\label{tab:diffs}
\centering
\begin{tabular}{l c c c c}
\hline\hline
Parameter & $\eta.25$ & $\eta.50$ & $\eta.75$ & Unit\\
\hline
$T_{\rm eff,1}$ & 28 & 63 & 115 & K\\ \
        $T_{\rm eff,2}$ & 42 & 87 & 175 & K\\ \
        $\log g_{1}$ & 0.06 & 0.14 & 0.26 & dex\\ \
        $\log g_{2}$ & 0.07 & 0.18 & 0.36 & dex\\ \
        $\mathrm{[Fe/H]}$ & 0.03 & 0.06 & 0.11 & dex\\ \
        $V_{r,1}$ & 0.3 & 0.9 & 3.0 & km\,s$^{-1}$\\ \
        $V_{r,2}$ & 4.9 & 26.8 & 75.4 & km\,s$^{-1}$\\ \
        $v_{\rm mic,1}$ & 0.03 & 0.06 & 0.10 & km\,s$^{-1}$\\ \
        $v_{\rm mic,2}$ & 0.03 & 0.06 & 0.12 & km\,s$^{-1}$\\ \
        $v_{\rm broad,1}$ & 0.2 & 0.9 & 2.5 & km\,s$^{-1}$\\ \
        $v_{\rm broad,2}$ & 0.1 & 0.7 & 2.2 & km\,s$^{-1}$\\ \
        $R_{1}$ & 0.01 & 0.02 & 0.05 & $R_{\odot}$\\ \
        $R_{2}$ & 0.01 & 0.03 & 0.07 & $R_{\odot}$\\ \
        $E(B-V)$ & 0.00 & 0.01 & 0.02 &\\

\hline
\end{tabular}
\end{table}

\subsubsection{Testing the co-evolution (same-age) assumption} \label{sec:ages}

We performed a cursory examination of the ages for the two components of our binary systems. The age estimates discussed here are derived following a Bayesian fitting of PARSEC isochrones introduced in \citet{2019A&A...622A..27H}, which has also recently been used by \citet{2019MNRAS.482..895S} to derive benchmark ages of Gaia benchmark stars. The input to the method are $\theta$ parameters $T_{\rm eff[1,2]}$, $\mathrm{[Fe/H]}$, and bolometric luminosities of the two components. The luminosities are computed using $T_{\rm eff[1,2]}$ and $R_{[1,2]}$, while the uncertainties for temperature and radii are compiled by summing the formal uncertainties and the median value from Table \ref{tab:diffs} in quadrature. The uncertainties of the luminosities are propagated in a Monte Carlo fashion given the uncertainties for temperature and radii. For each of the two stars in our final sample of binary systems, the method produces a PDF of the age estimate, and we extracted only the mode and the confidence limits of the PDF \citep[see section 3.6 of][confidence limits correspond to 1 $\sigma$ if the PDF is Gaussian]{2005A&A...436..127J}.

\begin{figure}[!htp]
   \centering
   \includegraphics[width=\linewidth]{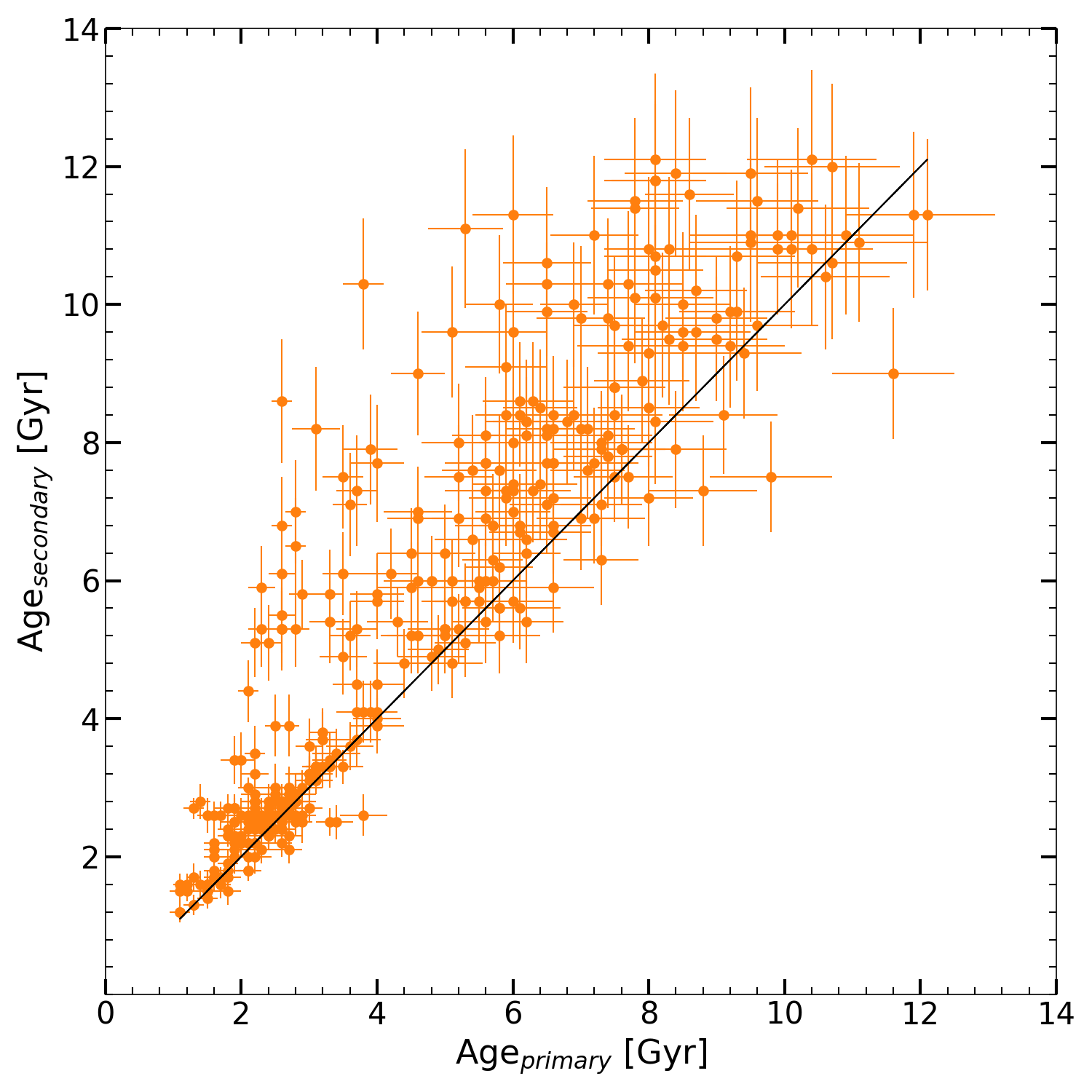}   
   \caption{Age comparison between the two components for binary systems with most reliable age estimates. The black line represents a one-to-one relation.}
   \label{fig:ages}
\end{figure} 

We used the average of the confidence limits to compute the relative uncertainty on the age estimate, and retained only those with a value below 10\%. As expected, these are mainly stars from the turn-off and sub-giant branch. The comparison of primary and secondary ages for this selection of binary systems is shown in Fig.\,\ref{fig:ages}. 

The ages of the primary and secondary star mostly agree given the individual age uncertainties, which is an encouraging confirmation of the validity of our results. However, we observe some outliers and a trend towards secondary stars being older than the primaries. This asymmetry can be explained by the potential systematic uncertainty in luminosity of the two components (i.e. one not reflected in the age uncertainties) and by considering how ages are computed. When we take two stars in a binary system whose true position is on the main-sequence, they will both move into the turn-off region   when we increase their luminosity. Their ages can be reliably determined there, and those results will appear in Fig.\,\ref{fig:ages}. However, in this case, the secondary will always fall on an isochrone of greater age than the primary. On the other hand, when we decrease the luminosity of both stars, they will move to a region of densely packed isochrones with very poor age determination and are then excluded from the selection shown in Fig.\,\ref{fig:ages}.

In our modelled binary systems, the scatter and a potential bias in luminosity towards greater values can come from several sources: a) uncertainty of parameters used for calculating luminosity, b) uncertainty of the parallax for binary sources \citep[see e.g. the recent study by ][]{2020arXiv200305467B}, c) underestimated Gaia DR2 global parallax zero-point correction and possible unknown local variations, and d) third-light contributions (e.g. foreground or background contamination, triple or higher order multiple systems). The latter is also supported by our close inspection of outliers in Fig.\,\ref{fig:ages}, where we identify a few resolved, nearly equal-mass triple systems, in which case the luminosity of an assumed binary system can be erroneously boosted by a factor of $\sim 1.5$. In this study, we apply a correction of +0.029 mas to the \textit{Gaia} DR2 parallax values, which slightly decreases derived luminosities, however, some authors \citep[see e.g. Figure~14 in ][]{2020MNRAS.tmp..522C} suggest larger corrections. Additionally, the complex morphology of isochrones can lead to several evolutionary stages being assigned to the same star, which can limit the accuracy of the age estimates presented here to $\pm 20 \%$ \citep[see e.g. discussions in ][]{2014EAS....65...99L,2019MNRAS.482..895S}.

\section{Discussion} \label{sec:discussion}

\subsection{Identification of binary stars}

The FGK binary stars investigated here represent systems with relatively short orbital periods and other properties that enable the detection of multiple lines in spectra, for instance, similar luminosities. We applied conservative selection criteria to obtain our final sample of binary stars, but the true number of binaries in the data is much greater.  Our detection methods might be further improved in order to extend the sample towards longer orbital periods and lower mass ratios (see also the methods in \citealt{2018MNRAS.476..528E} and Merle et al. 2019). 

Epoch photometry can be used to help identify binary star candidates in addition to spectroscopic identification. Eclipsing binaries, although much fewer in numbers and biased towards short-period systems, can be efficiently discovered by large-scale photometric surveys with high enough cadence. Facilities such as \textit{Gaia}, LSST \citep{ivezic2008lsst}, and TESS \citep{2015JATIS...1a4003R} promise to deliver orders-of-magnitude larger binary star samples than we currently have, but it will be a challenge to secure high-quality spectra for even a fraction of these detections in order to perform an analysis such as presented in this work.

Another detection approach is possible by considering high-quality astrometry, which enables the comparison of an astrometric parallax with the luminosity parallax. The latter can be inferred by knowing the absolute magnitude of the stellar system, which can be deduced based on the standard stellar parameters ($T_{\rm eff}$, log $g$, [Fe/H]) measured from e.g. assumed single-star spectra. The binarity would be revealed in case of a significant mismatch between the two parallaxes, and this approach is discussed and attempted in \cite{2019arXiv190809727X}. Similarly, \citet{2019arXiv190404841C} identified objects that are more luminous than predicted by their spectral type and \textit{Gaia} distance, and specifically focused on double and triple Sun-like stars. Alternatively, comparing the results of spectral analysis that incorporate either a single or a binary star model can be used to distinguish between the two realities based on which model fits the data better, as demonstrated by \cite{2018MNRAS.476..528E}. The approaches described in this paragraph demand that detailed spectral analysis is performed on the whole data-set before attempting to identify binarity, however.

When we consider relatively wider binary systems, a detection method based on comparing 6D phase-space information combined with chemistry can be undertaken to infer binarity \citep[see e.g.][]{2019MNRAS.482.5302S}. However, when the angular separation between the two stars is sufficient, such a study would then transition to separate observations rather than observing light of the two stars in a single spectrum. 

One potential contamination not considered here is chance alignments. Two stars can be positioned at angular distances smaller than the fibre aperture and thus be treated as a single target, although they can be spatially completely unrelated and thus can falsely pretend to represent a binary star. This might occur in GALAH, although efforts were made in the construction of the input catalogue (based on 2MASS) to avoid such contamination. A pilot study using a model of the Galaxy and observed fields of the GALAH survey suggests that these chance alignments are possible, especially towards lower Galactic latitudes, but more work is needed to properly account for them.

\subsection{Undetected binarity and the effect on stellar parameters}

In the context of Galactic archaeology, it is important to detect binary spectra, whether they result from chance alignments or actual gravitationally bound stars. Large-scale spectroscopic surveys should be aware of the influence of binarity on their analysis because some recent studies have shown that neglecting the proper treatment of binary spectra introduces systematic biases in the derived stellar parameters and elemental abundances that are not negligible.

\citet{2010ApJ...719..996S} reported the uncertainty from binarity in SEGUE \citep{2009AJ....137.4377Y} targets for spectra of S/N $\geq$ 25 to be $\sim 80$ K in temperature and $\sim 0.1$ dex in [Fe/H], with a smaller impact on $\log g$, while a study by \citet{10.1093/mnras/stx2758} estimated typical systematic errors of 300 K in temperature, 0.1 dex in [Fe/H], and 0.1 dex in $\log g$ for APOGEE-like spectra of solar-type stars. A proper investigation of this effect is therefore essential in order to make the findings of current and future projects dedicated to Galactic archaeology more reliable \citep[see also][]{2019arXiv190809727X}.

\subsection{Systematically lower metallicity of the final sample} \label{sec:diffusion}

\begin{figure}[!htp]
   \centering
   \includegraphics[width=\linewidth]{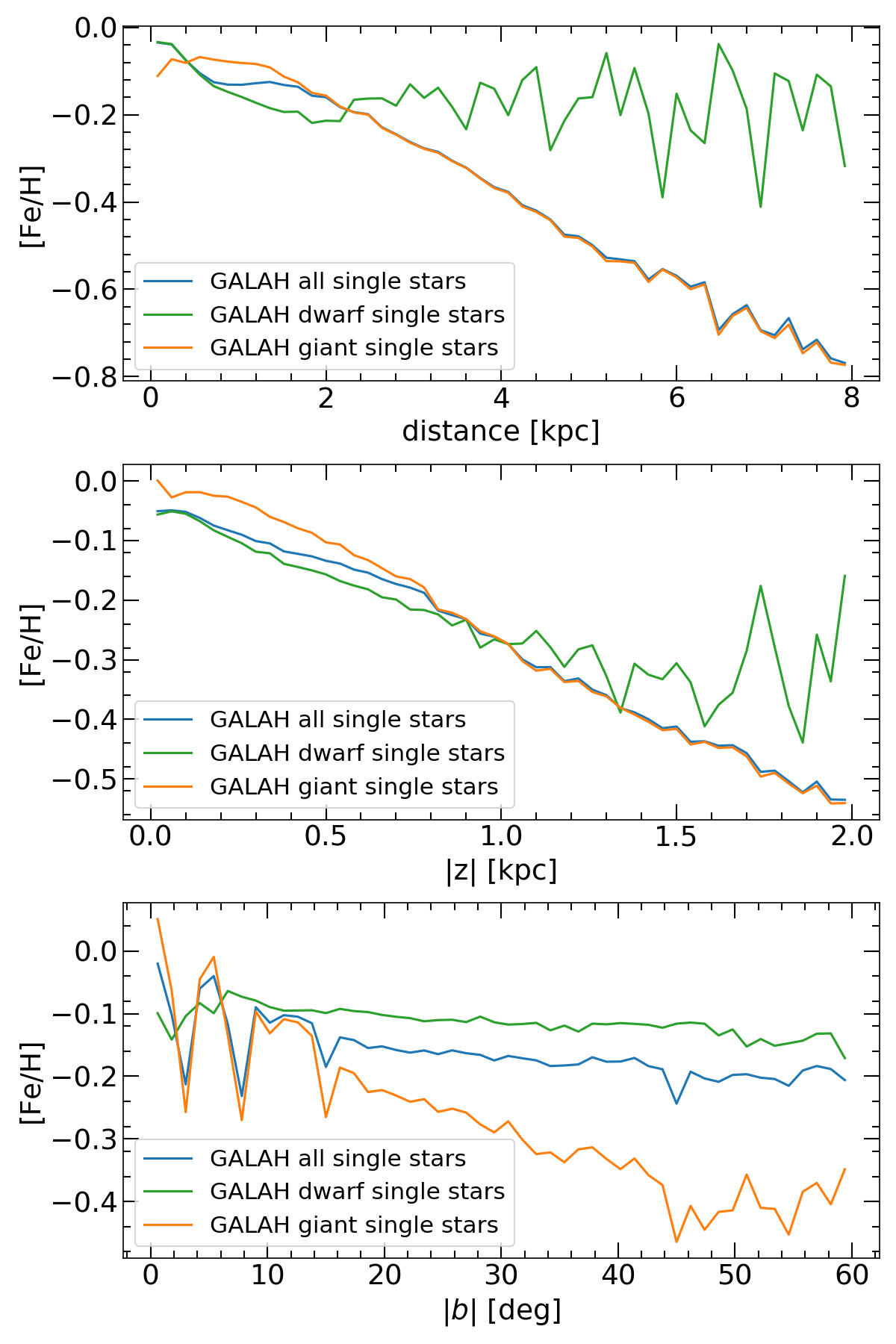}             
   \caption{Mean metallicity of GALAH DR2 single stars binned by distance (top), distance from the Galactic plane (middle), and absolute Galactic latitude (bottom). The GALAH DR2 single-star samples are those with {\sl flag\_cannon} = 0 \citep{2018MNRAS.478.4513B}, and the dwarf-giant distinction following \citet{2018MNRAS.481..645Z}. The sawtooth-shaped part of the lines indicates very few data points in the corresponding bins.}
   \label{fig:feh_dist_binned}
\end{figure}

Figure \ref{fig:mdf} clearly shows an overall lower metallicity of the final sample of binary systems investigated in this work (median [Fe/H] $\sim 0.2$ dex lower compared to GALAH single dwarfs). To some extent, as indicated in Fig.\,\ref{fig:dist}, binary stars probe a more distant volume of space than single dwarf stars because their luminosity is intrinsically higher. Some change towards lower metallicities can be expected because of this: the trend of decreasing metallicity with distance (as well as with height above Galactic plane) is observed for GALAH stars (see Fig.\,\ref{fig:feh_dist_binned}). Furthermore, the distinction between GALAH single giant and dwarf stars adopted here \citep{2018MNRAS.481..645Z}, and the fact that there is a significant number of evolved systems in our binary star sample (see Figs. \ref{kiel1} and \ref{fig:cumu}), make a comparison between single dwarf stars and binary stars more difficult. These binary systems with evolved and giant stars push the distance limit even further. 

We plan to further investigate the comparison of metallicity values for single versus binary GALAH stars. The metallicity comparison can be performed per each small volume of physical space where both single and binary stars are located, while also examining the orbits of stars to determine their common origin. This analysis is defered to a future study because it will greatly benefit from the next \textit{Gaia} data release (eDR3), which will bring more accurate 6D phase-space information.

Another aspect of lower metallicity values of our binary stars has to do with the fitting of the model to data for single as compared to binary stars. In general, there is a potential bias in the spectroscopically derived metallicity because the fit suffers to some extent from the degeneracy between temperature and metallicity, for example. However, in the case of binary stars, this degeneracy should be alleviated by the assumed common metallicity, which is now constrained by two sets of spectral lines. We expect this to be the case even more so when the two stars in a binary system have significantly different temperatures. On the other hand, when we consider binary systems where the mass ratio of the two stars or their age is such that they are found in significantly different evolutionary stages, we would have to consider effects of atomic diffusion and radiative levitation that can challenge the common metallicity assumption.

\begin{figure}[!htp]
   \centering
   \includegraphics[width=\linewidth]{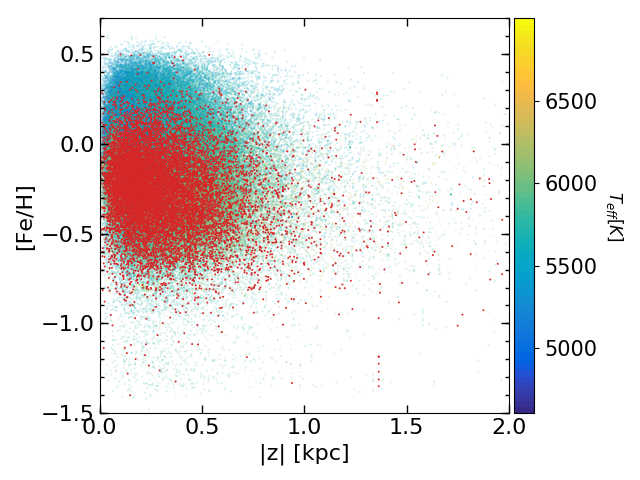}                        
   \caption{Relationship between [Fe/H] and distance from the Galactic plane for our final binary star sample as defined in Sect.~\ref{sec:selection} and GALAH DR2 single-star sample with {\sl flag\_cannon} = 0 \citep{2018MNRAS.478.4513B}, where GALAH single stars are colour-coded by temperature and the binary stars are plotted in red.}
   \label{fig:mdf_dist}
\end{figure}

It is also possible that our binary identification techniques are unable to identify a significant portion of metal-rich binary spectra. In order to investigate this, we studied the shape of the CCF for a few dozen representative binary spectra that were created by summing GALAH DR2 single-star spectra. These binary spectra were split into three metallicity bins ([Fe/H] < -0.5, -0.5 < [Fe/H] < -0.1, [Fe/H] > -0.1) defined with the help of Fig.\,\ref{fig:mdf_dist}, and the two stellar components put at $\Delta V_r$ of 15 km\,s$^{-1}$ and higher, but no obvious diagnostic issue of the resulting CCFs was found for any of the metallicity/$\Delta V_r$ combinations.

Finally, the change to lower metallicities might be attributed to the widely observed negative correlation between metallicity and binary fraction for short-period binary systems. From the observational side, this has recently been addressed by \citet{2019ApJ...875...61M}, who examined five different surveys and thoroughly accounted for their underlying selection biases. They measured the intrinsic occurrence rate of close solar-type binaries and showed that all five observational samples exhibit a quantitatively consistent anti-correlation between the close binary star fraction (a $<$ 10 AU; this type of binaries also make up our final sample) and metallicity. Furthermore, hydrodynamical simulations performed by \citet{2019MNRAS.484.2341B} agreed with this observed trend, with the explanation that the greater cooling rates at high gas densities that are due to the lower opacities at low metallicities increase the fragmentation on small spatial scales, leading to a higher ratio of close binary stars.

\subsection{Improvements of the model and the fitting procedure}

The spectroscopic model for the data we presented here is based on a data-driven technique, {\it The Cannon}, which is trained on a training set analysed by a model-driven method, SME. We argue that this combination might be ideal because model-driven techniques are considered to be theoretically accurate although generally more expensive (and thus more appropriate for smaller data-sets), while data-driven ones are fast and can account for systematics in the data, but can be limited by their complexity and availability of good-quality and sufficient training data. Selecting good training data is a complex process and subject of ongoing investigation, where unsupervised classification methods and data visualisation (see Sect.~\ref{sec:detect}) can help greatly in determining the types and number of distinct objects (spectra) in a given data-set. 

The spectral flux modelling can also be improved with readily available techniques that would allow for a higher complexity or flexibility of the spectroscopic model, such as the neural-networks-based {\it The Payne} \citep[developed as a model emulator by ][see their Fig.~1]{2019ApJ...879...69T} and its data-driven counterparts \citep[see e.g.][]{2019arXiv190809727X}. Additionally, there are other complex physical aspects of flux modelling that might be crucial for our work, such as proper treatment of the limb darkening and rotational velocity that it depends on, the effects of which can for example change the fitted $T_{\rm eff}$. Improvements in this direction can be achieved through higher complexity of the spectroscopic model, a better and larger training set, or even by post-processing of modelled spectra.

The photometric fit is strongly influenced by the \textit{Gaia} photometry, but because of the large width of the three \textit{Gaia} passbands, these are mostly important for constraining the total flux rather than the overall shape of the SED, and thus inclusion of more photometric data points (passbands) would benefit the accuracy of results. Furthermore, we here relied on the PySynPhot package to provide model-driven SEDs for our photometric model. However, in the future, a sufficient collection of observed SEDs might become available in order to train a data-driven model. Additionally, construction of SEDs provided by PySynPhot represents the majority of computational time in our method, thus it is possible that a method such as {\it The Cannon} or {\it The Payne} might perform better, having been trained either on model-driven or data-driven training data. 

Passband transmission functions are another aspect of the photometric model where uncertainties are sometimes not known or difficult to estimate. The revised passbands for \textit{Gaia} DR2 are an example of a step forward in this context \citep{2018A&A...617A.138W}. Moreover, synthetic photometry involving photometric systems that have been defined from the ground and at different locations, or photometry that is given in the Vega system, are often only accurate to about 5\,\% \citep{2013ascl.soft03023S}, which mostly pertains to the overall zero-point comparison between synthetic and observed magnitudes, rather than the shape of the SED. This might explain the discrepancy between the radii derived here for single stars and the seismic values (see Sect.~\ref{sec:validation}).

From a purely technical point of view, the convergence of our method to the best possible solution, or in other words, to the global minimum, is not strictly guaranteed. We chose the scheme of MCMC sampling of the posterior distribution function in order to explore the whole model parameter space. However, we implement some custom functionalities to make it more time efficient, and thus it can happen that in some cases, the convergence is not optimal. Solutions potentially include the use of Hamiltonian Monte Carlo (HMC; \citealt{neal2011mcmc}), which effectively reduces the number of Markov chain samples required for the same result, and other methods from the statistical modelling and data analysis platform (STAN; \citealt{Carpenter_stan:a}), such as variational inference for a much quicker first estimate of parameter distributions.

\subsection{Comparison to previous works}

This paper presents a complementary effort to previous studies of binary stars performed in the context of large all-sky surveys \citep{2009A&A...501..941H,2010AJ....140..184M,2014ApJ...788L..37G,2017A&A...608A..95M,2018AAS...23124416S,2018MNRAS.476..528E,2019AJ....157..196K}. The binary star samples investigated in those studies were connected to diverse observational strategies, instrumentation used by the corresponding surveys (e.g. the Geneva-Copenhagen Survey - GCS, \citealt{2009A&A...501..941H}; RAVE, \citealt{2006AJ....132.1645S}; Sloan Digital Sky Survey - SDSS, \citealt{2017AJ....154...28B}), the methods, and derived parameters, which makes them difficult to directly compare to our sample. Purely in terms of the number of identified and analysed binary systems, our sample of 12\,760 SB2s is a significant increase, for example, from  \citet{2010AJ....140..184M}, \citet{2011AJ....141..200M}, \citet{2017A&A...608A..95M}, or \citet[][$\sim$ 3000 binary stars]{2018MNRAS.476..528E}, mainly by virtue of the size and properties of the GALAH data-set. Additionally, our sample includes several dozen SB2 giant systems, which otherwise are a rare occurrence, with only a handful known in the literature \citep[see e.g. ][]{2017A&A...608A..95M}. 

The philosophy and approach of combining all available observational data in order to extract the maximum amount of information from studied objects have previously been proposed and developed for example by \cite{10.1093/mnras/stu1072}, and recently by \cite{2019arXiv190707690C}. In contrast to these studies of single stars, we avoid using models of stellar evolution (e.g. isochrone fitting) in order to obtain the properties of both stars in a binary system with the least amount of model-dependent assumptions as possible. However, the framework that we develop can be easily extended to include more theoretical background if needed. We also recognise the work by \cite{2016AJ....152..180S} on benchmarking astrometric parallaxes as a parallel effort to combine different observational data in the context of binary stars, albeit with a different prior or model configuration and desired outcomes.

\subsection{\textit{Gaia} revolution}

In the context of statistics of binary stars, it is important to mention \textit{Gaia} DR3, which is scheduled for the second half of 2021, in which astrometric solutions for binaries will be published, along with a predicted about$\text{}$ one million eclipsing binaries identified during processing of \textit{Gaia} DR2, and solutions for many spectroscopic binaries. \textit{Gaia} data in synergy with other data-sets of binary stars, such as the one presented in this work, have the potential to truly revolutionise the field of multiplicity statistics across the full range of primary masses, mass ratios, orbital periods, eccentricities, chemistry, ages and evolutionary states, and environments.

\section{Conclusions} \label{sec:conclusions}

We presented a general comprehensive method for the analysis of double-lined FGK binary stars, validated and applied on a large sample of identified SB2 systems. We identified binary stars in the GALAH spectroscopic sample using two different approaches: a sophisticated method based on the cross-correlation function, and a classification approach based on the dimensionality reduction technique t-SNE. By combining candidates from these two identifications, we obtained a sample of 19773 candidate SB2s out of 587\,153 spectra from the second GALAH internal data release.

To characterise the binary candidates, we combined photometric, spectroscopic, and astrometric data in a Bayesian framework and used an MCMC method to obtain the best-fit model parameters. We validated our method on a benchmark sample of 620 single stars to confirm that our derived parameters are consistent with those from external sources, such as the effective temperature from the IRFM method or stellar radius from asteroseismology. Following the analysis of candidate binary systems and an application of conservative selection criteria, we present a catalogue of 12\,760 reliable detections. We derived standard stellar parameters for the two stars in a binary system, their radii, and interstellar reddening. 

Based on the parameters derived here, we conclude that our catalogue represents a population of close binary systems (a $<$ 10 AU) with mass ratios $0.5 \leq q \leq 1$ that can be detected through double lines in their spectra. The results additionally indicate an overall lower metallicity of binary stars compared to single dwarf stars that are observed in the same magnitude-limited sample of the GALAH survey. Some of the parameters are used to construct cumulative distributions shown in Fig.\,\ref{fig:mdf}, which can provide further insight into the statistical properties of the analysed population of binary stars.

We aim to improve and employ our method for the future GALAH releases, and to apply it to forthcoming stellar spectroscopic surveys such as WEAVE \citep{2012SPIE.8446E..0PD} and 4MOST \citep{2012SPIE.8446E..0TD}.

\begin{acknowledgements}

We acknowledge helpful conversations with Paul McMillan and Thomas Bensby. We are grateful to the referee for the very useful comments which also encouraged us to investigate certain aspects of our analysis that improved the understanding and validity of our results.\\

We acknowledge the use of data reduction infrastructure developed within the GALAH Survey collaboration. The GALAH survey web site is www.galah-survey.org. This work is based on data acquired through the Australian Astronomical Observatory, under programmes: A/2014A/25, A/2015A/19, A2017A/18 (The GALAH survey); A/2015A/03, A/2015B/19, A/2016A/22, A/2016B/12, A/2017A/14 (The K2-HERMES K2-follow-up program); A/2016B/10 (The HERMES-TESS program); A/2015B/01 (Accurate physical parameters of Kepler K2 planet search targets); S/2015A/012 (Planets in clusters with K2). We acknowledge the traditional owners of the land on which the AAT stands, the Gamilaraay people, and pay our respects to elders past and present. 

This work has made use of data from the European Space Agency (ESA) mission \textit{Gaia} (https://www.cosmos.esa.int/gaia), processed by the \textit{Gaia} Data Processing and Analysis Consortium (DPAC, https://www.cosmos.esa.int/web/gaia/dpac/consortium). Funding for the DPAC has been provided by national institutions, in particular the institutions participating in the \textit{Gaia} Multilateral Agreement.\\

This research was made possible through the use of the AAVSO Photometric All-Sky Survey (APASS), funded by the Robert Martin Ayers Sciences Fund and NSF AST-1412587.\\

This publication makes use of data products from the Two Micron All Sky Survey, which is a joint project of the University of Massachusetts and the Infrared Processing and Analysis Center/California Institute of Technology, funded by the National Aeronautics and Space Administration and the National Science Foundation.\\

This publication makes use of data products from the Wide-field Infrared Survey Explorer, which is a joint project of the University of California, Los Angeles, and the Jet Propulsion Laboratory/California Institute of Technology, funded by the National Aeronautics and Space Administration.

This research has made use of the NASA/ IPAC Infrared Science Archive, which is operated by the Jet Propulsion Laboratory, California Institute of Technology, under contract with the National Aeronautics and Space Administration.\\

This work was supported by the Swedish strategic research programme eSSENCE. G.T. and S.F. were supported by the project grant ''The New Milky Way'' from the Knut and Alice Wallenberg foundation and by the grant 2016-03412 from the Swedish Research Council. T.M. is supported by a grant from the Fondation ULB. Y.S.T. is grateful to be supported by the NASA Hubble Fellowship grant HST-HF2-51425.001 awarded by the Space Telescope Science Institute. G.T., K.Č., and T.Z. acknowledge financial support of the Slovenian Research Agency (research core funding No. P1-0188 and project N1-0040).\\

The computations were partially performed on resources provided by the Swedish National Infrastructure for Computing (SNIC) at LUNARC partially funded by the Swedish Research Council through grant agreement no. 2016-07213 and partially funded by the Royal Fysiographic Society of Lund.\\

JDS acknowledges the support of the Australian Research Council through Discovery Project grant DP180101791. SB acknowledges funds from the Alexander von Humboldt Foundation in the framework of the Sofja Kovalevskaja Award endowed by the Federal Ministry of Education and Research. This research has been supported by the Australian Research Council (grants DP150100250 and DP160103747). This research was partly supported by the Australian Research Council Centre of Excellence for All Sky Astrophysics in 3 Dimensions (ASTRO 3D), through project number CE170100013.\\

This research has made use of NASA's Astrophysics Data System and the CDS services (Strasbourg, France).
 
\end{acknowledgements}

\bibliographystyle{aa}
\bibliography{sample}

\begin{appendix}

\section{The catalogue} \label{app:cat}

The binary star parameters derived in this work are presented in the catalogue, which is described in Table \ref{tab:cat}. The columns chi2\_binary\_pipeline, chi2\_binary\_pipeline\_spec, and chi2\_binary\_pipeline\_phot represent reduced $\chi^2$ values (average $\chi^2$ per data point) for all data points, only the spectroscopic, and only the photometric points, respectively. These can be used as quality indicators or the formal goodness of fit. If all the uncertainties of our analysis scheme were properly accounted for, all these values would be close to 1. 

These values, however, can be large for several reasons such as strong oversubtraction of the background light in the spectral reduction, unrecognised high values of noise in spectra, improper continuum normalisation and discordant wavelength solution in spectra, erroneous photometric data, underestimated observational uncertainties, issues with the spectroscopic or photometric model, and failed convergence of the solution. Conversely, the reduced $\chi^2$ values can also be too low, for example, because the observational uncertainties are overestimated. For example, the median of $\chi^2_{\rm spec}$ values for our results is $\sim$ 0.8, which implies that our spectroscopic observational uncertainties, as derived in \citet{2017MNRAS.464.1259K}, might be slightly overestimated. One reason for this might also be correlations between different contributions to the observational uncertainty budget.

It is beyond scope of this work to determine which of these issues influence our reduced $\chi^2$ values in a specific case, but we aim to conduct this proper analysis in the future. The user of the data presented in Table \ref{tab:cat} should be aware of the possible underlying factors. 

The reported radius of the stars in binary systems is directly related to the value of parallax, therefore we encourage users to investigate the quality of the parallax measurement when the radii values derived here are used. This can be aided by the RUWE parameter from the Gaia DR2 catalogue (included in Table \ref{tab:cat}).

\begin{table*}
\caption{Description of the catalogue of 12\,760 successfully analysed binary stars, available in electronic form at the CDS. The columns in this table that list direct products of the binary pipeline (from teff1 to r2) are placeholders for actual columns in the catalogue, where each placeholder contains five columns to report the 16th$^{\rm }$, 50th$^{\rm }$, and 84th$^{\rm }$ percentile along withe the mean and mode of the posterior distributions (e.g. teff1\_16, teff1\_50, teff1\_84, teff1\_mean, and teff1\_mode for the effective temperature of the primary component, see Sect.~\ref{sec:sample} for more details).}
\label{tab:cat}
\centering
\small
\begin{tabular}{l l l}
\hline\hline
Column & Unit & Description\\
\hline
sobject\_id & & Unique per-observation star ID (spectrum ID)\\
star\_id &  & 2MASS ID number by default, UCAC4 ID number if 2MASS unavailable (begins with UCAC4-)\\
gaia\_id & & \textit{Gaia} DR2 identifier\\
field\_id & & GALAH field identification number\\
raj2000 & deg & Right ascension from 2MASS, J2000\\
dej2000 & deg & Declination from 2MASS, J2000\\
Bmag & mag & B magnitude from APASS\\
e\_Bmag & mag & Uncertainty of Bmag\\
Vmag & mag & V magnitude from APASS\\
e\_Vmag & mag & Uncertainty of Vmag\\
GBPmag & mag & G$_{BP}$ magnitude from \textit{Gaia} DR2\\
e\_GBPmag & mag & Uncertainty of GBPmag\\
Gmag & mag & G magnitude from \textit{Gaia} DR2\\
e\_Gmag & mag & Uncertainty of Gmag\\
GRPmag & mag & G$_{RP}$ magnitude from \textit{Gaia} DR2\\
e\_GRPmag & mag & Uncertainty of GRPmag\\
Jmag & mag & J magnitude from 2MASS\\
e\_Jmag & mag & Uncertainty of Jmag\\
Hmag & mag & H magnitude from 2MASS\\
e\_Hmag & mag & Uncertainty of Hmag\\
Kmag & mag & Ks magnitude from 2MASS\\
e\_Kmag & mag & Uncertainty of Kmag\\
W1mag & mag & W1 magnitude from WISE\\
e\_W1mag & mag & Uncertainty of W1mag\\
W2mag & mag & W2 magnitude from WISE\\
e\_W2mag & mag & Uncertainty of W2mag\\
parallax & mas & Parallax from \textit{Gaia} DR2\\
e\_parallax & mas & Uncertainty of parallax\\
sn\_c1 & & S/N per pixel in the HERMES blue channel (GALAH spectral band 1)\\
sn\_c2 & & S/N per pixel in the HERMES green channel (GALAH spectral band 2)\\
sn\_c3 & & S/N per pixel in the HERMES red channel (GALAH spectral band 3)\\
sn\_c4 & & S/N per pixel in the HERMES IR channel (GALAH spectral band 4)\\
flag\_cannon & & Flags for spectrum information in a bitmask format (see \citealp{2018MNRAS.478.4513B})\\
teff1 & K & Effective temperature - primary component\\
teff2 & K & Effective temperature - secondary component\\
logg1 & dex & Surface gravity - primary component\\
logg2 & dex & Surface gravity - secondary component\\
feh & dex & Iron abundance ([Fe/H])\\
V1 & km\,s$^{-1}$ & RV - primary component\\
V2 & km\,s$^{-1}$ & RV - secondary component\\
ratio1 & & Ratio of spectral flux in GALAH band 1 (secondary/primary)\\
ratio2 & & Ratio of spectral flux in GALAH band 2 (secondary/primary)\\
ratio3 & & Ratio of spectral flux in GALAH band 3 (secondary/primary)\\
ratio4 & & Ratio of spectral flux in GALAH band 4 (secondary/primary)\\
vmic1 & km\,s$^{-1}$ & Microturbulence velocity - primary component\\
vmic2 & km\,s$^{-1}$ & Microturbulence velocity - secondary component\\
vsini1 & km\,s$^{-1}$ & Line-of-sight rotational velocity - primary component\\
vsini2 & km\,s$^{-1}$ & Line-of-sight rotational velocity - secondary component\\
reddening & & Interstellar reddening\\
r1 & $R_{\odot}$ & Stellar radius - primary component\\
r2 & $R_{\odot}$ & Stellar radius - secondary component\\
chi2\_binary\_pipeline & & $\chi^2_{\rm obj} \left( \chi^2_{\rm spec} + \chi^2_{\rm phot} + \chi^2_{\rm astro} \right)$, see Sect.~\ref{sec:model}\\
chi2\_binary\_pipeline\_spec & & $\chi^2_{\rm spec}$ (see Sect.~\ref{sec:model})\\
chi2\_binary\_pipeline\_phot & & $\chi^2_{\rm phot}$ (see Sect.~\ref{sec:model})\\
ruwe & & Renormalised unit weight error (RUWE, Gaia DR2)\\
\hline
\end{tabular}
\end{table*}

\section{Higher order of {\it The Cannon}}

\label{app:cannon}

\begin{figure*}[!htp]
   \centering
   \includegraphics[width=0.49\linewidth]{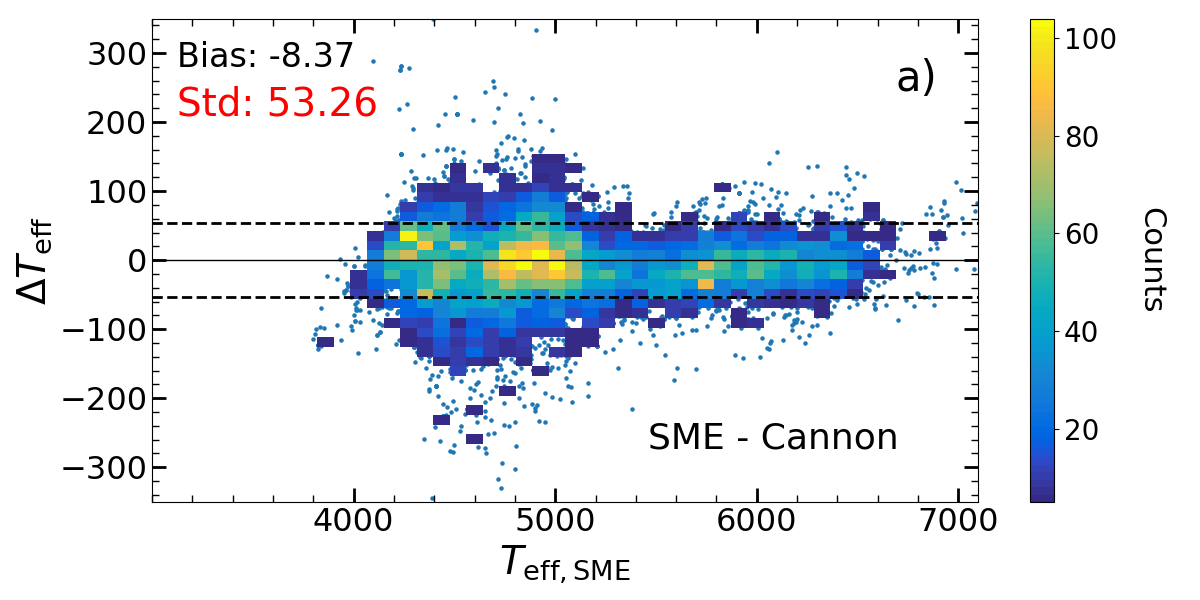}
   \includegraphics[width=0.49\linewidth]{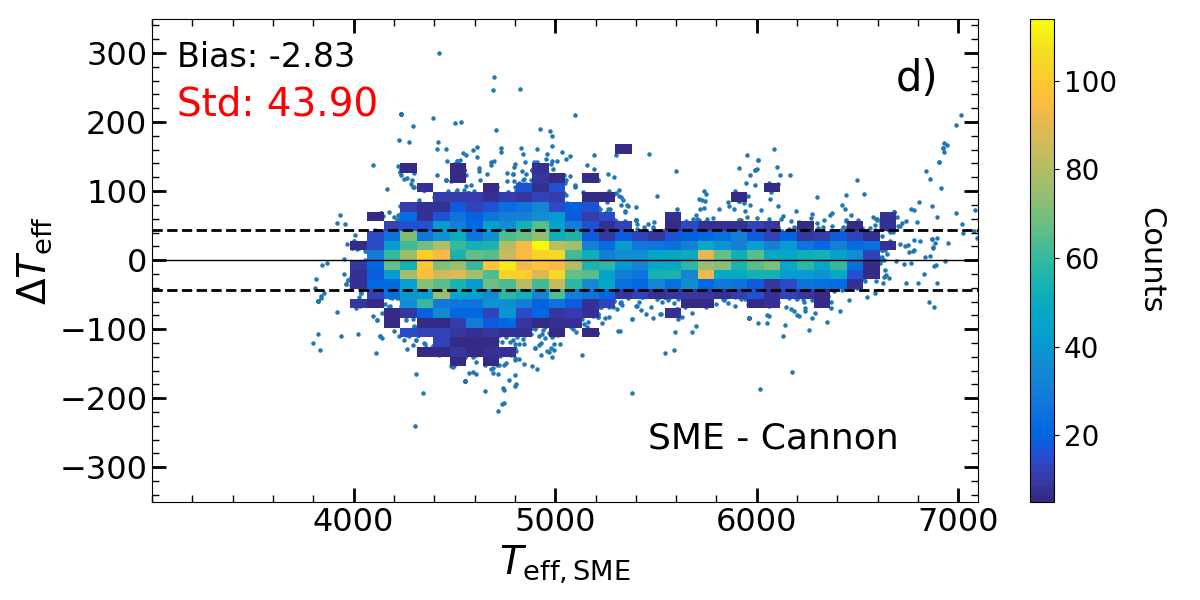}\\
   \includegraphics[width=0.49\linewidth]{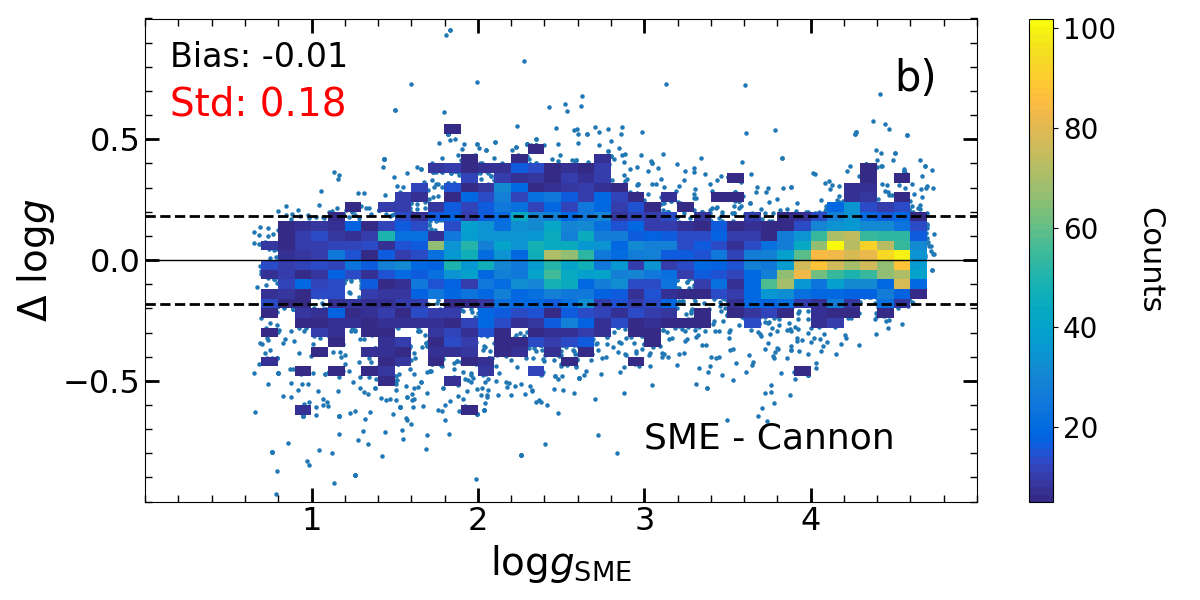}
   \includegraphics[width=0.49\linewidth]{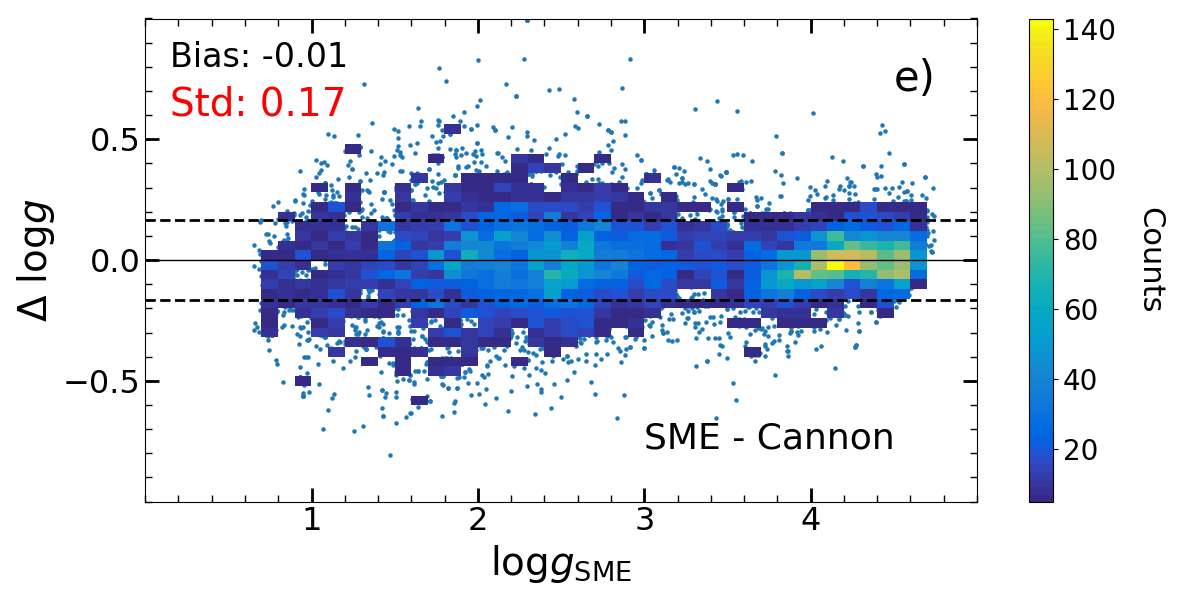}\\
   \includegraphics[width=0.49\linewidth]{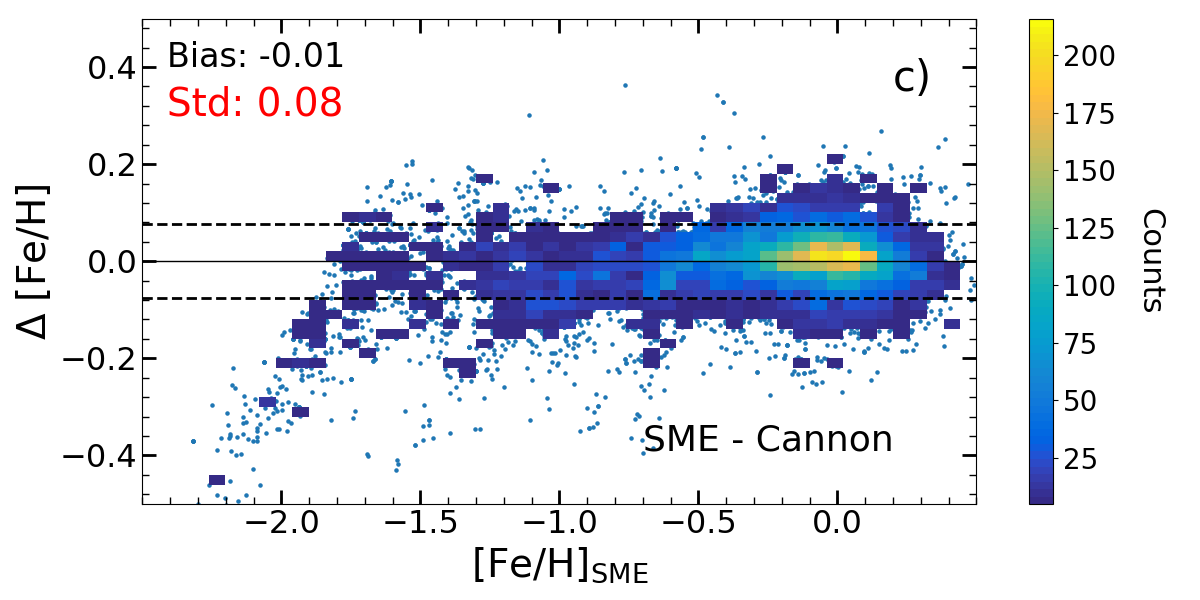}    
   \includegraphics[width=0.49\linewidth]{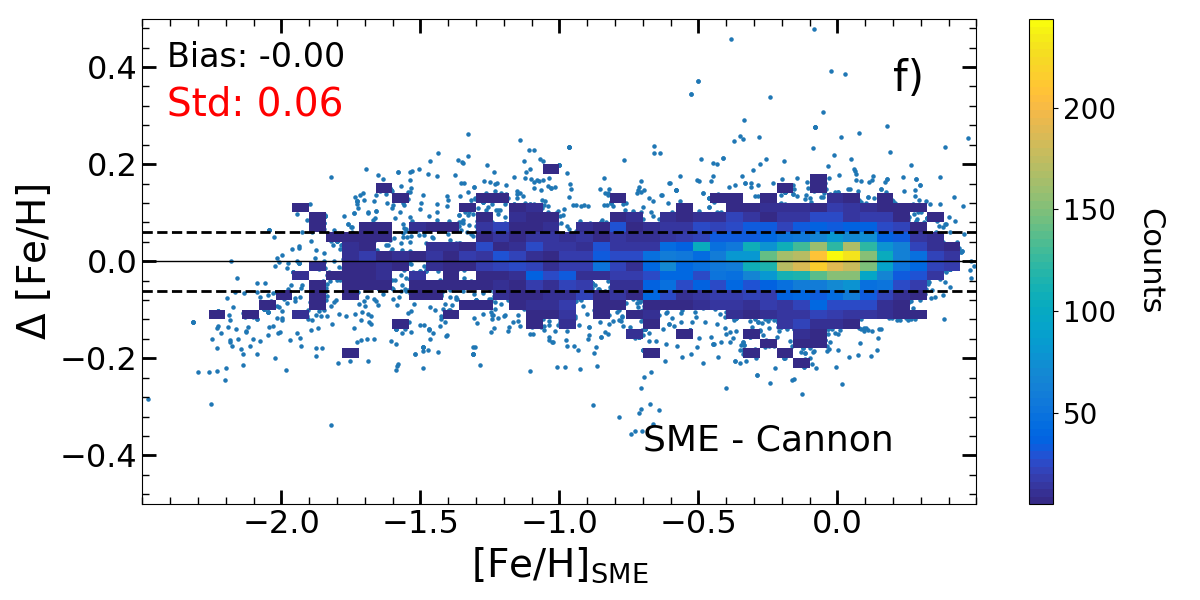} 
   \caption{Comparison of SME and {\it The Cannon} label values for the {\it \textup{training set}} (single-star spectra). Panels a), b), and c) are similar to Fig.~8 in \citet{2018MNRAS.478.4513B}, but we also include a comparison to the model generated with labels in the third order shown in panels d), e), and f).}
   \label{cannons}
\end{figure*}

Figure \ref{cannons} shows a comparison of SME derived labels and the labels reproduced by {\it The Cannon} interpolation for the {\it \textup{training set}}, and demonstrates that augmenting the labels to the third order improves {\it The Cannon} performance. This trend continues with the fourth order as expected because of the increased flexibility (complexity) of the model, but a more thorough investigation of {\it The Cannon} of higher orders is required to determine the limits of over-fitting, for instance, or issues with extrapolation in a model that is too complex.

We only show plots of the three basic stellar parameters in Fig.\,\ref{cannons}, but six parameters were used in generating {\it The Cannon} model ($T_{\rm eff}$, $\log g$, [Fe/H], $v_{\rm mic}$, $v_{\rm broad}$, and $A_{Ks}$), in accordance with the main GALAH pipeline. The values of the bias and standard deviation in Fig.\,\ref{cannons} (left column) are an indication of the internal precision of {\it The Cannon} model we used, and they agree with those reported in \cite{2018MNRAS.478.4513B} (see their Fig.~8) for all six parameters. We note that these values should always be appropriately incorporated in the presentation of final results obtained by inference that includes {\it The Cannon} method. 

\section{t-SNE map}
\label{app:tsne}

\begin{figure*}[!htp]
   \centering
   \includegraphics[width=\linewidth]{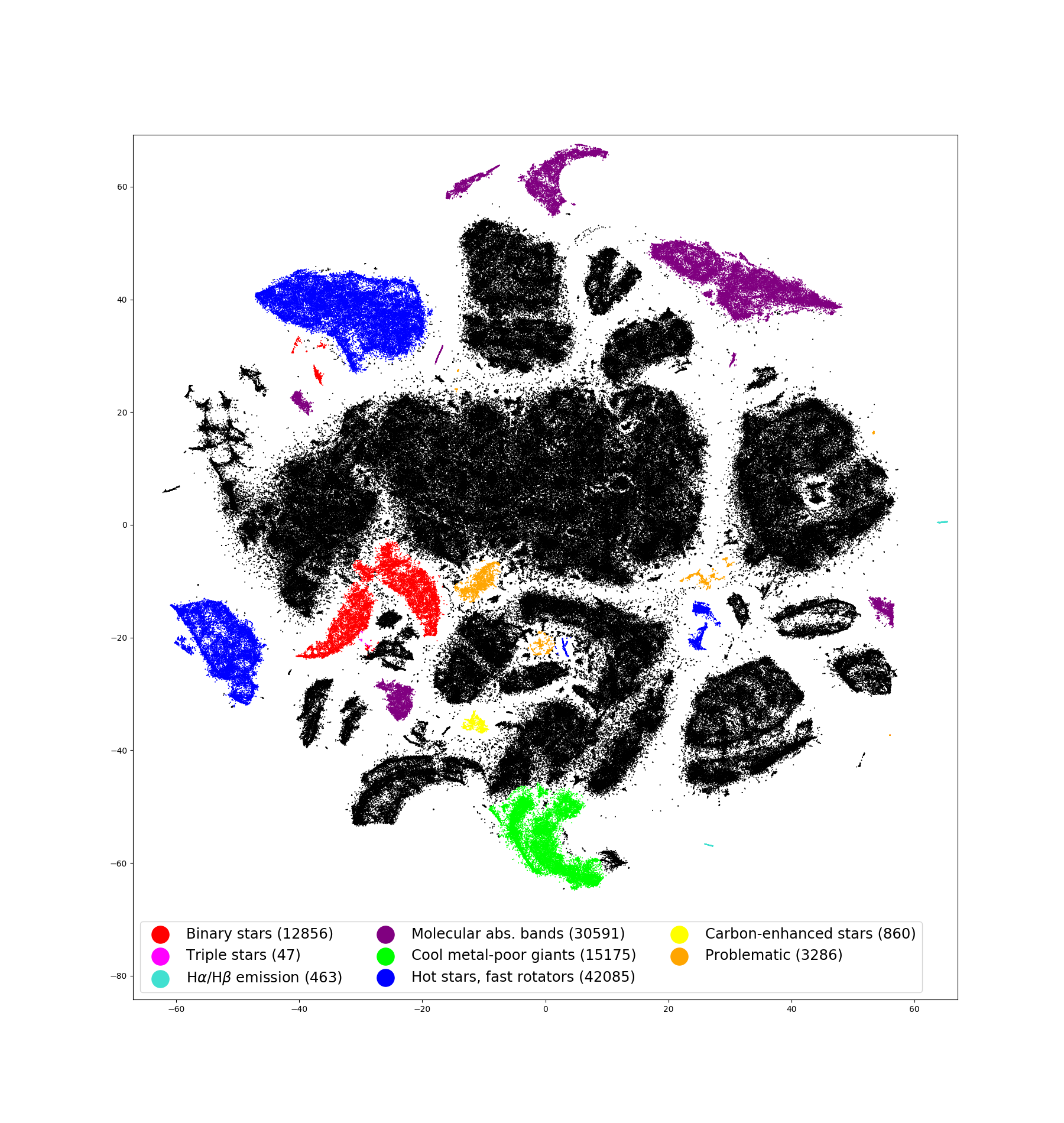}
   \caption{t-SNE projection map of 587\,153 GALAH spectra. The points (spectra) are colour-coded by classification category. The count of spectra for each category is given in the legend. Black points denote normal, well-behaved spectra.}
   \label{sol}
\end{figure*}

Figure \ref{sol} shows the t-SNE map (presented in \citealt{2018MNRAS.478.4513B}) that was used to identify binary star candidates in addition to the CCF method. Only a few dozen triple stars are flagged, and they are hardly visible next to the larger group of binary stars to the left, whereas additional smaller groups of binaries are located in the top left corner in the figure. 

\section{Formal uncertainties of the binary pipeline}
\label{app:unc}

\begin{figure*}[!htp]
   \centering
   \includegraphics[width=\linewidth]{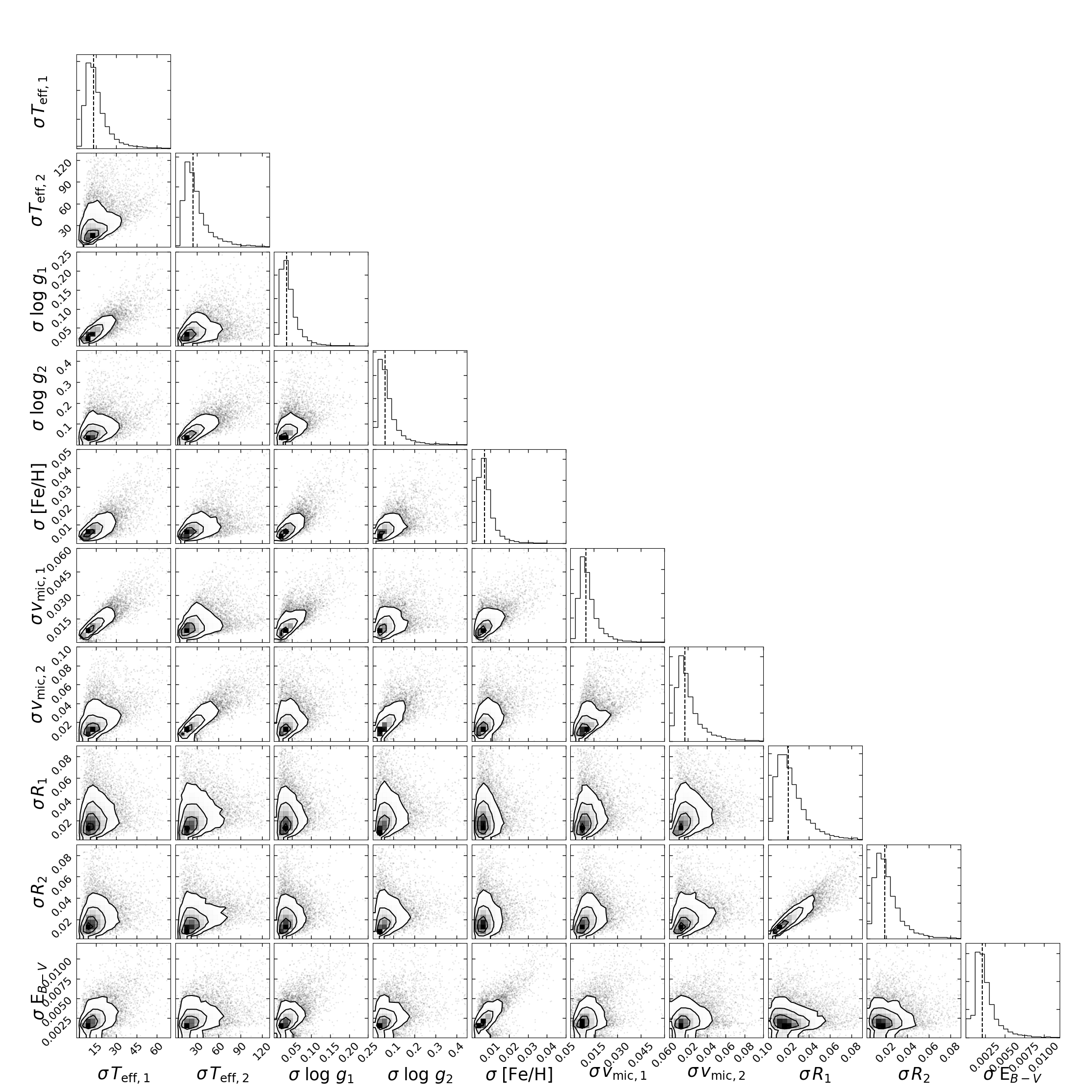}
   \caption{An ensemble of uncertainties of $\theta$ parameters for the final sample of 12\,760 analysed binary systems (see also Fig.\,\ref{mcmcres}). The uncertainty distributions are shown with diagonal panels, and the other panels indicate their correlations. The uncertainty units are those of the corresponding parameters (see Table~\ref{modpar}). The tails of the distributions are truncated for clarity.}
   \label{mcmceres}
\end{figure*}

When the results from our catalogue of binary star solutions are used, formal uncertainties (distributions shown in in Fig.~\ref{mcmceres}) should be taken in consideration together with validation tests in Sects. \ref{sec:validation} and \ref{sec:repeats}.

\section{Photometric validation tests}
\label{app:phot}

\begin{figure*}[!htp]
   \centering
   \includegraphics[width=\linewidth]{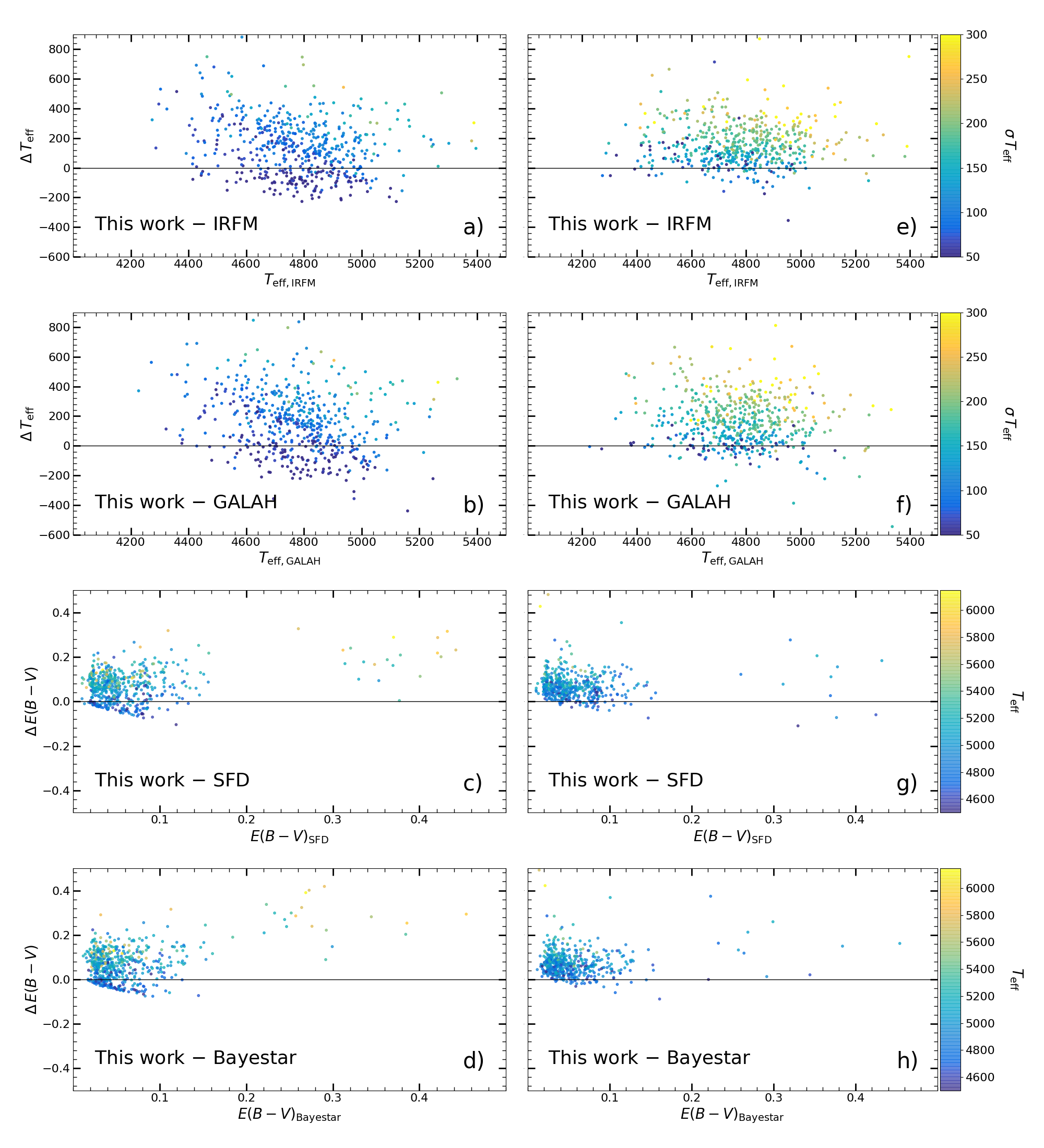}
   \caption{Comparison of $T_{\rm eff}$ and $E(B-V)$ derived here to the literature values (see text) for the benchmark validation sample. Panels e-h show results for the special case where we exclude \textit{Gaia} magnitudes from the fit. The colour-coding is by $T_{\rm eff}$ (panels c, d, g, and  h) and the formal uncertainty on $T_{\rm eff}$ (panels a, b, e, and f). The label ``GALAH'' stands for GALAH DR2 published values.}
   \label{fig:degent_ext}
\end{figure*}

Figure \ref{fig:degent_ext} shows results of the photometric validation for the benchmark validation sample of stars. The grid effects are especially clearly visible in the temperature comparison, where \textit{Gaia} magnitudes are included in the fit.

\end{appendix}

\end{document}